\documentclass[12pt]{article}
\usepackage{amsmath,amsthm}
\usepackage{eucal}
\usepackage{eufrak}

\DeclareMathAlphabet{\bb}{U}{msb}{m}{n}
\gdef\C{\bb C}
\gdef\dZ{\bb Z}

\gdef\dS{\bb S}
\gdef\R{\bb R}
\gdef\K{\bb K}
\gdef\BH{\bb H}
\gdef\F{\bb F}
\gdef\dO{\bb O}

\DeclareMathOperator{\End}{End}
\DeclareMathOperator{\spin}{{\bf Spin}}
\DeclareMathOperator{\pin}{{\bf Pin}}
\DeclareMathOperator{\fD}{\mathfrak{D}}
\DeclareMathOperator{\Id}{Id}

\DeclareMathOperator{\Aut}{Aut}
\DeclareMathOperator{\sAut}{{\sf Aut}}
\DeclareMathOperator{\sExt}{{\sf Ext}}
\DeclareMathOperator{\Ker}{Ker}

\DeclareMathOperator{\Ext}{Ext}

\newcommand{\s}{\!}

\newcommand{\cA}{\mathcal{A}}
\newcommand{\cB}{\mathcal{B}}

\newcommand{\cD}{\mathcal{D}}
\newcommand{\cE}{\mathcal{E}}
\newcommand{\cN}{\mathcal{N}}

\newcommand{\M}{{\bf\sf M}}

\newcommand{\sA}{{\sf A}}

\newcommand{\sI}{{\sf I}}
\newcommand{\sW}{{\sf W}}
\newcommand{\sE}{{\sf E}}
\newcommand{\sC}{{\sf C}}
\newcommand{\sF}{{\sf F}}
\newcommand{\sT}{{\sf T}}
\newcommand{\sS}{{\sf S}}

\newcommand{\sK}{{\sf K}}

\newcommand{\bi}{{\bf i}}
\newcommand{\bj}{{\bf j}}
\newcommand{\bk}{{\bf k}}
\newcommand{\bx}{{\bf x}}

\newcommand{\bZ}{{\bf Z}}

\newcommand{\fC}{\mathfrak{C}}
\newcommand{\fR}{\mathfrak{R}}
\newcommand{\fH}{\mathfrak{H}}
\newcommand{\fO}{\mathfrak{O}}

\newcommand{\Lip}{\boldsymbol{\Gamma}}

\newcommand{\cl}{C\kern -0.2em \ell}

\newcommand{\p}{\prime}
\newcommand{\e}{\mbox{\bf e}}

\newtheorem{theorem}{Theorem}

\newtheorem{prop}{Proposition}

\begin{document}
\title{UNIVERSAL COVERINGS OF THE ORTHOGONAL GROUPS}
\author{V.~V. Varlamov\\
{\small\it Department of Mathematics, Siberia State University of Industry,}\\
{\small\it Kirova 42, Novokuznetsk 654007, Russia}}
\date{}
\maketitle
\begin{abstract}
Universal coverings of the orthogonal groups and their extensions are studied
in terms of Clifford-Lipschitz groups.
An algebraic description of basic discrete symmetries (space inversion $P$,
time reversal $T$, charge conjugation $C$ and their combinations $PT$, $CP$,
$CT$, $CPT$) is given. Discrete subgroups $\{1,\,P,\,T,\,PT\}$ of
orthogonal groups of multidimensional spaces over the fields of real and
complex numbers are considered in terms of fundamental automorphisms of
Clifford algebras. The fundamental automorphisms form a finite group of
order 4. The charge conjugation is represented by a 
complex conjugation pseudoautomorphism
of the Clifford algebra. Such a description allows one to extend the
automorphism group. It is shown that an extended automorphism group
($CPT$-group) forms a finite group of order 8. The group structure and
isomorphisms between the extended automorphism groups and finite groups
are studied in detail. It is proved that there exist 64 different
realizations of $CPT$-group. An extension of universal coverings
(Clifford-Lipschitz groups) of the orthogonal groups is given in terms
of $CPT$-structures which include well-known Shirokov-D\c{a}browski
$PT$-structures as a particular case.
Quotient Clifford-Lipschitz groups and quotient representations are
introduced. It is shown that a complete classification of the quotient
groups depends on the structure of various subgroups of the
extended automorphism group.
\end{abstract}

\section{Introduction}
From historical point of view Clifford algebras have essentially geometric
origin \cite{3}, because they are the synthesis of Hamilton quaternion
calculus \cite{Ham} and Grassmann {\it Ausdehnungslehre}
\cite{Grass}, and by this reason they called by Clifford as
{\it geometric algebras} \cite{Cliff2}. Further, Lipschitz
\cite{Lips} showed that Clifford algebras closely related with the study
of the rotation groups of multidimensional spaces. After fundamental works of
Cartan \cite{Car08}, Witt \cite{Wit37} and Chevalley \cite{Che54}
the Clifford algebra theory takes its modern form
\cite{Cru91,Port95,Lou97}. 
The $\pin$ and $\spin$ groups (Clifford-Lipschitz groups) widely used in
algebraic topology \cite{BH,Hae56,AtBSh,Kar68,Kar,KT89},
in the definition of pinor and spinor structures on the riemannian 
manifolds \cite{Mil63,Ger68,Ish78,Wh78,DT86,DP87,LM89,DR89,Cru91,AlCh94,
Ch94a,Ch94,AlCh96},
spinor bundles \cite{RF90,RO90,FT96,Fr98,FT99},
and also have great importance in the theory of the Dirac operator on
manifolds \cite{Bau81,Bar91,Tr92,Fr97,Amm98}.
The Clifford-Lipschitz groups also intensively used in theoretical
physics \cite{CDD82,DW90,FRO90a,RS93,DWGK,BDGK,Ch97}.

Universal coverings of the orthogonal groups allow us to describe discrete
symmetries on an equal footing, from the one group theoretical viewpoint.
Importance of discrete transformations is well--known, many textbooks
on quantum theory began with description of the discrete symmetries, and
famous L\"{u}ders--Pauli $CPT$--Theorem is a keystone of quantum field
theory. 

However, usual practice of definition of the discrete symmetries
from the analysis of relativistic wave equations does not give a full and
consistent theory of the discrete transformations. In the standard approach,
except a well studied case of the spin $j=1/2$ (Dirac equation), a situation
with the discrete symmetries remains unclear for the fields of higher spin
$j>1/2$. It is obvious that the main reason of this is an absence of a fully
adequate formalism for description of higher--spin fields (all widely
accepted higher--spin formalisms such as Rarita--Schwinger approach \cite{RS41},
Bargmann--Wigner \cite{BW48} and Gel'fand--Yaglom \cite{GY48} multispinor
theories, and also Joos--Weinberg $2(2j+1)$--component formalism 
\cite{Joo62,Wein} have many intrinsic contradictions and difficulties).
Moreover, Lee and Wick \cite{LW66} claimed that ``the situation is clearly
an unsatisfactory one from a fundamental point of view".
The first attempt of going out from this situation was initiated by
Gel'fand, Minlos and Shapiro in 1958 \cite{GMS}. In the
Gel'fand--Minlos--Shapiro approach the discrete symmetries are represented
by outer involutory automorphisms of the Lorentz group (there are also other
realizations of the discrete symmetries via the outer automorphisms, see
\cite{Mic64,Kuo71,Sil92}).
At present, the
Gel'fand--Minlos--Shapiro ideas have been found further development in the
works of Buchbinder, Gitman and Shelepin \cite{BGS00,GS01}, where the
discrete symmetries are represented by both outer and inner automorphisms of
the Poincar\'{e} group.

In 1909, Minkowski showed \cite{Min} that a causal structure of the world
is described by a 4--dimensional pseudo--Euclidean geometry. In accordance
with \cite{Min} the quadratic form $x^2+y^2+z^2-c^2t^2$ remains invariant
under the action of linear transformations of the four variables $x,y,z$ and 
$t$,
which form a general Lorentz group $G$. As known, the general Lorentz group
$G$ consists of a proper orthochronous Lorentz group $G_0$ and three reflections
(discrete transformations) $P,\,T,\,PT$, where $P$ and $T$ are space and
time reversal, and $PT$ is a so--called full reflection. The discrete
transformations $P,\,T$ and $PT$ added to an identical transformation
form a finite group. Thus, the general Lorentz group may be represented by
a semidirect product $G_0\odot\{1,P,T,PT\}$. 

In 1958, Shirokov pointed out \cite{Shi58,Shi60} that an universal covering of the
inhomogeneous Lorentz group has eight inequivalent realizations. Later on,
in the eighties this idea was applied to a general orthogonal group
$O(p,q)$ by D\c{a}browski \cite{Dab88}.
As known, the orthogonal
group $O(p,q)$ of the real space $\R^{p,q}$ is represented by the semidirect
product of a connected component $O_0(p,q)$ and a discrete subgroup
$\{1,P,T,PT\}$.
Further, a double covering of the orthogonal group $O(p,q)$ is a
Clifford--Lipschitz group $\pin(p,q)$ which is completely constructed within
a Clifford algebra $\cl_{p,q}$. In accordance with squares of elements of the
discrete subgroup ($a=P^2,\,b=T^2,\,c=(PT)^2$) there exist eight double
coverings (D\c{a}browski groups \cite{Dab88}) of the orthogonal group
defining by the signatures $(a,b,c)$, where $a,b,c\in\{-,+\}$. Such in brief is
a standard description scheme of the discrete transformations.
However, in this scheme there is one essential flaw. Namely, the
Clifford--Lipschitz group is an intrinsic notion of the algebra $\cl_{p,q}$
(a set of the all invertible elements of $\cl_{p,q}$), whereas the discrete
subgroup is introduced into the standard scheme in an external way, and the
choice of the signature $(a,b,c)$ of the discrete subgroup is not
determined by the signature of the space $\R^{p,q}$. Moreover, it is suggest
by default that for any signature $(p,q)$ of the vector space there exist
the all eight kinds of the discrete subgroups. It is obvious that a
consistent description of the universal coverings of $O(p,q)$ in terms of
the Clifford--Lipschitz groups $\pin(p,q)\subset\cl_{p,q}$ can be obtained
only in the case when the discrete subgroup $\{1,P,T,PT\}$ is also defined
within the algebra $\cl_{p,q}$. Such a description has been given in the
works \cite{Var99,Var00,Var03}, where the discrete symmetries are
represented by fundamental automorphisms of the Clifford algebras.
So, the space inversion $P$, time reversal $T$ and their
combination $PT$ correspond to an automorphism $\star$
(involution), an antiautomorphism $\widetilde{\phantom{cc}}$ (reversion) and
an antiautomorphism $\widetilde{\star}$ (conjugation), respectively.
The fundamental automorphisms of the Clifford algebras
are compared to elements of the finite group formed by the discrete
transformations. In turn, a set of the fundamental automorphisms,
added by an identical automorphism, forms a finite group 
$\Aut(\cl)$, for which in virtue of the Wedderburn--Artin Theorem
there exists a matrix representation. In such a way, an isomorphism
$\{1,P,T,PT\}\simeq\Aut(\cl)$ plays a central role and allows us
to use methods of the
Clifford algebra theory at the study of a group theoretical structure
of the discrete transformations. First of all, it allows one to classify the
discrete groups into Abelian 
$\dZ_2\otimes\dZ_2$, $\dZ_4$ and non--Abelian
$D_4$, $Q_4$ finite groups, and also to establish a dependence between the
finite groups and signature of the spaces in case of real numbers.
It is shown that
the division ring structure of $\cl_{p,q}$ imposes hard restrictions on
existence and choice of the discrete subgroup, and the signature
$(a,b,c)$ depends upon the signature of the underlying space $\R^{p,q}$.
Moreover, this description allows us to incorporate the
Gel'fand--Minlos--Shapiro automorphism theory into 
Shirokov--D\c{a}browski scheme and further to unite them on the basis
of the Clifford algebra theory.

Other important discrete symmetry is the charge conjugation $C$. In contrast
with the transformations $P$, $T$, $PT$, the operation $C$ is not
space--time discrete symmetry. This transformation is firstly appearred
on the representation spaces of the Lorentz group and its nature is
strongly different from other discrete symmetries. For that reason in this
work the charge conjugation $C$ is represented by a
pseudoautomorphism $\cA\rightarrow\overline{\cA}$ which is not fundamental
automorphism of the Clifford algebra. All spinor representations of the
pseudoautomorphism $\cA\rightarrow\overline{\cA}$ are given in
Theorem \ref{tpseudo}. An introduction of the transformation
$\cA\rightarrow\overline{\cA}$ allows us to extend the automorphism group
$\Aut(\cl)$ of the Clifford algebra. It is shown that automorphisms
$\cA\rightarrow\cA^\star$, $\cA\rightarrow\widetilde{\cA}$,
$\cA\rightarrow\widetilde{\cA^\star}$, $\cA\rightarrow\overline{\cA}$,
$\cA\rightarrow\overline{\cA^\star}$, 
$\cA\rightarrow\overline{\widetilde{\cA}}$ and
$\cA\rightarrow\overline{\widetilde{\cA^\star}}$ form a finite group of
order 8 (an extended automorphism group $\Ext(\cl)=\{\Id,\star,
\widetilde{\phantom{cc}},\widetilde{\star},\overline{\phantom{cc}},
\overline{\star},\overline{\widetilde{\phantom{cc}}},
\overline{\widetilde{\star}}\}$). The group $\Ext(\cl)$ is isomorphic
to a $CPT$-group $\{1,P,T,PT,C,CP,CT,CPT\}$. There exist isomorphisms
between $\Ext(\cl)$ and finite groups of order 8. A full number of different
realizations of $\Ext(\cl)$ is equal to 64. This result allows us to define
extended universal coverings ($CPT$-structures) of the orthogonal groups.
It is shown that the eight Shirokov-D\c{a}browski $PT$-structures present
a particular case of general $CPT$-structures.

Quotient representations of the group $\pin(n,\C)$ and $\pin(p,q)$
correspond to the types $n\equiv 1\pmod{2}$
($\F=\C$) and $p-q\equiv 1,5\pmod{8}$ ($\F=\R$), respectively. An explicit form
of the quotient representations is given in
Theorem \ref{tfactor}. It is proved that a classification of the quotient
groups depends on the structure of different subgroups of $\Ext(\cl)$.

\section{Clifford algebras and discrete symmetries}
In this section we will consider some basic facts concerning Clifford algebras
\index{algebra!Clifford}
which we will widely use below.
Let $\F$ be a field of characteristic 0 $(\F=\R,\,\F=\C)$, 
\index{field!of characteric 0}
where
$\R$ and $\C$ are the fields of real and complex numbers, respectively.
A Clifford algebra $\cl$ over a field $\F$ 
is an algebra with
$2^n$ basis elements\index{element!basis}: $\e_0$
(unit of the algebra) $\e_1,\e_2,\ldots,\e_n$ and products of the one--index
elements $\e_{i_1i_2\ldots i_k}=\e_{i_1}\e_{i_2}\ldots\e_{i_k}$.
Over the field $\F=\R$ the Clifford algebra is denoted as $\cl_{p,q}$, where
the indices
$p$ and $q$ correspond to the indices of the 
quadratic form\index{form!quadratic}
\[
Q=x^2_1+\ldots+x^2_p-\ldots-x^2_{p+q}
\]
of a vector space\index{space!vector}
$V$ associated with $\cl_{p,q}$. The multiplication law
of $\cl_{p,q}$ is defined by a following rule:
\begin{equation}\label{e1}
\e^2_i=\sigma(p-i)\e_0,\quad\e_i\e_j=-\e_j\e_i,
\end{equation}
where
\begin{equation}\label{e2}
\sigma(n)=\left\{\begin{array}{rl}
-1 & \mbox{if $n\leq 0$},\\
+1 & \mbox{if $n>0$}.
\end{array}\right.
\end{equation}
The square of a volume element\index{element!volume}
$\omega=\e_{12\ldots n}$ ($n=p+q$) plays an
important role in the theory of Clifford algebras,
\begin{equation}\label{e3}
\omega^2=\left\{\begin{array}{rl}
-1 & \mbox{if $p-q\equiv 2,3,6,7\pmod{8}$},\\
+1 & \mbox{if $p-q\equiv 0,1,4,5\pmod{8}$}.
\end{array}\right.
\end{equation}  
A center\index{center}
$\bZ_{p,q}$ of the algebra $\cl_{p,q}$ consists of the unit $\e_0$ 
and the volume element $\omega$. The element $\omega=\e_{12\ldots n}$ 
belongs to a center when $n$ is odd. Indeed,
\begin{eqnarray}
\e_{12\ldots n}\e_i&=&(-1)^{n-i}\sigma(q-i)\e_{12\ldots i-1 i+1\ldots n},
\nonumber\\
\e_i\e_{12\ldots n}&=&(-1)^{i-1}\sigma(q-i)\e_{12\ldots i-1 i+1\ldots n},
\nonumber
\end{eqnarray}
therefore, $\omega\in\bZ_{p,q}$ if and only if $n-i\equiv i-1\pmod{2}$, 
that is, $n$ is odd. Further, using (\ref{e3}) we obtain
\begin{equation}\label{e4}
\bZ_{p,q}=\left\{\begin{array}{rl}
\phantom{1,}1 & \mbox{if $p-q\equiv 0,2,4,6\pmod{8}$},\\
1,\omega & \mbox{if $p-q\equiv 1,3,5,7\pmod{8}$}.
\end{array}\right.
\end{equation}

An arbitrary element $\cA$ of the algebra $\cl_{p,q}$ is represented by a
following formal polynomial
\begin{gather}
\cA=a^0\e_0+\sum^n_{i=1}a^i\e_i+\sum^n_{i=1}\sum^n_{j=1}a^{ij}\e_{ij}+
\ldots+\sum^n_{i_1=1}\cdots\sum^n_{i_k=1}a^{i_1\ldots i_k}\e_{i_1\ldots i_k}+
\nonumber\\
+\ldots+a^{12\ldots n}\e_{12\ldots n}=\sum^n_{k=0}a^{i_1i_2\ldots i_k}
\e_{i_1i_2\ldots i_k}.\nonumber
\end{gather}

In Clifford algebra $\cl$ there exist four fundamental automorphisms.\\[0.2cm]
1) {\bf Identity}: An automorphism $\cA\rightarrow\cA$ and 
$\e_{i}\rightarrow\e_{i}$.\\
This automorphism is an identical automorphism of the algebra $\cl$. 
$\cA$ is an arbitrary element of $\cl$.\\[0.2cm]
2) {\bf Involution}: An automorphism $\cA\rightarrow\cA^\star$ and 
$\e_{i}\rightarrow-\e_{i}$.\\
In more details, for an arbitrary element $\cA\in\cl$ there exists a
decomposition
$
\cA=\cA^{\p}+\cA^{\p\p},
$
where $\cA^{\p}$ is an element consisting of homogeneous odd elements, and
$\cA^{\p\p}$ is an element consisting of homogeneous even elements,
respectively. Then the automorphism
$\cA\rightarrow\cA^{\star}$ is such that the element
$\cA^{\p\p}$ is not changed, and the element $\cA^{\p}$ changes sign:
$
\cA^{\star}=-\cA^{\p}+\cA^{\p\p}.
$
If $\cA$ is a homogeneous element, then
\begin{equation}\label{auto16}
\cA^{\star}=(-1)^{k}\cA,
\end{equation}
where $k$ is a degree of the element. It is easy to see that the
automorphism $\cA\rightarrow\cA^{\star}$ may be expressed via the volume
element $\omega=\e_{12\ldots p+q}$:
\begin{equation}\label{auto17}
\cA^{\star}=\omega\cA\omega^{-1},
\end{equation}
where
$\omega^{-1}=(-1)^{\frac{(p+q)(p+q-1)}{2}}\omega$. When $k$ is odd, the basis
elements 
$\e_{i_{1}i_{2}\ldots i_{k}}$ the sign changes, and when $k$ is even, the sign
is not changed.\\[0.2cm]
3) {\bf Reversion}: An antiautomorphism $\cA\rightarrow\widetilde{\cA}$ and
$\e_i\rightarrow\e_i$.\\
The antiautomorphism $\cA\rightarrow\widetilde{\cA}$ is a reversion of the
element $\cA$, that is the substitution of each basis element
$\e_{i_{1}i_{2}\ldots i_{k}}\in\cA$ by the element
$\e_{i_{k}i_{k-1}\ldots i_{1}}$:
\[
\e_{i_{k}i_{k-1}\ldots i_{1}}=(-1)^{\frac{k(k-1)}{2}}
\e_{i_{1}i_{2}\ldots i_{k}}.
\]
Therefore, for any $\cA\in\cl_{p,q}$ we have
\begin{equation}\label{auto19}
\widetilde{\cA}=(-1)^{\frac{k(k-1)}{2}}\cA.
\end{equation}
4) {\bf Conjugation}: An antiautomorphism $\cA\rightarrow\widetilde{\cA^\star}$
and $\e_i\rightarrow-\e_i$.\\
This antiautomorphism is a composition of the antiautomorphism
$\cA\rightarrow\widetilde{\cA}$ with the automorphism
$\cA\rightarrow\cA^{\star}$. In the case of a homogeneous element from
the formulae (\ref{auto16}) and (\ref{auto19}), it follows
\begin{equation}\label{20}
\widetilde{\cA^{\star}}=(-1)^{\frac{k(k+1)}{2}}\cA.
\end{equation}
As known, the complex algebra $\C_n$ is associated with a complex vector
space $\C^n$. Let $n=p+q$, then an extraction operation of the real subspace
$\R^{p,q}$ in $\C^n$  forms the foundation of definition of the discrete
transformation known in physics as
{\it a charge conjugation} $C$. Indeed, let
$\{\e_1,\ldots,\e_n\}$ be an orthobasis in the space $\C^n$, $\e^2_i=1$.
Let us remain the first $p$ vectors of this basis unchanged, and other $q$
vectors multiply by the factor $i$. Then the basis
\begin{equation}\label{6.23}
\left\{\e_1,\ldots,\e_p,i\e_{p+1},\ldots,i\e_{p+q}\right\}
\end{equation}
allows one to extract the subspace $\R^{p,q}$ in $\C^n$. Namely,
for the vectors $\R^{p,q}$ we take the vectors of
$\C^n$ which decompose on the basis
(\ref{6.23}) with real coefficients. In such a way we obtain a real vector
space $\R^{p,q}$ endowed (in general case) with a non--degenerate
quadratic form\index{form!quadratic}
\[
Q(x)=x^2_1+x^2_2+\ldots+x^2_p-x^2_{p+1}-x^2_{p+2}-\ldots-x^2_{p+q},
\]
where $x_1,\ldots,x_{p+q}$ are coordinates of the vector $\bx$ 
in the basis (\ref{6.23}).
It is easy to see that the extraction of
$\R^{p,q}$ in $\C^n$ induces an extraction of
{\it a real subalgebra}\index{subalgebra!real}
$\cl_{p,q}$ in $\C_n$. Therefore, any element
$\cA\in\C_n$ can be unambiguously represented in the form
\[
\cA=\cA_1+i\cA_2,
\]
where $\cA_1,\,\cA_2\in\cl_{p,q}$. The one-to-one mapping
\begin{equation}\label{6.24}
\cA\longrightarrow\overline{\cA}=\cA_1-i\cA_2
\end{equation}
transforms the algebra $\C_n$ into itself with preservation of addition
and multiplication operations for the elements $\cA$; the operation of
multiplication of the element $\cA$ by the number transforms to an operation
of multiplication by the complex conjugate number.
Any mapping of $\C_n$ satisfying these conditions is called
{\it a pseudoautomorphism}.\index{pseudoautomorphism}
Thus, the extraction of the subspace
$\R^{p,q}$ in the space $\C^n$ induces in the algebra $\C_n$ 
a pseudoautomorphism $\cA\rightarrow\overline{\cA}$ \cite{Rash,Ras58}.

One of the most fundamental theorems in the theory of associative algebras
is
\begin{theorem}[{\rm Wedderburn--Artin}]
Any finite--dimensional associative simple algebra $\mathfrak{A}$ over the
field $\F$ is isomorphic to a full matrix algebra $\M_n(\K)$, where a
natural number $n$ defined unambiguously, and a division ring $\K$ defined
with an accuracy of isomorphism.
\end{theorem}
In accordance with this theorem all properties of the initial algebra
$\mathfrak{A}$ are isomorphically transferred to the matrix algebra 
$\M_n(\K)$. Later on we will widely use this theorem. In turn, for the
Clifford algebra $\cl_{p,q}$ over the field $\F=\R$ we have an isomorphism
$\cl_{p,q}\simeq\End_{\K}(I_{p,q})\simeq\M_{2^m}(\K)$, where $m=\frac{p+q}{2}$,
$I_{p,q}=\cl_{p,q}f$ is a minimal left ideal of $\cl_{p,q}$, and
$\K=f\cl_{p,q}f$ is a division ring of $\cl_{p,q}$. The primitive idempotent
of the algebra $\cl_{p,q}$ has a form
\[
f=\frac{1}{2}(1\pm\e_{\alpha_1})\frac{1}{2}(1\pm\e_{\alpha_2})\cdots\frac{1}{2}
(1\pm\e_{\alpha_k}),
\]
where $\e_{\alpha_1},\e_{\alpha_2},\ldots,\e_{\alpha_k}$ are commuting
elements with square 1 of the canonical basis of $\cl_{p,q}$ generating
a group of order $2^k$. The values of $k$ are defined by a formula
$k=q-r_{q-p}$, where $r_i$ are the Radon--Hurwitz numbers \cite{Rad22,Hur23},
values of which form a cycle of the period 8: $r_{i+8}=r_i+4$. The values of
all $r_i$ are
\begin{center}
\begin{tabular}{lcccccccc}
$i$  & 0 & 1 & 2 & 3 & 4 & 5 & 6 & 7\\ \hline
$r_i$& 0 & 1 & 2 & 2 & 3 & 3 & 3 & 3
\end{tabular}.
\end{center}
The all Clifford algebras $\cl_{p,q}$ over the field $\F=\R$ are divided
into eight different types with a following division ring structure:\\[0.3cm]
{\bf I}. Central simple algebras.
\begin{description}
\item[1)] Two types $p-q\equiv 0,2\pmod{8}$ with a division ring 
$\K\simeq\R$.
\item[2)] Two types $p-q\equiv 3,7\pmod{8}$ with a division ring
$\K\simeq\C$.
\item[3)] Two types $p-q\equiv 4,6\pmod{8}$ with a division ring
$\K\simeq\BH$.
\end{description}
{\bf II}. Semi--simple algebras.
\begin{description}
\item[4)] The type $p-q\equiv 1\pmod{8}$ with a double division ring
$\K\simeq\R\oplus\R$.
\item[5)] The type $p-q\equiv 5\pmod{8}$ with a double quaternionic 
division ring $\K\simeq\BH\oplus\BH$.
\end{description}
The Table 1 (Budinich--Trautman Periodic Table \cite{BT88})
explicitly shows a distribution of the real Clifford algebras in dependence
on the division ring structure\index{structure!division ring}, 
here ${}^2\R(n)=\R(n)\oplus\R(n)$ and
${}^2\BH(n)=\BH(n)\oplus\BH(n)$.
\begin{figure}\label{Periodic}
\begin{center}{\small {\bf Table 1.} Distribution of the real Clifford algebras.}
\end{center}
{\renewcommand{\arraystretch}{1.2}
\begin{tabular}{c|ccccccccl}
p  &  0 & 1 & 2 & 3 & 4 & 5 & 6 & 7 & \ldots\\ \hline
q &      &   &   &   &   &   &   &   &\\
0 &   $\R$&${}^2\R$&$\R(2)$&$\C(2)$&$\BH(2)$&${}^2\BH(2)$&$\BH(4)$&$\C(8)$&
$\ldots$\\
1&$\C$&$\R(2)$&${}^2\R(2)$&$\R(4)$&$\C(4)$&$\BH(4)$&${}^2\BH(4)$&$\BH(8)$&
$\ldots$\\
2&$\BH$&$\C(2)$&$\R(4)$&${}^2\R(4)$&$\R(8)$&$\C(8)$&$\BH(8)$&${}^2\BH(8)$&
$\ldots$\\
3&${}^2\BH$&$\BH(2)$&$\C(4)$&$\R(8)$&${}^2\R(8)$&$\R(16)$&$\C(16)$&$\BH(16)$&
$\ldots$\\
4&$\BH(2)$&${}^2\BH(2)$&$\BH(4)$&$\C(8)$&$\R(16)$&${}^2\R(16)$&$\R(32)$&
$\C(32)$&$\ldots$\\
5&$\C(4)$&$\BH(4)$&${}^2\BH(4)$&$\BH(8)$&$\C(16)$&$\R(32)$&${}^2\R(32)$&
$\R(64)$&$\ldots$\\
6&$\R(8)$&$\C(8)$&$\BH(8)$&${}^2\BH(8)$&$\BH(16)$&$\C(32)$&$\R(64)$&
${}^2\R(64)$&$\ldots$\\
7&${}^2\R(8)$&$\R(16)$&$\C(16)$&$\BH(16)$&${}^2\BH(16)$&$\BH(32)$&$\C(64)$&
$\R(128)$&$\ldots$\\
$\vdots$&$\vdots$&$\vdots$&$\vdots$&$\vdots$&$\vdots$&$\vdots$&$\vdots$&
$\vdots$
\end{tabular}}
\end{figure}

Over the field $\F=\C$ there is an isomorphism $\C_n\simeq\M_{2^{n/2}}(\C)$
and there are two different types of complex Clifford algebras $\C_n$:
$n\equiv 0\pmod{2}$ and $n\equiv 1\pmod{2}$.

When $\cl_{p,q}$ is simple, then the map
\begin{equation}\label{Simple}
\cl_{p,q}\overset{\gamma}{\longrightarrow}\End_{\K}(\dS),\quad
u\longrightarrow\gamma(u),\quad \gamma(u)\psi=u\psi
\end{equation}\begin{sloppypar}\noindent
gives an irreducible and faithful representation
of $\cl_{p,q}$ in the
spinspace
$\dS_{2^m}(\K)\simeq I_{p,q}=\cl_{p,q}f$, where $\psi\in\dS_{2^m}$,
$m=\frac{p+q}{2}$.\end{sloppypar}

On the other hand, when $\cl_{p,q}$ is semi-simple, then the map
\begin{equation}\label{Semi-Simple}
\cl_{p,q}\overset{\gamma}{\longrightarrow}\End_{\K\oplus\hat{\K}}
(\dS\oplus\hat{\dS}),\quad u\longrightarrow\gamma(u),\quad
\gamma(u)\psi=u\psi
\end{equation}
gives a faithful but reducible representation of $\cl_{p,q}$ in the double
spinspace\index{spinspace!double}
$\dS\oplus\hat{\dS}$, where $\hat{\dS}=\{\hat{\psi}|\psi\in\dS\}$.
In this case, the ideal $\dS\oplus\hat{\dS}$ possesses a right
$\K\oplus\hat{\K}$-linear structure, $\hat{\K}=\{\hat{\lambda}|\lambda\in\K\}$,
and $\K\oplus\hat{\K}$ is isomorphic to the double division ring
$\R\oplus\R$ when $p-q\equiv 1\pmod{8}$ or to $\BH\oplus\BH$ when
$p-q\equiv 5\pmod{8}$. The map $\gamma$ in (\ref{Simple}) and
(\ref{Semi-Simple}) defines the so called {\it left-regular} spinor
representation\index{representation!spinor}
of $\cl(Q)$ in $\dS$ and $\dS\oplus\hat{\dS}$, respectively.
Furthermore, $\gamma$ is {\it faithful} which means that $\gamma$ is an
algebra monomorphism. In (\ref{Simple}), $\gamma$ is {\it irreducible}
which means that $\dS$ possesses no proper (that is, $\neq 0,\,\dS$)
invariant subspaces\index{subspace!invariant}
under the left action of $\gamma(u)$, $u\in\cl_{p,q}$.
Representation $\gamma$ in (\ref{Semi-Simple}) is therefore
{\it reducible} since $\{(\psi,0)|\psi\in\dS\}$ and
$\{(0,\hat{\psi})|\hat{\psi}\in\hat{\dS}\}$ are two proper subspaces of
$\dS\oplus\hat{\dS}$ invariant under $\gamma(u)$ 
(see \cite{Lou97,Cru91,Port95}).
\subsection{Salingaros groups}\index{group!Salingaros}
\label{Sec:1.3}
The structure of the Clifford algebras admits a very elegant description
in terms of finite groups \cite{Sal81a,Sal82,Sal84}. In accordance with
the multiplication law (\ref{e1}) the basis elements of $\cl_{p,q}$ form
a finite group\index{group!finite} of order $2^{n+1}$,
\begin{equation}\label{FG}
G(p,q)=\left\{\pm 1,\,\pm\e_i,\,\pm\e_i\e_j,\,\pm\e_i\e_j\e_k,\,\ldots,\,
\pm\e_1\e_2\cdots\e_n\right\}\quad(i<j<k<\ldots).
\end{equation}

Salingaros showed \cite{Sal81a,Sal82} that there exist five distinct types
of finite groups (\ref{FG}) that arise from Clifford algebras.
In \cite{Sal81a,Sal82} they were called `vee groups' and were labelled as
\begin{equation}\label{FG2}
N_{\text{odd}},\;N_{\text{even}},\;\Omega_{\text{odd}},\;\Omega_{\text{even}},\;
S_k.
\end{equation}
The odd $N$--groups correspond to real spinors, for example, $N_1$ is related
to real 2--spinors, and $N_3$ is the group of the real Majorana matrices.
The even $N$--groups define the quaternionic groups\index{group!quaternionic}. 
The $S$--groups
are the `spinor groups'\index{group!spinor}
($S_k=N_{2k}\otimes\C\simeq N_{2k-1}\otimes\C$):
$S_1$ is the group of the complex Pauli matrices, and $S_2$ is the group of the
Dirac matrices. Furthermore, the $\Omega$--groups are double copies of the
$N$--groups and can be written as a direct product of the $N$--groups with
the group of two elements $\dZ_2$:
\begin{equation}\label{FG3}
\Omega_k=N_k\otimes\dZ_2.
\end{equation}

Let us consider now several simplest examples of the groups (\ref{FG2}).
First of all, a finite group corresponding to the Clifford algebra $\cl_{0,0}$
with an arbitrary element $\cA=a^0$ and division ring $\K\simeq\R$
($p-q\equiv 0\pmod{8}$) is a cyclic group $\dZ_2=\{1,-1\}$ with the
following multiplication table
\[
{\renewcommand{\arraystretch}{1.4}
\begin{tabular}{|c||c|c|}\hline
 & $1$ & $-1$\\ \hline\hline  
$1$ & $1$ & $-1$ \\ \hline
$-1$& $-1$& $1$ \\ \hline
\end{tabular}.
}
\]
It is easy to see that in accordance with (\ref{FG2}) the finite group
corresponding to $\cl_{0,0}$ is $N_0$--group ($N_0=\dZ_2$).

A further Clifford algebra is $\cl_{1,0}$: $\cA=a^0+a^1\e_1,\,\e^2_1=1$,
$\K\simeq\R\oplus\R$, $p-q\equiv 1\pmod{8}$. In this case the basis elements
of $\cl_{1,0}$ form the Gauss--Klein four--group 
$\dZ_2\otimes\dZ_2=\{1,-1,\e_1,-\e_1\}$. The multiplication table of
$\dZ_2\otimes\dZ_2$ has a form
\[
{\renewcommand{\arraystretch}{1.4}
\begin{tabular}{|c||c|c|c|c|}\hline
  & $1$ & $-1$ & $\e_1$ & $-\e_1$ \\ \hline\hline
$1$ & $1$ & $-1$ & $\e_1$ & $-\e_1$ \\ \hline
$-1$ & $-1$ & $1$ & $-\e_1$ & $\e_1$ \\ \hline
$\e_1$ & $\e_1$ & $-\e_1$ & $1$ & $-1$ \\ \hline
$-\e_1$ & $-\e_1$ & $\e_1$ & $-1$ & $1$ \\ \hline
\end{tabular}.
}
\]
In accordance with (\ref{FG3}) we have here a first $\Omega$--group:
$\Omega_0=N_0\otimes\dZ_2=N_0\otimes N_0=\dZ_2\otimes\dZ_2$.

The algebra $\cl_{0,1}$ with $\cA=a^0+a^1\e_1$, $\e^2_1=-1$
($\K\simeq\C,\,p-q\equiv 7\pmod{8}$) corresponds to 
the complex group\index{group!complex}
$\dZ_4=\{1,-1,\e_1,-\e_1\}$ with the multiplication table
\[
{\renewcommand{\arraystretch}{1.4}
\begin{tabular}{|c||c|c|c|c|}\hline
  & $1$ & $-1$ & $\e_1$ & $-\e_1$ \\ \hline\hline
$1$ & $1$ & $-1$ & $\e_1$ & $-\e_1$ \\ \hline
$-1$ & $-1$ & $1$ & $-\e_1$ & $\e_1$ \\ \hline
$\e_1$ & $\e_1$ & $-\e_1$ & $-1$ & $1$ \\ \hline
$-\e_1$ & $-\e_1$ & $\e_1$ & $1$ & $-1$ \\ \hline
\end{tabular}.
}
\]
It is easy to see that in accordance with Salingaros classification this
group is a first $S$--group: $S_0=\dZ_4$.

The three finite groups considered previously are Abelian groups. All other
Salingaros groups (\ref{FG2}) are non--Abelian. The first non--Abelian
Salingaros group correspond to the algebra $\cl_{2,0}$ with an arbitrary
element $\cA=a^0+a^1\e_1+a^2\e_2+a^{12}\e_{12}$ and $\e^2_1=\e^2_2=1,\,
\e_{12}^2=-1$ ($p-q\equiv 2\pmod{8},\,\K\simeq\R$). The basis elements of
$\cl_{2,0}$ form a dihedral group\index{group!dihedral}
$D_4=\{1,-1,\e_1,-\e_1,\e_2,-\e_2,
\e_{12},-\e_{12}\}$ with the table
\[
{\renewcommand{\arraystretch}{1.4}
\begin{tabular}{|c||c|c|c|c|c|c|c|c|}\hline
  & $1$ & $-1$ & $\e_1$ & $-\e_1$ & $\e_2$ & $-\e_2$ & $\e_{12}$ & $-\e_{12}$
\\ \hline\hline
$1$ & $1$ & $-1$ & $\e_1$ & $-\e_1$ & $\e_2$ & $-\e_2$ & $\e_{12}$ & $-\e_{12}$
\\ \hline
$-1$ & $-1$ & $1$ & $-\e_1$ & $\e_1$ & $-\e_2$ & $\e_2$ & $-\e_{12}$ & $\e_{12}$
\\ \hline
$\e_1$ & $\e_1$ & $-\e_1$ & $1$ & $-1$ & $\e_{12}$ & $-\e_{12}$ &$\e_2$&$-\e_2$
\\ \hline
$-\e_1$& $-\e_1$& $\e_1$ & $-1$ & $1$ &$-\e_{12}$ &$\e_{12}$ & $-\e_2$&$\e_2$
\\ \hline
$\e_2$ & $\e_2$ & $-\e_2$&$-\e_{12}$ &$\e_{12}$ & $1$ &$-1$& $-\e_1$ &$\e_1$
\\ \hline
$-\e_2$& $-\e_2$& $\e_2$ &$\e_{12}$ & $-\e_{12}$& $-1$ &$1$ & $\e_1$ & $-\e_1$
\\ \hline
$\e_{12}$& $\e_{12}$&$-\e_{12}$&$-\e_2$ &$\e_2$ &$\e_1$ &$-\e_1$ &$-1$ &$1$
\\ \hline
$-\e_{12}$&$-\e_{12}$&$\e_{12}$&$\e_2$ & $-\e_2$&$-\e_1$&$\e_1$ &$1$ &$-1$
\\ \hline
\end{tabular}.
}
\]
It is a first $N_{\text{odd}}$--group: $N_1=D_4$. It is easy to verify
that we come to the same group $N_1=D_4$ for the algebra $\cl_{1,1}$ with the
ring $\K\simeq\R$, $p-q\equiv 0\pmod{8}$.

The following non--Abelian finite group we obtain for the algebra
$\cl_{0,2}$ with a quaternionic ring $\K\simeq\BH$, $p-q\equiv 6\pmod{8}$.
In this case the basis elements of $\cl_{0,2}$ form a quaternionic group
$Q_4=\{\pm 1,\pm\e_1,\pm\e_2,\pm\e_{12}\}$ with the multiplication table
\[
{\renewcommand{\arraystretch}{1.4}
\begin{tabular}{|c||c|c|c|c|c|c|c|c|}\hline
  & $1$ & $-1$ & $\e_1$ & $-\e_1$ & $\e_2$ & $-\e_2$ & $\e_{12}$ & $-\e_{12}$
\\ \hline\hline
$1$ & $1$ & $-1$ & $\e_1$ & $-\e_1$ & $\e_2$ & $-\e_2$ & $\e_{12}$ & $-\e_{12}$
\\ \hline
$-1$ & $-1$ & $1$ & $-\e_1$ & $\e_1$ & $-\e_2$ & $\e_2$ & $-\e_{12}$ & $\e_{12}$
\\ \hline
$\e_1$ & $\e_1$ & $-\e_1$ & $-1$ & $1$ & $\e_{12}$ & $-\e_{12}$ &$-\e_2$&$\e_2$
\\ \hline
$-\e_1$& $-\e_1$& $\e_1$ & $1$ & $-1$ &$-\e_{12}$ &$\e_{12}$ & $\e_2$&$-\e_2$
\\ \hline
$\e_2$ & $\e_2$ & $-\e_2$&$-\e_{12}$ &$\e_{12}$ & $-1$ &$1$& $\e_1$ &$-\e_1$
\\ \hline
$-\e_2$& $-\e_2$& $\e_2$ &$\e_{12}$ & $-\e_{12}$& $1$ &$-1$ & $-\e_1$ & $\e_1$
\\ \hline
$\e_{12}$& $\e_{12}$&$-\e_{12}$&$\e_2$ &$-\e_2$ &$-\e_1$ &$\e_1$ &$-1$ &$1$
\\ \hline
$-\e_{12}$&$-\e_{12}$&$\e_{12}$&$-\e_2$ & $\e_2$&$\e_1$&$-\e_1$ &$1$ &$-1$
\\ \hline
\end{tabular}.
}
\]
It is a first $N_{\text{even}}$--group: $N_2=Q_4$.

Now we can to establish a relationship between the finite groups and
division ring structures of $\cl_{p,q}$. It is easy to see that the five
distinct types of Salingaros groups correspond to the five division rings
of the real Clifford algebras as follows
\begin{eqnarray}
N_{\text{odd}}&\leftrightarrow&\cl_{p,q},\;p-q\equiv 0,2\pmod{8},\;\K\simeq\R;
\nonumber\\
N_{\text{even}}&\leftrightarrow&\cl_{p,q},\;p-q\equiv 4,6\pmod{8},\;
\K\simeq\BH;\nonumber\\
\Omega_{\text{odd}}&\leftrightarrow&\cl_{p,q},\;p-q\equiv 1\pmod{8},\;
\K\simeq\R\oplus\R;\nonumber\\
\Omega_{\text{even}}&\leftrightarrow&\cl_{p,q},\;p-q\equiv 5\pmod{8},\;
\K\simeq\BH\oplus\BH;\nonumber\\
S_k&\leftrightarrow&\cl_{p,q},\;p-q\equiv 3,7\pmod{8},\;\K\simeq\C.\nonumber
\end{eqnarray}
Therefore, the Periodic Table can be rewritten in terms
of finite group structure (see the Table 2).
\begin{figure}
\begin{center}{\small {\bf Table 2.} Finite group structure of the real Clifford
algebras.}\end{center}
\begin{center}
{\renewcommand{\arraystretch}{1.2}
\begin{tabular}{c|cccccccccc}
  & p & 0 & 1 & 2 & 3 & 4 & 5 & 6 & 7 & \ldots\\ \hline
q &   &   &   &   &   &   &   &   &   &\\
0 &   &$N_1$&$\Omega_0$&$N_1$&$S_1$&$N_4$&$\Omega_4$&$N_6$&$S_3$&
$\ldots$\\
1&&$S_0$&$N_1$&$\Omega_1$&$N_3$&$S_2$&$N_6$&$\Omega_6$&$N_8$&
$\ldots$\\
2&&$N_2$&$S_1$&$N_3$&$\Omega_3$&$N_5$&$S_3$&$N_8$&$\Omega_8$&
$\ldots$\\
3&&$\Omega_2$&$N_4$&$S_2$&$N_5$&$\Omega_5$&$N_7$&$S_4$&$N_{10}$&
$\ldots$\\
4&&$N_4$&$\Omega_4$&$N_6$&$S_3$&$N_7$&$\Omega_7$&$N_9$&
$S_5$&$\ldots$\\
5&&$S_2$&$N_6$&$\Omega_6$&$N_8$&$S_4$&$N_9$&$\Omega_9$&
$N_{11}$&$\ldots$\\
6&&$N_5$&$S_3$&$N_8$&$\Omega_8$&$N_{10}$&$S_5$&$N_{11}$&
$\Omega_{11}$&$\ldots$\\
7&&$\Omega_5$&$N_7$&$S_4$&$N_{10}$&$\Omega_{10}$&$N_{12}$&$S_6$&
$N_{13}$&$\ldots$\\
$\vdots$&&$\vdots$&$\vdots$&$\vdots$&$\vdots$&$\vdots$&$\vdots$&$\vdots$&
$\vdots$
\end{tabular}}
\end{center}
\end{figure}
Further, in accordance with (\ref{e4}) a center $\bZ_{p,q}$ of the algebra
$\cl_{p,q}$ consists of the unit if $p-q\equiv 0,2,4,6\pmod{8}$ and the
elements $1,\;\omega=\e_{12\ldots n}$ if $p-q\equiv 1,3,5,7\pmod{8}$.
Let $\bZ(p,q)\subset\cl_{p,q}$ be a center of the finite group (\ref{FG}).
In such a way, we have three distinct realizations of the center
$\bZ(p,q)$:
\begin{eqnarray}
\bZ(p,q)&=&\{1,-1\}\simeq\dZ_2\;\;\text{if}\;p-q\equiv 0,2,4,6\pmod{8};
\nonumber\\
\bZ(p,q)&=&\{1,-1,\omega,-\omega\}\simeq\dZ_2\otimes\dZ_2\;\;\text{if}\;
p-q\equiv 1,5\pmod{8};\nonumber\\
\bZ(p,q)&=&\{1,-1,\omega,-\omega\}\simeq\dZ_4\;\;\text{if}\;
p-q\equiv 3,7\pmod{8}.\nonumber
\end{eqnarray}
The Abelian groups $\bZ(p,q)$ are the subgroups of the Salingaros groups
(\ref{FG2}). Namely, $N$--groups have the center $\dZ_2$,
$\Omega$--groups have the center $\dZ_2\otimes\dZ_2$, and $S$-group has
the center $\dZ_4$.

The following Theorem presents a key result in the group structure of 
$\cl_{p,q}$.
\begin{theorem}[{\rm Salingaros \cite{Sal81a}}]
The factor group $G(p,q)/\bZ(p,q)$ is the Abelian group 
$\left(\dZ_2\right)^{\otimes 2k}=\dZ_2\otimes\dZ_2\otimes\cdots\otimes\dZ_2$
($2k$ times):
\[
\frac{G(p,q)}{\bZ(p,q)}:\;\;
\frac{N_{2k-1}}{\dZ_2}\simeq\frac{N_{2k}}{\dZ_2}\simeq
\frac{\Omega_{2k-1}}{\dZ_2\otimes\dZ_2}\simeq
\frac{\Omega_{2k}}{\dZ_2\otimes\dZ_2}\simeq
\frac{S_k}{\dZ_4}\simeq\left(\dZ_2\right)^{\otimes 2k}.
\]
\end{theorem}
This Theorem allows one to identify the Salingaros groups with extraspecial
groups\index{group!extraspecial} \cite{Sal81a,Sal84,Bra85}. 
As known, a finite group $G$ is called
an extraspecial 2-group if $G$ is of order $2^n$ and $G/\bZ(G)$ is the
Abelian group $\dZ_2\otimes\cdots\otimes\dZ_2$ ($n-1$ times). Further, if
$G$ is the extraspecial 2-group of order $2^{2k+1}$, then
\begin{eqnarray}
G&\simeq&D_4\circ\cdots\circ D_4\quad (k\;\,\text{times}),\;\;\text{or}
\nonumber\\
G&\simeq&Q_4\circ D_4\circ\cdots\circ D_4\quad (k-1\;\,\text{times}),
\nonumber
\end{eqnarray}
where $\circ$ means a central product\index{product!central}
of two groups: that is, the direct
product with centers identified. For example, the direct product
$Q_4\otimes D_4$ has the resulting group of order $8\times 8=64$, and its
center is the direct product\index{product!tensor}
of the two individual centers and is equal to
$\dZ_2\otimes\dZ_2$. In contrast, the central product $Q_4\circ D_4$
amalgamates the $\dZ_2$ center of $Q_4$ with the $\dZ_2$ of $D_4$ to give
the center of $Q_4\circ D_4$ as $\dZ_2$. Therefore, the central product
$Q_4\circ D_4$ is of order $32$. In the case where one center is a subgroup
of the other center, they both amalgamate into the larger center.

In terms of the extraspecial groups all the Salingaros groups take the form:
\begin{eqnarray}
N_{2k-1}&\simeq&\left(N_1\right)^{\circ k},\nonumber\\
N_{2k}&\simeq&N_2\circ\left(N_1\right)^{\circ(k-1)},\nonumber\\
\Omega_{2k-1}&\simeq&N_{2k-1}\circ\left(\dZ_2\otimes\dZ_2\right),\nonumber\\
\Omega_{2k}&\simeq&N_{2k}\circ\left(\dZ_2\otimes\dZ_2\right),\nonumber\\
S_k&\simeq&N_{2k-1}\circ\dZ_4\simeq N_{2k}\circ\dZ_4.\nonumber
\end{eqnarray}
It should be noted here that a comprehensive consideration of the finite
groups (\ref{FG}) over a Galouis field\index{field!Galouis}
$GF(2)=\F_2=\{0,1\}$ was given by
Shaw \cite{Sha94}.

In the following sections we will consider finite groups that arise from
the fundamental automorphisms of the Clifford algebras. It will be shown
that these groups (both Abelian and non--Abelian) are the subgroups 
(except some extended automorphism groups) of the
Salingaros groups. Moreover, it will be shown also that the finite groups
of automorphisms, $\Aut(\cl_{p,q})$ and $\Ext(\cl_{p,q})$, 
form a natural basis for description
of the discrete symmetries in quantum field theory.

\subsection{Finite groups and fundamental automorphisms}
In virtue of the Wedderburn--Artin theorem the all fundamental automorphisms
of $\cl$ are transferred to the matrix algebra. Matrix (spinor) representations of the
fundamental automorphisms of $\C_n$ were first obtained by Rashevskii in 1955
\cite{Rash}: 1) Involution: $\sA^\star=\sW\sA\sW^{-1}$, where $\sW$ is a
matrix of the automorphism $\star$ (matrix representation of the volume
element $\omega$); 2) Reversion: $\widetilde{\sA}=\sE\sA^{\sT}\sE^{-1}$, where
$\sE$ is a matrix of the antiautomorphism $\widetilde{\phantom{cc}}$
satisfying the conditions $\cE_i\sE-\sE\cE^{\sT}_i=0$ and 
$\sE^{\sT}=(-1)^{\frac{m(m-1)}{2}}\sE$, here $\cE_i=\gamma(\e_i)$ are matrix
representations of the units of the algebra $\cl$; 3) Conjugation:
$\widetilde{\sA^\star}=\sC\sA^{\sT}\sC^{-1}$, where $\sC=\sE\sW^{\sT}$ 
is a matrix of
the antiautomorphism $\widetilde{\star}$ satisfying the conditions
$\sC\cE^{\sT}_i+\cE_i\sC=0$ and
$\sC^{\sT}=(-1)^{\frac{m(m+1)}{2}}\sC$.

A spinor representation of the pseudoautomorphism
$\cA\rightarrow\overline{\cA}$ of the algebra $\C_n$ when $n\equiv 0\s\pmod{2}$
is defined as follows.
In the spinor representation the every element $\cA\in\C_n$ should be
represented by some matrix $\sA$, and the pseudoautomorphism (\ref{6.24})
takes a form of the pseudoautomorphism of the full 
matrix algebra\index{algebra!matrix}
$\M_{2^{n/2}}$:
\[
\sA\longrightarrow\overline{\sA}.
\]\begin{sloppypar}\noindent
On the other hand, a transformation replacing the matrix $\sA$ by the
complex conjugate matrix, $\sA\rightarrow\dot{\sA}$, is also some
pseudoautomorphism of the algebra $\M_{2^{n/2}}$. The composition of the two
pseudoautomorpisms $\dot{\sA}\rightarrow\sA$ and
$\sA\rightarrow\overline{\sA}$, $\dot{\sA}\rightarrow\sA\rightarrow
\overline{\sA}$, is an internal automorphism\index{automorphism!internal}
$\dot{\sA}\rightarrow\overline{\sA}$ of the full matrix algebra $\M_{2^{n/2}}$:
\end{sloppypar}
\begin{equation}\label{6.25}
\overline{\sA}=\Pi\dot{\sA}\Pi^{-1},
\end{equation}
where $\Pi$ is a matrix of the pseudoautomorphism 
$\cA\rightarrow\overline{\cA}$ in the spinor representation.
The sufficient condition for definition of the pseudoautomorphism
$\cA\rightarrow\overline{\cA}$ is a choice of the matrix
$\Pi$ in such a way that the transformation 
$\sA\rightarrow\Pi\dot{\sA}\Pi^{-1}$ transfers into itself the matrices
$\cE_1,\ldots,\cE_p,i\cE_{p+1},\ldots,i\cE_{p+q}$
(the matrices of the spinbasis of $\cl_{p,q}$), that is,
\begin{equation}\label{6.26}
\cE_i\longrightarrow\cE_i=\Pi\dot{\cE}_i\Pi^{-1}\quad
(i=1,\ldots,p+q).
\end{equation}
\begin{theorem}\label{tpseudo}
Let $\C_n$ be a complex Clifford algebra for $n\equiv 0\s\pmod{2}$
and let $\cl_{p,q}\subset\C_n$ be its subalgebra with a real division ring
$\K\simeq\R$ when $p-q\equiv 0,2\s\pmod{8}$ and quaternionic division ring
$\K\simeq\BH$ when $p-q\equiv 4,6\s\pmod{8}$, $n=p+q$. Then in dependence
on the division ring structure of the real subalgebra $\cl_{p,q}$ the matrix
$\Pi$ of the pseudoautomorphism $\cA\rightarrow\overline{\cA}$ 
has the following form:\\[0.2cm]
1) $\K\simeq\R$, $p-q\equiv 0,2\s\pmod{8}$.\\[0.1cm]
The matrix $\Pi$ for any spinor representation over the ring $\K\simeq\R$
is proportional to the unit matrix.\\[0.2cm]
2) $\K\simeq\BH$, $p-q\equiv 4,6\s\pmod{8}$.\\[0.1cm]
$\Pi=\cE_{\alpha_1\alpha_2\cdots\alpha_a}$ when 
$a\equiv 0\s\pmod{2}$ and
$\Pi=\cE_{\beta_1\beta_2\cdots\beta_b}$ when $b\equiv 1\s\pmod{2}$,
where $a$ complex matrices $\cE_{\alpha_t}$ 
and $b$ real matrices $\cE_{\beta_s}$ form a basis of the spinor
representation of the algebra $\cl_{p,q}$ over the ring $\K\simeq\BH$,
$a+b=p+q,\,0<t\leq a,\,0<s\leq b$. At this point,
\begin{eqnarray}
\Pi\dot{\Pi}&=&\phantom{-}\sI\quad\text{if $a,b\equiv 0,1\s\pmod{4}$},
\nonumber\\
\Pi\dot{\Pi}&=&-\sI\quad\text{if $a,b\equiv 2,3\s\pmod{4}$},\nonumber
\end{eqnarray}
where $\sI$ is the unit matrix.
\end{theorem}
\begin{proof}\begin{sloppypar}\noindent
The algebra $\C_n$ ($n\equiv 0\s\pmod{2}$, $n=p+q$) in virtue of
$\C_n=\C\otimes\cl_{p,q}$ and definition of the division ring
$\K\simeq f\cl_{p,q}f$ 
($f$ is a primitive idempotent of the algebra $\cl_{p,q}$)
has four different real subalgebras: $p-q\equiv 0,2\s\pmod{8}$
for the real division ring $\K\simeq\R$ and $p-q\equiv 4,6\s\pmod{8}$ for
the quaternionic division ring $\K\simeq\BH$.\\[0.2cm]
1) $\K\simeq\R$.\\[0.1cm]
Since for the types $p-q\equiv 0,2\s\pmod{8}$ there is an isomorphism
$\cl_{p,q}\simeq\M_{2^{\frac{p+q}{2}}}(\R)$ (Wedderburn--Artin Theorem), then
all the matrices $\cE_i$ of the spinbasis of $\cl_{p,q}$ are real and
$\dot{\cE}_i=\cE_i$. Therefore, in this case the condition (\ref{6.26})
can be written as follows\end{sloppypar}
\[
\cE_i\longrightarrow\cE_i=\Pi\cE_i\Pi^{-1},
\]
whence $\cE_i\Pi=\Pi\cE_i$. Thus, for the algebras $\cl_{p,q}$ of the types
$p-q\equiv 0,2\s\pmod{8}$ the matrix $\Pi$ of the pseudoautomorphism
$\cA\rightarrow\overline{\cA}$ commutes with all the matrices $\cE_i$.
It is easy to see that
$\Pi\sim\sI$.\\[0.2cm]
2) $\K\simeq\BH$.\\[0.1cm]
In turn, for the quaternionic types $p-q\equiv 4,6\s\pmod{8}$ there is an
isomorphism $\cl_{p,q}\simeq\M_{2^{\frac{p+q}{2}}}(\BH)$. Therefore, among
the matrices of the spinbasis of the algebra $\cl_{p,q}$ there are matrices
$\cE_\alpha$ satisfying the condition $\dot{\cE}_\alpha=-\cE_\alpha$. 
Let $a$ be a quantity of the complex matrices, then the spinbasis of $\cl_{p,q}$
is divided into two subsets. The first subset
$\{\dot{\cE}_{\alpha_t}=-\cE_{\alpha_t}\}$ contains complex matrices,
$0<t\leq a$, and the second subset
$\{\dot{\cE}_{\beta_s}=\cE_{\beta_s}\}$ contains real matrices,
$0<s\leq p+q-a$. In accordance with a spinbasis structure of the algebra
$\cl_{p,q}\simeq\M_{2^{\frac{p+q}{2}}}(\BH)$ the condition (\ref{6.26})
can be written as follows
\[
\cE_{\alpha_t}\longrightarrow-\cE_{\alpha_t}=\Pi\cE_{\alpha_t}\Pi^{-1},\quad
\cE_{\beta_s}\longrightarrow\cE_{\beta_s}=\Pi\cE_{\beta_s}\Pi^{-1}.
\]
Whence
\begin{equation}\label{6.27}
\cE_{\alpha_t}\Pi=-\Pi\cE_{\alpha_t},\quad
\cE_{\beta_s}\Pi=\Pi\cE_{\beta_s}.
\end{equation}
Thus, for the quaternionic types $p-q\equiv 4,6\s\pmod{8}$ the matrix
$\Pi$ of the pseudoautomorphism $\cA\rightarrow\overline{\cA}$ anticommutes
with a complex part of the spinbasis of $\cl_{p,q}$ and commutes with
a real part of the same spinbasis. From (\ref{6.27}) it follows that a
structure of the matrix $\Pi$ is analogous to the structure of
the matrices $\sE$ and $\sC$ of the antiautomorphisms
$\cA\rightarrow\widetilde{\cA}$ and
$\cA\rightarrow\widetilde{\cA^\star}$, correspondingly 
(see Theorem \ref{tautr}), that is, the matrix
$\Pi$ of the pseudoautomorphism $\cA\rightarrow\overline{\cA}$ of the algebra
$\C_n$ is a product of only complex matrices, or only real matrices
of the spinbasis of the subalgebra $\cl_{p,q}$.

So, let $0<a<p+q$ and let $\Pi=\cE_{\alpha_1\alpha_2\cdots
\alpha_a}$ be a matrix of $\cA\rightarrow\overline{\cA}$,
then permutation conditions of the matrix $\Pi$ 
with the matrices $\cE_{\beta_s}$
of the real part ($0<s\leq p+q-a$) and with the matrices
$\cE_{\alpha_t}$ of the complex part ($0<t\leq a$) have the form
\begin{equation}\label{6.28}
\Pi\cE_{\beta_s}=(-1)^a\cE_{\beta_s}\Pi,
\end{equation}
\begin{eqnarray}
\Pi\cE_{\alpha_t}&=&(-1)^{a-t}\sigma(\alpha_t)\cE_{\alpha_1\alpha_2
\cdots\alpha_{t-1}\alpha_{t+1}\cdots\alpha_a},\nonumber\\
\cE_{\alpha_t}\Pi&=&(-1)^{t-1}\sigma(\alpha_t)\cE_{\alpha_1\alpha_2
\cdots\alpha_{t-1}\alpha_{t+1}\cdots\alpha_a},\label{6.29}
\end{eqnarray}
that is, when $a\equiv 0\s\pmod{2}$ the matrix $\Pi$ commutes with the real
part and anticommutes with the complex part of the spinbasis of $\cl_{p,q}$.
Correspondingly, when $a\equiv 1\s\pmod{2}$ the matrix $\Pi$ anticommutes
with the real part and commutes with the complex part. Further, let
$\Pi=\cE_{\beta_1\beta_2\cdots\beta_{p+q-a}}$ be a product of the
real matrices, then
\begin{eqnarray}
\Pi\cE_{\beta_s}&=&(-1)^{p+q-a-s}\sigma(\beta_s)\cE_{\beta_1\beta_2
\cdots\beta_{s-1}\beta_{s+1}\cdots\beta_{p+q-a}},\nonumber\\
\cE_{\beta_s}\Pi&=&(-1)^{s-1}\sigma(\beta_s)\cE_{\beta_1\beta_2
\cdots\beta_{s-1}\beta_{s+1}\cdots\beta_{p+q-a}},\label{6.30}
\end{eqnarray}
\begin{equation}\label{6.31}
\Pi\cE_{\alpha_t}=(-1)^{p+q-a}\cE_{\alpha_t}\Pi,
\end{equation}
that is, when $p+q-a\equiv 0\s\pmod{2}$ the matrix $\Pi$ anticommutes with
the real part and commutes with the complex part 
of the spinbasis of $\cl_{p,q}$. Correspondingly, when
$p+q-a\equiv 1\s\pmod{2}$ the matrix $\Pi$ commutes with the real part and
anticommutes with the complex part.

The comparison of the conditions (\ref{6.28})--(\ref{6.29}) 
with the condition (\ref{6.27}) shows that the matrix
$\Pi=\cE_{\alpha_1\alpha_2\cdots\alpha_a}$ exists only at
$a\equiv 0\s\pmod{2}$, that is, $\Pi$ is a product of the complex matrices
$\cE_{\alpha_t}$ of the even number. In turn, a comparison of
(\ref{6.30})--(\ref{6.31}) with
(\ref{6.27}) shows that the matrix $\Pi=\cE_{\beta_1\beta_2\cdots
\beta_{p+q-a}}$ exists only at $p+q-a\equiv 1\s\pmod{2}$, that is,
$\Pi$ is a product of the real matrices $\cE_{\beta_s}$ of the odd number.

Let us calculate now the product $\Pi\dot{\Pi}$. 
Let $\Pi=\cE_{\beta_1\beta_2\cdots
\beta_{p+q-a}}$ be a product of the $p+q-a$ real matrices.
Since $\dot{\cE}_{\beta_s}=\cE_{\beta_s}$, then
$\dot{\Pi}=\Pi$ and $\Pi\dot{\Pi}=\Pi^2$. Therefore,
\begin{equation}\label{6.32}
\Pi\dot{\Pi}=(\cE_{\beta_1\beta_2\cdots\beta_{p+q-a}})^2=
(-1)^{\frac{(p+q-a)(p+q-a-1)}{2}}\cdot\sI.
\end{equation}
Further, let $\Pi=\cE_{\alpha_1\alpha_2\cdots\alpha_a}$ be a
product of the $a$ complex matrices. Then
$\dot{\cE}_{\alpha_t}=-\cE_{\alpha_t}$ and $\dot{\Pi}=(-1)^a\Pi=\Pi$, since
$a\equiv 0\s\pmod{2}$. Therefore,
\begin{equation}\label{6.33}
\Pi\dot{\Pi}=(\cE_{\alpha_1\alpha_2\cdots\alpha_a})^2=
(-1)^{\frac{a(a-1)}{2}}\cdot\sI.
\end{equation}
Let $p+q-a=b$ be a quantity of the real matrices $\cE_{\beta_s}$ of the
spinbasis of $\cl_{p,q}$, then $p+q=a+b$. Since $p+q$ is always even number
for the quaternionic types $p-q\equiv 4,6\s\pmod{8}$, then $a$ and $b$ 
are simultaneously even or odd numbers. Thus, from (\ref{6.32}) and (\ref{6.33})
it follows
\[
\Pi\dot{\Pi}=\begin{cases}
\phantom{-}\sI,& \text{if $a,b\equiv 0,1\s\!\!\pmod{4}$},\\
-\sI,& \text{if $a,b\equiv 2,3\s\!\!\pmod{4}$},
\end{cases}
\]
which required to be proved.
\end{proof}
\begin{sloppypar}
In the present form of quantum field theory complex fields correspond
to charged particles. Thus, the extraction of the subalgebra $\cl_{p,q}$ with
the real ring $\K\simeq\R$ in $\C_n$, $p-q\equiv 0,2\s\pmod{8}$,
corresponds to physical fields describing {\it truly neutral particles}
such as photon and neutral mesons ($\pi^0,\,\eta^0,\,\rho^0,\,
\omega^0,\,\varphi^0,\,K^0$). In turn, the subalgebras $\cl_{p,q}$ with the
ring $\K\simeq\BH$, $p-q\equiv 4,6\s\pmod{8}$ correspond to charged or
neutral fields.
\end{sloppypar}

In the paper \cite{Var99} it has been shown that space
reversal $P$, time reversal $T$ and combination $PT$ are correspond 
respectively to the fundamental automorphisms 
$\cA\rightarrow\cA^\star$, $\cA\rightarrow\widetilde{\cA}$ and
$\cA\rightarrow\widetilde{\cA^{\star}}$. 
In like manner, charge conjugation $C$ is naturally included into a
general scheme by means of a complex conjugation pseudoautomorphism
$\cA\rightarrow\overline{A}$.
\begin{prop}\label{prop1}
Let $\cl_{p,q}$ ($p+q=2m$) be a Clifford algebra over the field $\F=\R$ and
let $\pin(p,q)$ be an universal covering of the orthogonal group $O(p,q)=O_0(p,q)
\odot\{1,P,T,PT\}\simeq O_0(p,q)\odot(\dZ_2\otimes\dZ_2)$ of transformations
of the space $\R^{p,q}$, where $\{1,P,T,PT\}\simeq\dZ_2\otimes\dZ_2$ is a
group of discrete transformations of $\R^{p,q}$, $\dZ_2\otimes\dZ_2$ is the
Gauss--Klein group. Then there is an isomorphism between the group
$\{1,P,T,PT\}$ and an automorphism group\index{group!automorphism}
$\{\Id,\star,\widetilde{\phantom{cc}},
\widetilde{\star}\}$ of the algebra $\cl_{p,q}$. In this case, space
inversion $P$, time reversal $T$ and combination $PT$ are correspond 
to the fundamental automorphisms $\cA\rightarrow\cA^\star,\,
\cA\rightarrow\widetilde{\cA}$ and $\cA\rightarrow\widetilde{\cA^\star}$.
\end{prop} 
An equivalence of the multiplication tables of the groups
$\{1,P,T,PT\}$ and 
$\Aut(\cl)=\{\Id,\star,\widetilde{\phantom{cc}},\widetilde{\star}\}$
proves this isomorphism
(in virtue of the commutativity $\widetilde{(\cA^\star)}=
(\widetilde{\cA})^\star$ and the 
involution conditions $(\star)^2=(\widetilde{\phantom{cc}
})^2=\Id$):
\[
{\renewcommand{\arraystretch}{1.4}
\begin{tabular}{|c||c|c|c|c|}\hline
        & $\Id$ & $\star$ & $\widetilde{\phantom{cc}}$ & $\widetilde{\star}$\\ \hline\hline
$\Id$   & $\Id$ & $\star$ & $\widetilde{\phantom{cc}}$ & $\widetilde{\star}$\\ \hline
$\star$ & $\star$ & $\Id$ & $\widetilde{\star}$ & $\widetilde{\phantom{cc}}$\\ \hline
$\widetilde{\phantom{cc}}$ & $\widetilde{\phantom{cc}}$ &$\widetilde{\star}$
& $\Id$ & $\star$ \\ \hline
$\widetilde{\star}$ & $\widetilde{\star}$ & $\widetilde{\phantom{cc}}$ &
$\star$ & $\Id$\\ \hline
\end{tabular}
\;\sim\;
\begin{tabular}{|c||c|c|c|c|}\hline
    & $1$ & $P$ & $T$ & $PT$\\ \hline\hline
$1$ & $1$ & $P$ & $T$ & $PT$\\ \hline
$P$ & $P$ & $1$ & $PT$& $T$\\ \hline
$T$ & $T$ & $PT$& $1$ & $P$\\ \hline
$PT$& $PT$& $T$ & $P$ & $1$\\ \hline
\end{tabular}.
}
\]\begin{sloppypar}\noindent
Further, in the case $P^2=T^2=(PT)^2=\pm 1$ and $PT=-TP$ there is an
isomorphism between the group $\{1,P,T,PT\}$ and an automorphism group
$\sAut(\cl)=\{\sI,\sW,\sE,\sC\}$. So, for the Dirac algebra $\C_4$ in the
canonical $\gamma$--basis there exists a standard (Wigner) representation
$P=\gamma_0$ and $T=\gamma_1\gamma_3$ \cite{BLP89}, therefore,
$\{1,P,T,PT\}=\{1,\gamma_0,\gamma_1\gamma_3,\gamma_0\gamma_1\gamma_3\}$.
On the other hand, in the $\gamma$--basis an automorphism group
of $\C_4$ has a form $\sAut(\C_4)=\{\sI,\sW,\sE,\sC\}=
\{\sI,\gamma_0\gamma_1\gamma_2\gamma_3,\gamma_1\gamma_3,\gamma_0\gamma_2\}$.
It has been shown \cite{Var99} that $\{1,P,T,PT\}=
\{1,\gamma_0,\gamma_1\gamma_3,\gamma_0\gamma_1\gamma_3\}\simeq\sAut(\C_4)
\simeq\dZ_4$, where $\dZ_4$ is a complex group with the signature
$(+,-,-)$. 
Generalizations of these results for the Clifford algebras over the
fields $\F=\C$ and $\F=\R$
are contained in the following two theorems:\end{sloppypar}
\begin{theorem}[{\rm\cite{Var99}}]\label{taut}
Let $\sAut=\{\sI,\,\sW,\,\sE,\,\sC\}$ be 
the automorphism group\index{group!automorphism} of the algebra
$\C_{p+q}$ $(p+q=2m)$, where 
$\sW=\cE_{12\cdots m m+1 m+2\cdots p+q}$,
and $\sE=\cE_{12\cdots m}$, $\sC=\cE_{m+1 m+2\cdots p+q}$ if
$m\equiv 1\pmod{2}$, and $\sE=\cE_{m+1 m+2\cdots p+q}$, 
$\sC=\cE_1\cE_2\cdots
\cE_m$ if $m\equiv 0\pmod{2}$. Let $\sAut_-$ and $\sAut_+$ be the automorphism 
groups, in which the all elements correspondingly commute
$(m\equiv 0\pmod{2})$ and anticommute $(m\equiv 1\pmod{2})$.
Then over the field $\F=\C$ there are only two non--isomorphic groups:
$\sAut_-\simeq\dZ_2\otimes\dZ_2$ for the signature $(+,\,+,\,+)$ if
$n\equiv 0,1\pmod{4}$ and
$\sAut_+\simeq Q_4/\dZ_2$ for the signature
 $(-,\,-,\,-)$ if
$n\equiv 2,3\pmod{4}$.
\end{theorem}
\begin{theorem}[{\rm\cite{Var00}}]\label{tautr}\begin{sloppypar}\noindent
Let $\cl_{p,q}$ be a Clifford algebra over a field $\F=\R$ and let
$\sAut(\cl_{p,q})=\{\sI,\sW,\sE,\sC\}$ be a group of fundamental
automorphisms\index{automorphism!fundamental}
of the algebra $\cl_{p,q}$. Then for eight types of the 
algebras $\cl_{p,q}$ there exist, depending upon a division ring structure
of $\cl_{p,q}$, following isomorphisms between finite groups and groups
$\sAut(\cl_{p,q})$ with different signatures
$(a,b,c)$, where $a,b,c\in\{-,+\}$:\\[0.2cm]
1) $\K\simeq\R$, types $p-q\equiv 0,2\pmod{8}$.\\
If $\sE=\cE_{p+1 p+2\cdots p+q}$ and $\sC=\cE_{12\cdots p}$,
then Abelian groups $\sAut_-(\cl_{p,q})\simeq\dZ_2\otimes\dZ_2$
with the signature $(+,+,+)$ and $\sAut_-(\cl_{p,q})\simeq\dZ_4$ with the
signature
$(+,-,-)$ exist at $p,q\equiv 0\pmod{4}$ and $p,q\equiv 2\pmod{4}$, 
respectively,
for the type $p-q\equiv 0\pmod{8}$, and also Abelian groups
$\sAut_-(\cl_{p,q})\simeq\dZ_4$ with the signature $(-,-,+)$ and
$\sAut_-(\cl_{p,q})
\simeq\dZ_4$ with the signature $(-,+,-)$ exist 
at $p\equiv 0\pmod{4},\,
q\equiv 2\pmod{4}$ and $p\equiv 2\pmod{4},\,q\equiv 0\pmod{4}$ for the type
$p-q\equiv 2\pmod{8}$, respectively.\\
If $\sE=\cE_{12\cdots p}$ and $\sC=\cE_{p+1 p+2\cdots p+q}$,
then non--Abelian groups $\sAut_+(\cl_{p,q})\simeq D_4/\dZ_2$ with the
signature $(+,-,+)$ and $\sAut_+(\cl_{p,q})\simeq D_4/\dZ_2$ with the
signature
$(+,+,-)$ exist at $p,q\equiv 3\pmod{4}$ and $p,q\equiv 1\pmod{4}$, 
respectively,
for the type $p-q\equiv 0\pmod{8}$, and also non--Abelian groups
$\sAut_+(\cl_{p,q})\simeq Q_4/\dZ_2$ with $(-,-,-)$ and 
$\sAut_+(\cl_{p,q})\simeq
D_4/\dZ_2$ with $(-,+,+)$ exist at $p\equiv 3\pmod{4},\,q\equiv 1
\pmod{4}$ and $p\equiv 1\pmod{4},\,q\equiv 3\pmod{4}$ for the type
$p-q\equiv 2\pmod{8}$, respectively.\\[0.2cm]
2) $\K\simeq\BH$, types $p-q\equiv 4,6\pmod{8}$.\\
If $\sE=\cE_{j_1j_2\cdots j_k}$ is a product of $k$
skewsymmetric matrices (among which $l$ matrices have a square $+\sI$
and $t$ matrices have a square $-\sI$)
and $\sC=\cE_{i_1i_2\cdots i_{p+q-k}}$ is a product of $p+q-k$
symmetric matrices (among which $h$ matrices have a square $+\sI$ and
$g$ have a square $-\sI$),
then at $k\equiv 0\pmod{2}$ for the type $p-q\equiv 4\pmod{8}$ there exist
Abelian groups $\sAut_-(\cl_{p,q})\simeq\dZ_2\otimes\dZ_2$ with $(+,+,+)$
and $\sAut_-(\cl_{p,q})\simeq\dZ_4$ with $(+,-,-)$ if
$l-t,\,h-g\equiv 0,1,4,5\pmod{8}$ and
$l-t,\,h-g\equiv 2,3,6,7\pmod{8}$, respectively. And also at
$k\equiv 0\pmod{2}$ for the type $p-q\equiv 6\pmod{8}$ there exist
$\sAut_-(\cl_{p,q})\simeq\dZ_4$ with $(-,+,-)$ and 
$\sAut_-(\cl_{p,q})\simeq\dZ_4$
with $(-,-,+)$ if $l-t\equiv 0,1,4,5\pmod{8},\,
h-g\equiv 2,3,6,7\pmod{8}$ and $l-t\equiv 2,3,6,7\pmod{8},\,
h-g\equiv 0,1,4,5\pmod{8}$,respectively.\\
Inversely, if $\sE=\cE_{i_1i_2\cdots i_{p+q-k}}$ is a product of
$p+q-k$ symmetric matrices and 
$\sC=\cE_{j_1j_2\cdots j_k}$ is a product of $k$ skewsymmetric
matrices, then at $k\equiv 1\pmod{2}$
for the type $p-q\equiv 4\pmod{8}$ there exist non--Abelian groups
$\sAut_+(\cl_{p,q})
\simeq D_4/\dZ_2$ with $(+,-,+)$ and $\sAut_+(\cl_{p,q})\simeq D_4/\dZ_2$ with
$(+,+,-)$ if $h-g\equiv 2,3,6,7\pmod{8},\,l-t\equiv
0,1,4,5\pmod{8}$ and $h-g\equiv 0,1,4,5\pmod{8},\,l-t\equiv 2,3,6,7\pmod{8}$,
respectively.
And also at $k\equiv 1\pmod{2}$ for the type $p-q\equiv 6\pmod{8}$ there exist
$\sAut_+(\cl_{p,q})\simeq Q_4/\dZ_2$ with $(-,-,-)$ and $\sAut_+(\cl_{p,q})
\simeq D_4/\dZ_2$ with $(-,+,+)$ if $h-g,
\,l-t\equiv 2,3,6,7\pmod{8}$ and $h-g,\,l-t\equiv 0,1,4,5
\pmod{8}$, respectively.\\[0.2cm]
3) $\K\simeq\R\oplus\R,\,\K\simeq\BH\oplus\BH$, types $p-q\equiv 1,5\pmod{8}$.\\
For the algebras $\cl_{0,q}$ of the types $p-q\equiv 1,5\pmod{8}$ there exist
Abelian automorphism groups with the signatures
$(-,-,+)$, $(-,+,-)$ and non--Abelian automorphism groups with the signatures
$(-,-,-)$, $(-,+,+)$. Correspondingly, for the algebras $\cl_{p,0}$ of the
types $p-q\equiv 1,5\pmod{8}$ there exist Abelian groups with
$(+,+,+)$, $(+,-,-)$ and non--Abelian groups with $(+,-,+)$,
$(+,+,-)$. In general case for $\cl_{p,q}$, the types $p-q\equiv 1,5\pmod{8}$
admit all eight automorphism groups.\\[0.2cm]
4) $\K=\C$, types $p-q\equiv 3,7\pmod{8}$.\\
The types $p-q\equiv 3,7\pmod{8}$ admit the Abelian group $\sAut_-(\cl_{p,q})
\simeq\dZ_2\otimes\dZ_2$ with the signature $(+,+,+)$ if $p\equiv 0\pmod{2}$ and
$q\equiv 1\pmod{2}$, and also non--Abelian group 
$\sAut_+(\cl_{p,q})\simeq
Q_4/\dZ_2$ with the signature $(-,-,-)$ if $p\equiv 1\pmod{2}$ and
$q\equiv 0\pmod{2}$. \end{sloppypar}
\end{theorem} 

\section{Extended automorphism groups}
An introduction of the pseudoautomorphism $\cA\rightarrow\overline{\cA}$
allows us to extend the automorphism set of the complex Clifford algebra
$\C_n$. Namely, we add to the four fundamental automorphisms
$\cA\rightarrow\cA$, $\cA\rightarrow\cA^\star$, 
$\cA\rightarrow\widetilde{\cA}$, $\cA\rightarrow\widetilde{\cA^\star}$
the pseudoautomorphism $\cA\rightarrow\overline{\cA}$ and
following three combinations:\\
1) A pseudoautomorphism $\cA\rightarrow\overline{\cA^\star}$. This 
transformation is a composition of the pseudoautomorphism
$\cA\rightarrow\overline{\cA}$ with 
the automorphism $\cA\rightarrow\cA^\star$.\\
2) A pseudoantiautomorphism\index{pseudoantiautomorphism}
$\cA\rightarrow\overline{\widetilde{\cA}}$.
This transformation is a composition of $\cA\rightarrow\overline{\cA}$ with
the antiautomorphism $\cA\rightarrow\widetilde{\cA}$.\\
3) A pseudoantiautomorphism $\cA\rightarrow\overline{\widetilde{\cA^\star}}$
(a composition of $\cA\rightarrow\overline{\cA}$ with the antiautomorphism
$\cA\rightarrow\widetilde{\cA^\star}$).

Thus, we obtain an automorphism set of $\C_n$ consisting of the
eight transformations. Let us show that the set
$\{\Id,\,\star,\,\widetilde{\phantom{cc}},\,\widetilde{\star},\,
\overline{\phantom{cc}},\,\overline{\star},\,
\overline{\widetilde{\phantom{cc}}},\,\overline{\widetilde{\star}}\}$
forms a finite group of order 8 and 
let us give a physical interpretation of this
group.
\begin{prop}\label{prop2}\begin{sloppypar}\noindent
Let $\C_n$ be a Clifford algebra over the field $\F=\C$ and let
$\Ext(\C_n)=
\{\Id,\,\star,\,\widetilde{\phantom{cc}},\,\widetilde{\star},\,
\overline{\phantom{cc}},\,\overline{\star},\,
\overline{\widetilde{\phantom{cc}}},\,\overline{\widetilde{\star}}\}$
be an extended automorphism group\index{group!automorphism!extended}
of the algebra $\C_n$. Then there is
an isomorphism between $\Ext(\C_n)$ and the full $CPT$--group
of the discrete transformations,
$\Ext(\C_n)\simeq\{1,\,P,\,T,\,PT,\,C,\,CP,\,CT,\,CPT\}\simeq
\dZ_2\otimes\dZ_2\otimes\dZ_2$. In this case, space inversion $P$, time
reversal $T$, full reflection $PT$, charge conjugation $C$, transformations
$CP$, $CT$ and the full $CPT$--transformation correspond to the automorphism
$\cA\rightarrow\cA^\star$, antiautomorphisms $\cA\rightarrow\widetilde{\cA}$,
$\cA\rightarrow\widetilde{\cA^\star}$, pseudoautomorphisms
$\cA\rightarrow\overline{\cA}$, $\cA\rightarrow\overline{\cA^\star}$,
pseudoantiautomorphisms $\cA\rightarrow\overline{\widetilde{\cA}}$ and
$\cA\rightarrow\overline{\widetilde{\cA^\star}}$, respectively.\end{sloppypar}
\end{prop}
\begin{proof}\begin{sloppypar}\noindent
The group $\{1,\,P,\,T,\,PT,\,C,\,CP,\,CT,\,CPT\}$ at the conditions
$P^2=T^2=(PT)^2=C^2=(CP)^2=(CT)^2=(CPT)^2=1$ and commutativity of all the
elements forms an Abelian group of order 8, which is isomorphic to a cyclic
group $\dZ_2\otimes\dZ_2\otimes\dZ_2$. 
The multiplication table
of this group has a form\end{sloppypar}
\begin{center}{\renewcommand{\arraystretch}{1.4}
\begin{tabular}{|c||c|c|c|c|c|c|c|c|}\hline
     & $1$  & $P$  & $T$  & $PT$ & $C$  & $CP$ & $CT$ & $CPT$ \\ \hline\hline
$1$  & $1$  & $P$  & $T$  & $PT$ & $C$  & $CP$ & $CT$ & $CPT$ \\ \hline
$P$  & $P$  & $1$  & $PT$ & $T$  & $CP$ & $C$  & $CPT$& $CT$\\ \hline
$T$  & $T$  & $PT$ & $1$  & $P$  & $CT$ & $CPT$& $C$  & $CP$\\ \hline
$PT$ & $PT$ & $T$  & $P$  & $1$  & $CPT$& $CT$ & $CP$ & $C$\\ \hline
$C$  & $C$  & $CP$ & $CT$ & $CPT$& $1$  & $P$  & $T$  & $PT$\\ \hline
$CP$ & $CP$ & $C$  & $CPT$& $CT$ & $P$  & $1$  & $PT$ & $T$\\ \hline
$CT$ & $CT$ & $CPT$& $C$  & $CP$ & $T$  & $PT$ & $1$  & $P$\\ \hline
$CPT$& $CPT$& $CT$ & $CP$ & $C$  & $PT$ & $T$  & $P$  & $1$\\ \hline
\end{tabular}.
}
\end{center}
In turn, for the extended automorphism group
$\{\Id,\,\star,\,\widetilde{\phantom{cc}},\,\widetilde{\star},\,
\overline{\phantom{cc}},\,\overline{\star},\,
\overline{\widetilde{\phantom{cc}}},\,\overline{\widetilde{\star}}\}$
in virtue of commutativity $\widetilde{\left(\cA^\star\right)}=
\left(\widetilde{\cA}\right)^\star$, 
$\overline{\left(\cA^\star\right)}=\left(\overline{\cA}\right)^\star$,
$\overline{\left(\widetilde{\cA}\right)}=
\widetilde{\left(\overline{\cA}\right)}$,
$\overline{\left(\widetilde{\cA^\star}\right)}=
\widetilde{\left(\overline{\cA}\right)^\star}$ and an involution property
$\star\star=\widetilde{\phantom{cc}}\widetilde{\phantom{cc}}=
\overline{\phantom{cc}}\;\overline{\phantom{cc}}=\Id$ we have a following
multiplication table
\begin{center}{\renewcommand{\arraystretch}{1.4}
\begin{tabular}{|c||c|c|c|c|c|c|c|c|}\hline
  & $\Id$ & $\star$ & $\widetilde{\phantom{cc}}$ & $\widetilde{\star}$ &
$\overline{\phantom{cc}}$ & $\overline{\star}$ & 
$\overline{\widetilde{\phantom{cc}}}$ &
$\overline{\widetilde{\star}}$ \\ \hline\hline
$\Id$ & $\Id$ & $\star$ & $\widetilde{\phantom{cc}}$ & $\widetilde{\star}$ &
$\overline{\phantom{cc}}$ & $\overline{\star}$ & 
$\overline{\widetilde{\phantom{cc}}}$ &
$\overline{\widetilde{\star}}$ \\ \hline
$\star$ & $\star$ & $\Id$ & $\widetilde{\star}$ & $\widetilde{\phantom{cc}}$ &
$\overline{\star}$ & $\overline{\phantom{cc}}$ &
$\overline{\widetilde{\star}}$ & $\overline{\widetilde{\phantom{cc}}}$\\ \hline
$\widetilde{\phantom{cc}}$ & 
$\widetilde{\phantom{cc}}$ & $\overline{\star}$ & $\Id$ &
$\star$ & $\overline{\widetilde{\phantom{cc}}}$ & $\overline{\widetilde{\star}}$ &
$\overline{\phantom{cc}}$ & $\overline{\star}$\\ \hline
$\widetilde{\star}$ & $\widetilde{\star}$ & $\widetilde{\phantom{cc}}$ &
$\star$ & $\Id$ & $\overline{\widetilde{\star}}$ & 
$\overline{\widetilde{\phantom{cc}}}$ &
$\overline{\star}$ & $\overline{\phantom{cc}}$\\ \hline
$\overline{\phantom{cc}}$ & $\overline{\phantom{cc}}$ & $\overline{\star}$ &
$\overline{\widetilde{\phantom{cc}}}$ & $\overline{\widetilde{\star}}$ & $\Id$ &
$\star$ & $\widetilde{\phantom{cc}}$ & $\widetilde{\star}$\\ \hline
$\overline{\star}$ & $\overline{\star}$ & $\overline{\phantom{cc}}$ &
$\overline{\widetilde{\star}}$ & 
$\overline{\widetilde{\phantom{cc}}}$ & $\star$ &
$\Id$ & $\widetilde{\star}$ & $\widetilde{\phantom{cc}}$\\ \hline
$\overline{\widetilde{\phantom{cc}}}$ & 
$\overline{\widetilde{\phantom{cc}}}$ &
$\overline{\widetilde{\star}}$ & 
$\overline{\phantom{cc}}$ & $\overline{\star}$ &
$\widetilde{\phantom{cc}}$ & $\widetilde{\star}$ & $\Id$ & $\star$\\ \hline
$\overline{\widetilde{\star}}$ & $\overline{\widetilde{\star}}$ &
$\overline{\widetilde{\phantom{cc}}}$ & $\overline{\star}$ & 
$\overline{\phantom{cc}}$ &
$\widetilde{\star}$ & $\widetilde{\phantom{cc}}$ & $\star$ & $\Id$\\ \hline
\end{tabular}.
}
\end{center}
The identity of multiplication tables proves the group isomorphism
\begin{multline}
\{1,\,P,\,T,\,PT,\,C,\,CP,\,CT,\,CPT\}\simeq\\
\{\Id,\,\star,\,\widetilde{\phantom{cc}},\,\widetilde{\star},\,
\overline{\phantom{cc}},\,\overline{\star},\,
\overline{\widetilde{\phantom{cc}}},\,\overline{\widetilde{\star}}\}\simeq
\dZ_2\otimes\dZ_2\otimes\dZ_2.\nonumber
\end{multline}
\end{proof}

Further, in the case of $P^2=T^2=\ldots=(CPT)^2=\pm 1$ and anticommutativity
of the elements we have an isomorphism between the $CPT$--group and a group
$\sExt(\C_n)$. The elements of $\sExt(\C_n)$ are spinor
representations of the automorphisms of the algebra $\C_n$. As mentioned
previously,
the Wedderburn--Artin Theorem allows us to define any spinor representaions for
the automorphisms of $\C_n$.
\subsection{Pseudoautomorphism $\cA\rightarrow\overline{\cA^\star}$}
Let us find a spinor representation of 
the pseudoautomorphism\index{pseudoautomorphism}
$\cA\rightarrow\overline{\cA^\star}$. The transformation
$\cA\rightarrow\overline{\cA^\star}$ is a composition of the
pseudoautomorphism $\cA\rightarrow\overline{\cA}$ and the automorphism
$\cA\rightarrow\cA^\star$. Under action of $\cA\rightarrow\cA^\star$ we have
$\e_i\rightarrow-\e_i$, where $\e_i$ are the units of $\cl_{p,q}$.
In turn, under action of $\cA\rightarrow\overline{\cA}$ the units $\e_i$
remain unaltered, $\e_i\rightarrow\e_i$. Therefore, under action of the
pseudoautomorphism $\cA\rightarrow\overline{\cA^\star}$ we obtain
$\e_i\rightarrow-\e_i$.

As it shown previously, the transformations $\cA\rightarrow\cA^\star$
and $\cA\rightarrow\overline{\cA}$ in the spinor representation are defined
by the expressions $\sA^\star=\sW\sA\sW^{-1}$ and 
$\overline{\sA}=\Pi\dot{\sA}\Pi^{-1}$. The order of the composition of these
transformations is not important ($\overline{\cA^\star}=
\left(\overline{\cA}\right)^\star=\overline{\left(\cA^\star\right)}$).
Indeed, if $\sW$ is a real matrix, then
\[
\overline{\sA^\star}=\sW\Pi\dot{\sA}\Pi^{-1}\sW^{-1}=
\Pi\left(\sW\sA\sW^{-1}\right)^{\cdot}\Pi^{-1},
\]
or
\begin{equation}\label{CPT1}
\overline{\sA^\star}=(\sW\Pi)\dot{\sA}(\sW\Pi)^{-1}=
(\Pi\sW)\dot{\sA}(\Pi\sW)^{-1}.
\end{equation}
Otherwise, we have $\overline{\left(\sA^\star\right)}=
\Pi\dot{\sW}\dot{\sA}\dot{\sW}^{-1}\Pi^{-1}$. Let us assume that $\sW$ is a
complex matrix, then $\dot{\sW}=-\sW$ and, therefore,
$\overline{\left(\sA^\star\right)}=\Pi(-\sW)\dot{\sA}(-\sW^{-1})\Pi^{-1}=
(\Pi\sW)\dot{\sA}(\Pi\sW)^{-1}$. Thus, the relation (\ref{CPT1}) is always
fulfilled.

Let $\sK=\Pi\sW$ be a matrix of the pseudoautomorphism
$\cA\rightarrow\overline{\cA^\star}$. Then (\ref{CPT1}) can be written as
follows
\begin{equation}\label{CPT2}
\overline{\sA^\star}=\sK\dot{\sA}\sK^{-1}.
\end{equation}
Since under action of the pseudoautomorphism 
$\cA\rightarrow\overline{\cA^\star}$ we have $\e_i\rightarrow-\e_i$,
in the spinor representation we must demand
$\cE_i\rightarrow-\cE_i$ also, or
\begin{equation}\label{CPT3}
\cE_i\longrightarrow -\cE_i=\sK\dot{\cE}_i\sK^{-1}.
\end{equation}
In the case of real subalgebras $\cl_{p,q}$ with the ring $\K\simeq\R$ we have
$\dot{\cE}_i=\cE_i$ and the relation (\ref{CPT3}) takes a form
\[
\cE_i\longrightarrow -\cE_i=\sK\cE_i\sK^{-1},
\]
whence
\[
\cE_i\sK=-\sK\cE_i,
\]
that is, the matrix $\sK$ is always anticommutes with the matrices of the
spinbasis. However, for the ring $\K\simeq\R$ the matrix $\Pi$ of
$\cA\rightarrow\overline{\cA}$ is proportional to the unit matrix,
$\Pi\sim\sI$ (Theorem \ref{tpseudo}). Therefore, in this case we have
$\sK\sim\sW$.

In the case of real subalgebras $\cl_{p,q}$ with the quaternionic ring
$\K\simeq\BH$ the spinbasis is divided into two parts:
a complex part $\left\{\dot{\cE}_{\alpha_t}=-\cE_{\alpha_t}\right\}$,
$(0<t\leq a)$, where $a$ is a number of the complex matrices of the spinbasis,
and a real part $\left\{\dot{\cE}_{\beta_s}=\cE_{\beta_s}\right\}$,
where $p+q-a$ is a number of the real matrices, $(0<s\leq p+q-a)$.
Then, in accordance with the spinbasis structure of the algebra
$\cl_{p,q}\simeq\M_{2^{\frac{p+q}{2}}}(\BH)$, the relation (\ref{CPT3})
can be written as follows
\[
\cE_{\alpha_t}\longrightarrow\cE_{\alpha_t}=\sK\cE_{\alpha_t}\sK^{-1},\quad
\cE_{\beta_s}\longrightarrow -\cE_{\beta_s}=\sK\cE_{\beta_s}\sK^{-1}.
\]
Whence
\begin{equation}\label{CPT4}
\cE_{\alpha_t}\sK=\sK\cE_{\alpha_t},\quad
\cE_{\beta_s}\sK=-\sK\cE_{\beta_s}.
\end{equation}
Thus, for the quaternionic types $p-q\equiv 4,6\pmod{8}$ the matrix $\sK$ of
the pseudoautomorphism $\cA\rightarrow\overline{\cA^\star}$ commutes with
the complex part and anticommutes with the real part of the spinbasis of
$\cl_{p,q}$. Hence it follows that a structure of the matrix $\sK$ is
analogous to the structure of the matrix $\Pi$ of the pseudoautomorphism
$\cA\rightarrow\overline{\cA}$ (see Theorem \ref{tpseudo}), that is,
the matrix $\sK$ of $\cA\rightarrow\overline{\cA^\star}$ is a product of
only complex or only real matrices.

So, let $0< a\leq p+q$ and let 
$\sK=\cE_{\alpha_1\alpha_2\cdots\alpha_a}$ be the matrix of the
pseudoautomorphism $\cA\rightarrow\overline{\cA^\star}$, then permutation
conditions of the matrix $\sK$ with the matrices $\cE_{\beta_s}$ of the
real part ($0< s\leq p+q-a$) and the matrices $\cE_{\alpha_t}$ of the
complex part ($0< t \leq a$) have the form
\begin{gather}
\sK\cE_{\beta_s}=(-1)^a\cE_{\beta_s}\sK,\label{CPT5}\\
\sK\cE_{\alpha_t}=(-1)^{a-t}\sigma(\alpha_t)
\cE_{\alpha_1\alpha_2\cdots
\alpha_{t-1}\alpha_{t+1}\cdots\alpha_a},\nonumber\\
\cE_{\alpha_t}\sK=(-1)^{t-1}\sigma(\alpha_t)
\cE_{\alpha_1\alpha_2\cdots
\alpha_{t-1}\alpha_{t+1}\cdots\alpha_a},\label{CPT6}
\end{gather}
that is, at $a\equiv 0\pmod{2}$ $\sK$ commutes with the real part and
anticommutes with the complex part of the spinbasis. Correspondingly,
at $a\equiv 1\pmod{2}$ $\sK$ anticommutes with the real and commutes with
the complex part. Further, let
$\sK=\cE_{\beta_1\beta_2\cdots\beta_{p+q-a}}$ be a product of the
real matrices of the spinbasis, then
\begin{eqnarray}
\sK\cE_{\beta_s}&=&(-1)^{p+q-a-s}\sigma(\beta_s)
\cE_{\beta_1\beta_2\cdots\beta_{s-1}\beta_{s+1}\cdots
\beta_{p+q-a}},\nonumber\\
\cE_{\beta_s}\sK&=&(-1)^{s-1}\sigma(\beta_s)\cE_{\beta_1\beta_2\cdots
\beta_{s-1}\beta_{s+1}\cdots\beta_{p+q-a}},\label{CPT7}
\end{eqnarray}
\begin{equation}\label{CPT8}
\sK\cE_{\alpha_t}=(-1)^{p+q-a}\cE_{\alpha_t}\sK,
\end{equation}
that is, at $p+q-a\equiv 0\pmod{2}$ the matrix $\sK$ anticommutes with the
real part and commutes with the complex part of the spinbasis.
Correspondingly, at $p+q-a\equiv 1\pmod{2}$ $\sK$ commutes with the real
and anticommutes with the complex part.
\begin{sloppypar}
A comparison of the conditions (\ref{CPT5})--(\ref{CPT6}) with (\ref{CPT4})
shows that the matrix 
$\sK=\cE_{\alpha_1\alpha_2\cdots\alpha_a}$ exists only if
$a\equiv 1\pmod{2}$. In turn, a comparison of the conditions
(\ref{CPT7})--(\ref{CPT8}) with (\ref{CPT4}) shows that the matrix
$\sK=\cE_{\beta_1\beta_2\cdots\beta_{p+q-a}}$ exists only if
$p+q-a\equiv 0\pmod{2}$.\end{sloppypar}

Let us find now squares of the matrix $\sK$. In accordance with obtained
conditions there exist two possibilities:\\
1) $\sK=\cE_{\alpha_1\alpha_2\cdots\alpha_a}$,
$a\equiv 1\pmod{2}$.
\[
\sK^2=\begin{cases}
+\sI, & \text{if $a_+-a_-\equiv 1,5\pmod{8}$},\\
-\sI, & \text{if $a_+-a_-\equiv 3,7\pmod{8}$},
\end{cases}
\]\begin{sloppypar}\noindent
where $a_+$ and $a_-$ are numbers of matrices with `$+$'- and `$-$'-squares in
the product $\cE_{\alpha_1\alpha_2\cdots\alpha_a}$.\\
2) $\sK=\cE_{\beta_1\beta_2\cdots\beta_{p+q-a}}$,
$p+q-a\equiv 0\pmod{2}$.\end{sloppypar}
\[
\sK^2=\begin{cases}
+\sI, & \text{if $b_+-b_-\equiv 0,4\pmod{8}$},\\
-\sI, & \text{if $b_+-b_-\equiv 2,6\pmod{8}$},
\end{cases}
\]\begin{sloppypar}\noindent
where $b_+$ and $b_-$ are numbers of matrices with `$+$'- and `$-$'-squares in
the product $\cE_{\beta_1\beta_2\cdots\beta_{p+q-a}}$,
respectively.\end{sloppypar}
\subsection{Pseudoantiautomorphism 
$\cA\rightarrow\overline{\widetilde{\cA}}$}
The pseudoantiautomorphism\index{pseudoantiautomorphism}
$\cA\rightarrow\overline{\widetilde{\cA}}$ is the
composition of the pseudoautomorphism $\cA\rightarrow\overline{\cA}$ with
the antiautomorphism $\cA\rightarrow\widetilde{\cA}$. Under action of
$\cA\rightarrow\widetilde{\cA}$ the units $\e_i$ remain unaltered,
$\e_i\rightarrow\e_i$. Analogously, under action of 
$\cA\rightarrow\overline{\cA}$ we have $\e_i\rightarrow\e_i$. Therefore,
under action of the pseudoantiautomorphism 
$\cA\rightarrow\overline{\widetilde{\cA}}$ the units $\e_i$ remain
unaltered also, $\e_i\rightarrow\e_i$.

The spinor representations of the transformations 
$\cA\rightarrow\widetilde{\cA}$ and $\cA\rightarrow\overline{\cA}$ are
defined by the expressions 
$\widetilde{\sA}=\sE\sA^{\sT}\sE^{-1}$ and
$\overline{\sA}=\Pi\dot{\sA}\Pi^{-1}$, respectively. Let us find a spinor
representation of the transformation 
$\cA\rightarrow\overline{\widetilde{\cA}}$. The order of the composition of
these transformations is not important,
$\overline{\widetilde{\cA}}=\widetilde{\left(\overline{\cA}\right)}=
\overline{\left(\widetilde{\cA}\right)}$. Indeed,
\[
\overline{\widetilde{\sA}}=\sE\left(\Pi\dot{\sA}\Pi^{-1}\right)^{\sT}\sE^{-1}=
\Pi\left(\sE\sA^{\sT}\sE^{-1}\right)^{\cdot}\Pi^{-1}
\]
or,
\begin{equation}\label{CPT9}
\overline{\widetilde{\sA}}=(\sE\Pi)\left(\dot{\sA}\right)^{\sT}(\sE\Pi)^{-1}=
(\Pi\sE)\left(\sA^{\sT}\right)^{\cdot}(\Pi\sE)^{-1},
\end{equation}\begin{sloppypar}\noindent
since $\Pi^{-1}=\Pi^{\sT}$ and
$\overline{\widetilde{\sA}}=\Pi\dot{\sE}\left(\sA^{\sT}\right)^{\cdot}
\dot{\sE}^{-1}\Pi^{-1}=\Pi\sE\left(\sA^{\sT}\right)^{\cdot}\sE^{-1}\Pi^{-1}$
in the case when $\dot{\sE}=\sE$ is a real matrix and
$\overline{\widetilde{\sA}}=\Pi\dot{\sE}\left(\sA^{\sT}\right)^\cdot
\dot{\sE}^{-1}\Pi^{-1}=\Pi(-\sE)\left(\sA^{\sT}\right)^\cdot(-\sE^{-1})\Pi^{-1}=
\Pi\sE\left(\sA^{\sT}\right)^\cdot(\Pi\sE)^{-1}$ in the case when
$\dot{\sE}=-\sE$ is a complex matrix. Let $\sS=\Pi\sE$ be a matrix of the
pseudoantiautomorphism $\cA\rightarrow\overline{\widetilde{\cA}}$ in the
spinor representation. Then (\ref{CPT9}) can be rewritten as follows\end{sloppypar}
\begin{equation}\label{CPT10}
\overline{\widetilde{\sA}}=\sS\left(\sA^{\sT}\right)^\cdot\sS^{-1}.
\end{equation}
Since under action of the transformation 
$\cA\rightarrow\overline{\widetilde{\cA}}$ we have $\e_i\rightarrow\e_i$,
in the spinor representation we must demand 
$\cE_i\rightarrow\cE_i$ also, or
\begin{equation}\label{CPT11}
\cE_i\longrightarrow\cE_i=\sS\dot{\cE}^{\sT}_i\sS^{-1}.
\end{equation}
In the case of real subalgebras $\cl_{p,q}$ with the ring $\K\simeq\R$ 
we have $\dot{\cE}_i=\cE_i$ and, therefore, the relation (\ref{CPT11})
takes a form
\begin{equation}\label{CPT12}
\cE_i\longrightarrow\cE_i=\sS\cE^{\sT}_i\sS^{-1}.
\end{equation}
Let $\left\{\cE_{\gamma_i}\right\}$ be a set of symmetric matrices
$\left(\cE^{\sT}_{\gamma_i}=\cE_{\gamma_i}\right)$ and let
$\left\{\cE_{\delta_j}\right\}$ be a set of skewsymmetric matrices
$\left(\cE^{\sT}_{\delta_j}=-\cE_{\delta_j}\right)$ of the spinbasis of
the algebra $\cl_{p,q}$. Then from the relation (\ref{CPT12}) it follows
\[
\cE_{\gamma_i}\longrightarrow\cE_{\gamma_i}=\sS\cE_{\gamma_i}\sS^{-1},\quad
\cE_{\delta_j}\longrightarrow\cE_{\delta_j}=-\sS\cE_{\delta_j}\sS^{-1}.
\]
Whence
\[
\cE_{\gamma_i}\sS=\sS\cE_{\gamma_i},\quad
\cE_{\delta_j}\sS=-\sS\cE_{\delta_j},
\]
that is, the matrix $\sS$ of the pseudoantiautomorphism
$\cA\rightarrow\overline{\widetilde{\cA}}$ in the case of $\K\simeq\R$
commutes with the symmetric part and anticommutes with the skewsymmetric
part of the spinbasis of $\cl_{p,q}$. In virtue of Theorem \ref{tpseudo},
over the ring $\K\simeq\R$
the matrix $\Pi$ of the pseudoautomorphism $\cA\rightarrow\overline{\cA}$
is proportional to the unit matrix, $\Pi\sim\sI$. Therefore, in this case
we have $\sS\sim\sE$ and an explicit form of $\sS$ coincides with $\sE$.

Further, in case of the quaternionic ring $\K\simeq\BH$,
$p-q\equiv 4,6\pmod{8}$, a spinbasis of $\cl_{p,q}$ contains both complex
matrices $\cE_{\alpha_t}$ and real matrices $\cE_{\beta_s}$, among which
there are symmetric and skewsymmetric matrices. It is obvious that the sets
of complex and real matrices do not coincide with the sets of symmetric and
skewsymmetric matrices. Let $\left\{\cE_{\alpha_t}\right\}$ be a complex
part of the spinbasis, then the relation (\ref{CPT11}) takes a form
\begin{equation}\label{CPT13}
\cE_{\alpha_t}\longrightarrow\cE_{\alpha_t}=-\sS\cE^{\sT}_{\alpha_t}\sS^{-1}.
\end{equation}
Correspondingly, let $\left\{\cE_{\alpha_\gamma}\right\}$ and
$\left\{\cE_{\alpha_\delta}\right\}$ be the sets of symmetric and
skewsymmetric matrices of the complex part. Then the relation (\ref{CPT13})
can be written as follows
\[
\cE_{\alpha_\gamma}\longrightarrow\cE_{\alpha_\gamma}=
-\sS\cE_{\alpha_\gamma}\sS^{-1},\quad
\cE_{\alpha_\delta}\longrightarrow\cE_{\alpha_\delta}=
\sS\cE_{\alpha\delta}\sS^{-1}.
\]
Whence
\begin{equation}\label{CPT13'}
\cE_{\alpha_\gamma}\sS=-\sS\cE_{\alpha_\gamma},\quad
\cE_{\alpha_\delta}\sS=\sS\cE_{\alpha_\delta}.
\end{equation}
Therefore, the matrix $\sS$ of the pseudoantiautomorphism
$\cA\rightarrow\overline{\widetilde{\cA}}$ anticommutes with the complex
symmetric matrices and commutes with the complex skewsymmetric matrices of
the spinbasis of $\cl_{p,q}$.

Let us consider now the real part $\left\{\cE_{\beta_s}\right\}$ of the
spinbasis of $\cl_{p,q}$, $p-q\equiv 4,6\pmod{8}$. In this case the
relation (\ref{CPT11}) takes a form
\begin{equation}\label{CPT14}
\cE_{\beta_s}\longrightarrow\cE_{\beta_s}=\sS\cE^{\sT}_{\beta_s}\sS^{-1}.
\end{equation}
Let $\left\{\cE_{\beta_\gamma}\right\}$ and $\left\{\cE_{\beta_\delta}\right\}$
be the sets of real symmetric and real skewsymmetric matrices, respectively.
Then the relation (\ref{CPT14}) can be written as follows
\[
\cE_{\beta_\gamma}\longrightarrow\cE_{\beta_\gamma}=
\sS\cE_{\beta_\gamma}\sS^{-1},\quad
\cE_{\beta_\delta}\longrightarrow\cE_{\beta_\delta}=
-\sS\cE_{\beta_\delta}\sS^{-1}.
\]
Whence
\begin{equation}\label{CPT14'}
\cE_{\beta_\gamma}\sS=\sS\cE_{\beta_\gamma},\quad
\cE_{\beta_\delta}\sS=-\sS\cE_{\beta_\delta}.
\end{equation}
Thus, the matrix $\sS$ of the transformation
$\cA\rightarrow\overline{\widetilde{\cA}}$ commutes with the real
symmetric matrices and anticommutes with the real skewsymmetric matrices
of the spinbasis of $\cl_{p,q}$.

Let us find now an explicit form of the matrix $\sS=\Pi\sE$. In accordance
with Theorem \ref{tpseudo} for the quaternionic types 
$p-q\equiv 4,6\pmod{8}$ the matrix $\Pi$ takes the two different forms:
1) $\Pi=\cE_{\alpha_1\alpha_2\cdots\alpha_a}$ is the product
of complex matrices at $a\equiv 0\pmod{2}$; 2)
$\Pi=\cE_{\beta_1\beta_2\cdots\beta_b}$ is the product of real
matrices at $b\equiv 1\pmod{2}$. In turn, for the matrix $\sE$ of the
antiautomorphism $\cA\rightarrow\widetilde{\cA}$ over the ring $\K\simeq\BH$
(see Theorem \ref{tautr}) we have the following two forms:
1) $\sE=\cE_{j_1j_2\cdots j_k}$ is the product of skewsymmetric
matrices at $k\equiv 0\pmod{2}$; 2) $\sE=\cE_{i_1i_2\cdots
i_{p+q-k}}$ is the product of symmetric matrices at $k\equiv 1\pmod{2}$.
Thus, in accordance with definition $\sS=\Pi\sE$ we have four different
products: 
$\sS=\cE_{\alpha_1\alpha_2\cdots\alpha_a}\cE_{j_1j_2\cdots j_k}$,
$\sS=\cE_{\alpha_1\alpha_2\cdots\alpha_a}\cE_{i_1i_2\cdots i_{p+q-k}}$,
$\sS=\cE_{\beta_1\beta_2\cdots\beta_b}\cE_{j_1j_2\cdots j_k}$,
$\sS=\cE_{\beta_1\beta_2\cdots\beta_b}\cE_{i_1i_2\cdots i_{p+q-k}}$.
It is obvious that in the given products there are identical matrices.

Let us examine the first product $\sS=\cE_{\alpha_1\alpha_2\cdots\alpha_a}
\cE_{j_1j_2\cdots j_k}$. Since in this case $\Pi$ contains all the complex
matrices of the spinbasis, among which there are symmetric and skewsymmetric
matrices, and $\sE$ contains all the skewsymmetric matrices of the spinbasis,
then $\Pi$ and $\sE$ contain a quantity of identical matrices
(complex skewsymmetric matrices). Let $m$ be a number of the complex
skewsymmetric matrices of the spinbasis of the algebra $\cl_{p,q}$,
$p-q\equiv 4,6\pmod{8}$. Then the product $\sS=\Pi\sE$ takes a form
\[
\cE_{\alpha_1\alpha_2\cdots\alpha_a j_1j_2\cdots j_k}=
(-1)^{\cN}\sigma(i_1)\sigma(i_2)\cdots\sigma(i_m)
\cE_{c_1c_2\cdots c_s},
\]\begin{sloppypar}\noindent
where the indices $c_1,\ldots, c_s$ present itself a totality of the
indices $\alpha_1,\ldots,\alpha_a, j_1,\ldots,j_k$ obtained after
removal of the indices occured twice; $\cN$ is a number of inversions.
\end{sloppypar}
First of all, let us remark that $\cE_{c_1c_2\cdots c_s}$ is
an even product, since the original product $\Pi\sE$ is the even product
also. Besides, the product $\cE_{c_1c_2\cdots c_s}$ contains
all the complex symmetric matrices and all the real skewsymmetric matrices
of the spinbasis.

Let us find now permutation conditions between the matrix
$\sS=\cE_{c_1c_2\cdots c_s}$ and the units of the spinbasis of
$\cl_{p,q}$, $p-q\equiv 4,6\pmod{8}$. For the complex symmetric matrices
$\cE_{\alpha_\gamma}$ and complex skewsymmetric matrices $\cE_{\alpha_\delta}$
we have
\begin{gather}
\sS\cE_{\alpha_\gamma}=(-1)^{s-\gamma}\cE_{c_1c_2\cdots c_{s-2}},
\nonumber\\
\cE_{\alpha_\gamma}\sS=(-1)^{\gamma-1}\cE_{c_1c_2\cdots c_{s-2}},
\label{CPT15}\\
\sS\cE_{\alpha_\delta}=(-1)^s\cE_{\alpha_\delta}\sS,
\label{CPT16}
\end{gather}
that is, the matrix $\sS$ always anticommutes with the complex symmetric
matrices and always commutes with the complex skewsymmetric matrices,
since $s\equiv 0\pmod{2}$. Further, for the real symmetric matrices
$\cE_{\beta_\gamma}$ and real skewsymmetric matrices $\cE_{\beta_\delta}$
we have
\begin{gather}
\sS\cE_{\beta_\gamma}=(-1)^s\cE_{\beta_\gamma}\sS,\label{CPT17}\\
\sS\cE_{\beta_\delta}=(-1)^{s-\delta}\cE_{c_1c_2\cdots c_{s-2}},
\nonumber\\
\cE_{\beta_\delta}\sS=(-1)^{\delta-1}\cE_{c_1c_2\cdots c_{s-2}}.
\label{CPT18}
\end{gather}
Therefore, $\sS$ always commutes with the real symmetric matrices and
always anticommutes with the real skewsymmetric matrices.

A comparison of the obtained conditions (\ref{CPT15})--(\ref{CPT18}) with
(\ref{CPT13'}) and (\ref{CPT14'}) shows that 
$\sS=\cE_{c_1c_2\cdots c_s}$ automatically satisfies the
conditions, which define the matrix of the pseudoantiautomorphism
$\cA\rightarrow\overline{\widetilde{\cA}}$.

Let examine the second product 
$\sS=\cE_{\alpha_1\alpha_2\cdots\alpha_a}\cE_{i_1i_2\cdots i_{p+q-k}}$.
In this case $\Pi$ contains all the complex matrices of the spinbasis,
among which there are symmetric and skewsymmetric matrices, and $\sE$
contains all the symmetric matrices of the spinbasis. Thus, $\Pi$ and $\sE$
contain a quantity of identical complex symmetric matrices.
Let $l$ be a number of the complex symmetric matrices, then for the
product $\sS=\Pi\sE$ we obtain
\[
\cE_{\alpha_1\alpha_2\cdots\alpha_ai_1i_2\cdots
i_{p+q-k}}=(-1)^{\cN}\sigma(i_1)\sigma(i_2)\cdots\sigma(i_l)
\cE_{d_1d_2\cdots d_g},
\]
where $d_1,\ldots, d_g$ present itself a totality of the indices
$\alpha_1,\ldots,\alpha_a,i_1,\ldots,i_{p+q-k}$ after removal of the indices
occured twice. The product $\sS=\cE_{d_1d_2\cdots d_g}$ is odd,
since the original product $\Pi\sE$ is odd also. Besides,
$\cE_{d_1d_2\cdots d_g}$ contains all the complex skewsymmetric
matrices and all the real symmetric matrices of the spinbasis.

Let us find permutation conditions of the matrix 
$\sS=\cE_{d_1 d_2\cdots d_g}$ with the units of the spinbasis of
$\cl_{p,q}$. For the complex part of the spinbasis we have
\begin{gather}
\sS\cE_{\alpha_\gamma}=(-1)^g\cE_{\alpha_\gamma}\sS,\label{CPT19}\\
\sS\cE_{\alpha_\delta}=(-1)^{g-\delta}\cE_{d_1d_2\cdots d_{g-2}},
\nonumber\\
\cE_{\alpha_\delta}\sS=(-1)^{\delta-1}\cE_{d_1d_2\cdots d_{g-2}},
\label{CPT20}
\end{gather}
that is, the matrix $\sS$ anticommutes with the complex symmetric
matrices and commutes with the complex skewsymmetric matrices, since
$g\equiv 1\pmod{2}$. For the real part of the spinbasis of $\cl_{p,q}$
we obtain
\begin{gather}
\sS\cE_{\beta_\gamma}=(-1)^{g-\gamma}\cE_{d_1d_2\cdots d_{g-2}},
\nonumber\\
\cE_{\beta_\gamma}\sS=(-1)^{\gamma-1}\cE_{d_1d_2\cdots d_{g-2}},
\label{CPT21}\\
\sS\cE_{\beta_\delta}=(-1)^g\cE_{\beta_\delta}\sS.\label{CPT22}
\end{gather}
Therefore, $\sS$ commutes with the real symmetric matrices and anticommutes
with the real skewsymmetric matrices.

Comparing the obtained conditions (\ref{CPT19})--(\ref{CPT22}) with
the conditions (\ref{CPT13'}) and (\ref{CPT14'}) we see that 
$\sS=\cE_{d_1d_2\cdots d_g}$ automatically satisfies the
conditions which define the matrix of the transformation
$\cA\rightarrow\overline{\widetilde{\cA}}$.

Let us consider now the third product 
$\sS=\cE_{\beta_1\beta_2\cdots\beta_b}\cE_{j_1j_2\cdots k}$.
In this product the matrix $\Pi$ of the pseudoautomorphism
$\cA\rightarrow\overline{\cA}$ contains all the real matrices of the
spinbasis, among which there are symmetric and skewsymmetric matrices,
and the matrix $\sE$ of $\cA\rightarrow\widetilde{\cA}$ contains all the
skewsymmetric matrices of the spinbasis, among which there are both the real
and complex matrices. Therefore, $\sS$ contains a quantity of identical
real skewsymmetric matrices. Let $u$ be a number of the real skewsymmetric
matrices of the spinbasis of the algebra $\cl_{p,q}$, $p-q\equiv 4,6\pmod{8}$,
then for the product $\sS$ we obtain
\[
\cE_{\beta_1\beta_2\cdots\beta_bj_1j_2\cdots
j_k}=(-1)^{\cN}\sigma(i_1)\sigma(i_2)\cdots\sigma(i_u)
\cE_{e_1e_2\cdots e_h},
\]\begin{sloppypar}\noindent
where the indices $e_1,e_2,\ldots, e_n$ present itself a totality of the
indices $\beta_1,\ldots,\beta_b,j_1,\ldots,j_k$ after removal of the
indices occurred twice. The product $\sS=\cE_{e_1e_2\cdots e_h}$
is odd, since the original product $\Pi\sE$ is odd also. It is easy to see
that $\cE_{e_1e_2\cdots e_h}$ contains all the real symmetric
matrices and all the complex skewsymmetric matrices of the spinbasis.
Therefore, the matrix $\sS=\cE_{e_1e_2\cdots e_h}$ is similar to
the matrix $\sS=\cE_{d_1d_2\cdots d_g}$, and its permutation
conditions with the units of the spinbasis of $\cl_{p,q}$ are equivalent to
the relations (\ref{CPT19})--(\ref{CPT22}).\end{sloppypar}

Finally, let us examine the fourth product
$\sS=\cE_{\beta_1\beta_2\cdots\beta_b}\cE_{i_1i_2\cdots i_{p+q-k}}$.
In turn, this product contains a quantity of identical real symmetric
matrices. Let $v$ be a number of the real symmetric matrices of the spinbasis
of $\cl_{p,q}$, then
\[
\cE_{\beta_1\beta_2\cdots\beta_b
i_1i_2\cdots i_{p+q-k}}=(-1)^{\cN}\sigma(i_1)\sigma(i_2)
\cdots\sigma(i_v)\cE_{f_1f_2\cdots f_w},
\]
where $f_1,\ldots,f_w$ present itself a totality of the indices
$\beta_1,\ldots,\beta_b,i_1,\ldots,i_{p+q-k}$ after removal of the indices
occurred twice. The product $\sS=\cE_{f_1f_2\cdots f_w}$ is even,
since the original product $\Pi\sE$ is even also. It is easy to see that
$\cE_{f_1f_2\cdots f_w}$ contains all the real skewsymmetric
matrices and all the complex symmetric matrices of the spinbasis.
Therefore, the matrix $\sS=\cE_{f_1f_2\cdots f_w}$ is similar to
the matrix $\sS=\cE_{c_1c_2\cdots c_s}$, and its permutation
conditions with the units of the spinbasis are equivalent to the relations
(\ref{CPT15})--(\ref{CPT18}).

Thus, from the four products we have only two non-equivalent products.
The squares of the non-equivalent matrices 
$\sS=\cE_{c_1c_2\cdots c_s}$ ($s\equiv 0\pmod{2}$) and
$\sS=\cE_{d_1d_2\cdots d_g}$ ($g\equiv 1\pmod{2}$) are
\begin{gather}
\sS^2=\left(\cE_{c_1c_2\cdots c_s}\right)^2=
\begin{cases}
+\sI, & \text{if $u+l\equiv 0,4\pmod{8}$},\\
-\sI, & \text{if $u+l\equiv 2,6\pmod{8}$};
\end{cases}\label{CPT22'}\\
\sS^2=\left(\cE_{d_1d_2\cdots d_g}\right)^2=
\begin{cases}
+\sI, & \text{if $m+v\equiv 1,5\pmod{8}$},\\
-\sI, & \text{if $m+v\equiv 3,7\pmod{8}$}.
\end{cases}\label{CPT22''}
\end{gather}
\subsection{Pseudoantiautomorphism 
$\cA\rightarrow\overline{\widetilde{\cA^\star}}$}
The pseudoantiautomorphism\index{pseudoantiautomorphism}
$\cA\rightarrow\overline{\widetilde{\cA^\star}}$,
that defines the $CPT$-transformation, is a composition of the
pseudoautomorphism $\cA\rightarrow\overline{\cA}$, antiautomorphism
$\cA\rightarrow\widetilde{\cA}$ and automorphism $\cA\rightarrow\cA^\star$.
Under action of the automorphism $\cA\rightarrow\cA^\star$ the units of
$\cl_{p,q}$ change the sign, $\e_i\rightarrow-\e_i$. In turn, under action
of the transformations $\cA\rightarrow\overline{\cA}$ and
$\cA\rightarrow\widetilde{\cA}$ the units remain unaltered,
$\e_i\rightarrow\e_i$. Therefore, under action of the pseudoantiautomorphism
$\cA\rightarrow\overline{\widetilde{\cA^\star}}$ the units change the sign,
$\e_i\rightarrow-\e_i$.

The spinor representations of the transformations $\cA\rightarrow\cA^\star$,
$\cA\rightarrow\widetilde{\cA}$ and $\cA\rightarrow\overline{\cA}$
have the form: $\sA^\star=\sW\sA\sW^{-1}$, 
$\widetilde{\sA}=\sE\sA^{\sT}\sE^{-1}$ and 
$\overline{\sA}=\Pi\dot{\sA}\Pi^{-1}$. Let us find a matrix of the
transformation $\cA\rightarrow\overline{\widetilde{\cA^\star}}$.
We will consider the pseudoantiautomorphism
$\cA\rightarrow\overline{\widetilde{\cA^\star}}$ as a composition of the
pseudoautomorphism $\cA\rightarrow\overline{\cA}$ with the antiautomorphism
$\cA\rightarrow\widetilde{\cA^\star}$. The spinor representation of
$\cA\rightarrow\widetilde{\cA^\star}$ is 
$\widetilde{\cA^\star}=\sC\sA^{\sT}\sC^{-1}$, where $\sC=\sE\sW$. Since
$\overline{\widetilde{\cA^\star}}=
\widetilde{\left(\overline{\cA}\right)^\star}=
\overline{\left(\widetilde{\cA^\star}\right)}$, then
\[
\overline{\widetilde{\sA^\star}}=
\sC\left(\Pi\dot{\sA}\Pi^{-1}\right)^{\sT}\sC^{-1}=
\Pi\left(\sC\sA^{\sT}\sC^{-1}\right)^{\cdot}\Pi^{-1},
\]
or
\begin{equation}\label{CPT23}
\overline{\widetilde{\sA^\star}}=(\sC\Pi)\dot{\sA}^{\sT}(\sC\Pi)^{-1}=
(\Pi\sC)\dot{\sA}^{\sT}(\Pi\sC)^{-1},
\end{equation}\begin{sloppypar}\noindent
since $\Pi^{-1}=\Pi^{\sT}$ and $\overline{\widetilde{\sA^\star}}=
\Pi\dot{\sC}\left(\sA^{\sT}\right)^{\cdot}\dot{\sC}^{-1}\Pi^{-1}=
\Pi\sC\dot{\sA}^{\sT}\sC^{-1}\Pi^{-1}$ in the case when $\dot{\sC}=\sC$ is
a real matrix, and also $\overline{\widetilde{\sA^\star}}=\Pi\dot{\sC}
\left(\sA^{\sT}\right)^{\cdot}\dot{\sC}^{-1}\Pi^{-1}=\Pi(-\sC)
\left(\sA^{\sT}\right)^{\cdot}\left(-\sC^{-1}\right)\Pi^{-1}=
\Pi\sC\dot{\sA}^{\sT}\sC^{-1}\Pi^{-1}$ in the case when $\dot{\sC}=-\sC$ is
a complex matrix.\end{sloppypar}

Let $\sF=\Pi\sC$ (or $\sF=\Pi\sE\sW$) be a matrix of the
pseudoantiautomorphism $\cA\rightarrow\overline{\widetilde{\cA^\star}}$.
Then the relation (\ref{CPT23}) can be written as follows
\begin{equation}\label{CPT24}
\overline{\widetilde{\sA^\star}}=\sF\dot{\sA}^{\sT}\sF^{-1}.
\end{equation}
Since under action of the transformation
$\cA\rightarrow\overline{\widetilde{\cA^\star}}$ we have
$\e_i\rightarrow-\e_i$, in the spinor representation we must demand
$\cE_i\rightarrow-\cE_i$ also, or
\begin{equation}\label{CPT25}
\cE_i\longrightarrow-\cE_i=\sF\dot{\cE}^{\sT}_i\sF^{-1}.
\end{equation}
In case of the real subalgebras $\cl_{p,q}$ with the ring $\K\simeq\R$,
$p-q\equiv 0,2\pmod{8}$, we have $\dot{\cE}_i=\cE_i$ for all matrices of
the spinbasis and, therefore, the relation (\ref{CPT25}) takes a form
\begin{equation}\label{CPT26}
\cE_{i}\longrightarrow-\cE_i=\sF\cE^{\sT}_i\sF^{-1}.
\end{equation}
Let $\left\{\cE_{\gamma_i}\right\}\cup\left\{\cE_{\delta_j}\right\}$ be
a spinbasis of the algebra $\cl_{p,q}$ over the ring $\K\simeq\R$,
($\cE^{\sT}_{\gamma_i}=\cE_{\gamma_i},\,\cE^{\sT}_{\delta_j}=-\cE_{\delta_j}$).
Then the relation (\ref{CPT26}) can be written in the form
\[
\cE_{\gamma_i}\longrightarrow-\cE_{\gamma_i}=\sF\cE_{\gamma_i}\sF^{-1},\quad
\cE_{\delta_j}\longrightarrow\cE_{\delta_j}=\sF\cE_{\delta_j}\sF^{-1}.
\]
Whence
\[
\cE_{\gamma_i}\sF=-\sF\cE_{\gamma_i},\quad
\cE_{\delta_j}\sF=\sF\cE_{\delta_j},
\]
that is, the matrix $\sF$ of the pseudoantiautomorphism
$\cA\rightarrow\overline{\widetilde{\cA^\star}}$ in case of the ring
$\K\simeq\R$ anticommutes with the symmetric part of the spinbasis of
$\cl_{p,q}$ and commutes with the skewsymmetric part. In virtue of
Theorem \ref{tpseudo} the matrix $\Pi$ of the pseudoautomorphism
$\cA\rightarrow\overline{\cA}$ over the ring $\K\simeq\R$ is proportional
to the unit matrix, $\Pi\simeq\sI$. Therefore, in this case $\sF\sim\sC$
($\sF\sim\sE\sW$) and an explicit form of the matrix $\sF$ coincides with
$\sC$ (see Theorem \ref{tautr}).

In case of the quaternionic ring $\K\simeq\BH$, $p-q\equiv 4,6\pmod{8}$,
the spinbasis of $\cl_{p,q}$ contains both complex matrices
$\cE_{\alpha_t}$ and real matrices $\cE_{\beta_r}$. For the complex part
the relation (\ref{CPT25}) takes a form
\[
\cE_{\alpha_t}\longrightarrow\cE_{\alpha_t}=\sF\cE^{\sT}_{\alpha_t}\sF^{-1},
\]
Or, taking into account complex symmetric and complex skewsymmetric
components of the spinbasis, we obtain
\[
\cE_{\alpha_\gamma}\longrightarrow\cE_{\alpha_\gamma}=
\sF\cE_{\alpha_\gamma}\sF^{-1},\quad
\cE_{\alpha_\delta}\longrightarrow\cE_{\alpha_\delta}=
-\sF\cE_{\alpha_\delta}\sF^{-1}.
\]
Whence
\begin{equation}\label{CPT27}
\cE_{\alpha_\gamma}\sF=\sF\cE_{\alpha_\gamma},\quad
\cE_{\alpha_\delta}\sF=-\sF\cE_{\alpha_\delta}.
\end{equation}
For the real part of the spinbasis of $\cl_{p,q}$ from (\ref{CPT25})
we obtain
\[
\cE_{\beta_r}\longrightarrow-\cE_{\beta_r}=\sF\cE^{\sT}_{\beta_r}\sF^{-1},
\]
or, taking into account symmetric and skewsymmetric components of the real
part of the spinbasis, we find
\[
\cE_{\beta_\gamma}\longrightarrow-\cE_{\beta_\gamma}=
\sF\cE_{\beta_\gamma}\sF^{-1},\quad
\cE_{\beta_\delta}\longrightarrow\cE_{\beta_\delta}=
\sF\cE_{\beta_\delta}\sF^{-1}.
\]
Whence
\begin{equation}\label{CPT28}
\cE_{\beta_\gamma}\sF=-\sF\cE_{\beta_\gamma},\quad
\cE_{\beta_\delta}\sF=\sF\cE_{\beta_\delta}.
\end{equation}
Therefore, the matrix $\sF$ of the pseudoantiautomorphism
$\cA\rightarrow\overline{\widetilde{\cA^\star}}$ commutes with the complex
symmetric and real skewsymmetric matrices, and also $\sF$ anticommutes
with the complex skewsymmetric and real symmetric matrices of the
spinbasis of $\cl_{p,q}$, $p-q\equiv 4,6\pmod{8}$.

Let us find an explicit form of the matrix $\sF=\Pi\sC$. It is easy to see
that in virtue of $\sF=\Pi\sC=(\Pi\sE)\sW=\sS\sW$ the matrix $\sF$ is a dual
with respect to the matrix $\sS$ of the pseudoantiautomorphism
$\cA\rightarrow\overline{\widetilde{\cA}}$. In accordance with Theorem
\ref{tpseudo} for the quaternionic types $p-q\equiv 4,6\pmod{8}$ the matrix
$\Pi$ has two different forms: $\Pi=\cE_{\alpha_1\alpha_2\cdots
\alpha_a}$ ($\dot{\cE}_{\alpha_t}=-\cE_{\alpha_t}$), $a\equiv 0\pmod{2}$;
$\Pi=\cE_{\beta_1\beta_2\cdots\beta_b}$ 
($\dot{\cE}_{\beta_r}=\cE_{\beta_r}$), $b\equiv 1\pmod{2}$. In turn,
for the quaternionic types the matrix $\sC$ of the antiautomorphism
$\cA\rightarrow\widetilde{\cA^\star}$ has the two forms (see Theorem \ref{tautr}):
1) $\sC=\cE_{i_1i_2\cdots i_{p+q-k}}$ is the product of all
symmetric matrices of the spinbasis of $\cl_{p,q}$ at $p+q-k\equiv 0\pmod{2}$;
2) $\sC=\cE_{j_1j_2\cdots j_k}$ is the product of all
skewsymmetric matrices of the spinbasis at $k\equiv 1\pmod{2}$. Thus,
in accordance with definition $\sF=\Pi\sC$ we have four products:
$\sF=\cE_{\alpha_1\alpha_2\cdots\alpha_a}\cE_{i_1i_2\cdots i_{p+q-k}}$,
$\sF=\cE_{\alpha_1\alpha_2\cdots\alpha_a}\cE_{j_1j_2\cdots j_k}$,
$\sF=\cE_{\beta_1\beta_2\cdots\beta_b}\cE_{i_1i_2\cdots i_{p+q-k}}$,
$\sF=\cE_{\beta_1\beta_2\cdots\beta_b}\cE_{j_1j_2\cdots j_k}$.

Let us examine the first product
$\sF=\cE_{\alpha_1\alpha_2\cdots\alpha_a}\cE_{i_1i_2\cdots i_{p+q-k}}$.
In this case $\Pi$ contains all the complex matrices of the spinbasis, among
which there are symmetric and skewsymmetric matrices. In turn, $\sC$ contains
all the symmetric matrices, among which there are both complex and real
matrices. It is obvious that in this case $\Pi\sC$ contains a quantity
of identical complex symmetric matrices. Therefore, the product $\sF$
consists of all the complex skewsymmetric matrices and all the real
symmetric matrices of the spinbasis. The product $\sF$ is even, since the
original product $\Pi\sC$ is even also. It is easy to see that $\sF$
coincides with the product $\cE_{d_1d_2\cdots d_g}$ at
$g\equiv 0\pmod{2}$.

Let us find permutation conditions of the matrix 
$\sF=\cE_{d_1d_2\cdots d_g}$ with the units of the spinbasis of
$\cl_{p,q}$, $p-q\equiv 4,6\pmod{8}$. For the complex and real parts
we obtain
\begin{gather}
\sF\cE_{\alpha_\gamma}=(-1)^g\cE_{\alpha_\gamma}\sF,\label{CPT29}\\
\sF\cE_{\alpha_\delta}=(-1)^{g-\delta}\cE_{d_1d_2\cdots d_{g-2}},
\nonumber\\
\cE_{\alpha_\delta}\sF=(-1)^{\delta-1}\cE_{d_1d_2\cdots d_{g-2}},
\label{CPT30}\\
\sF\cE_{\beta_\gamma}=(-1)^{g-\gamma}\cE_{d_1d_2\cdots d_{g-2}},
\nonumber\\
\cE_{\beta_\gamma}\sF=(-1)^{\gamma-1}\cE_{d_1d_2\cdots d_{g-2}},
\label{CPT31}\\
\sF\cE_{\beta_\delta}=(-1)^g\cE_{\beta_\delta}\sF.\label{CPT32}
\end{gather}
Therefore, since $g\equiv 0\pmod{2}$ the matrix $\sF$ always commutes
with the complex symmetric and real skewsymmetric matrices and always
anticommutes with the complex skewsymmetric and real symmetric matrices of
the spinbasis. A comparison of the permutation conditions 
(\ref{CPT29})--(\ref{CPT32}) with the conditions (\ref{CPT27})--(\ref{CPT28})
shows that $\sF=\cE_{d_1d_2\cdots d_g}$ at $g\equiv 0\pmod{2}$
automatically satisfies the conditions which define the matrix of the
pseudoantiautomorphism
$\cA\rightarrow\overline{\widetilde{\cA^\star}}$.

Let examine the second product
$\sF=\cE_{\alpha_1\alpha_2\cdots\alpha_a}\cE_{j_1j_2\cdots j_k}$.
This product contains all the complex part of the spinbasis and all the
skewsymmetric matrices. Therefore, in the product $\Pi\sC$ there is a
quantity of identical complex skewsymmetric matrices. The product
$\sF$ is odd and consists of all the complex symmetric and real skewsymmetric
matrices of the spinbasis. It is easy to see that in this case $\sF$
coincides with the product $\cE_{c_1c_2\cdots c_s}$ at
$s\equiv 1\pmod{2}$. Permutation conditions of the matrix
$\sF=\cE_{c_1c_2\cdots c_s}$ with the units of the spinbasis are
\begin{gather}
\sF\cE_{\alpha_\gamma}=(-1)^{s-\gamma}\cE_{c_1c_2\cdots c_{s-2}},
\nonumber\\
\cE_{\alpha_\gamma}\sF=(-1)^{\gamma-1}\cE_{c_1c_2\cdots c_{s-2}},
\label{CPT33}\\
\sF\cE_{\alpha_\delta}=(-1)^s\cE_{\alpha_\delta}\sF,\label{CPT34}\\
\sF\cE_{\beta_\gamma}=(-1)^s\cE_{\beta_\gamma}\sF,\label{CPT35}\\
\sF\cE_{\beta_\delta}=(-1)^{s-\delta}\cE_{c_1c_2\cdots c_{s-2}},
\nonumber\\
\cE_{\beta_\delta}\sF=(-1)^{\delta-1}\cE_{c_1c_2\cdots c_{s-2}}.
\label{CPT36}
\end{gather}
Therefore, since $s\equiv 1\pmod{2}$ the matrix $\sF$ always commutes with
the complex symmetric and real skewsymmetric matrices and always anticommutes
with the complex skewsymmetric and real symmetric matrices of the spinbasis.
Comparing the conditions (\ref{CPT33})--(\ref{CPT36}) with the conditions
(\ref{CPT27}) and (\ref{CPT28}) we see that
$\sF=\cE_{c_1c_2\cdots c_s}$ at $s\equiv 1\pmod{2}$ 
identically satisfies the conditions which define the matrix of the
transformation
$\cA\rightarrow\overline{\widetilde{\cA^\star}}$.

The third product 
$\sF=\Pi\sC=\cE_{\beta_1\beta_2\cdots\beta_b}\cE_{i_1i_2\cdots i_{p+q-k}}$
contains all the real part and all the symmetric matrices of the spinbasis.
Therefore, in the product $\Pi\sC$ there is a quantity of identical real
symmetric matrices. Thus, the product $\sF$ is odd and consists of all
the real skewsymmetric and complex symmetric matrices of the spinbasis.
It is easy to see that we came again to the matrix
$\sF=\cE_{c_1c_2\cdots c_s}$ ($s\equiv 1\pmod{2}$) with the
permutation conditions (\ref{CPT33})--(\ref{CPT36}).

Finally, the fourth product
$\sF=\cE_{\beta_1\beta_2\cdots\beta_b}\cE_{j_1j_2\cdots j_k}$
contains all the real part and all the skewsymmetric matrices of the spinbasis.
Therefore, in the product $\Pi\sC$ there is a quantity of identical real
skewsymmetric matrices. This product is equivalent to the matrix
$\sF=\cE_{d_1d_2\cdots d_g}$ ($g\equiv 0\pmod{2}$) with the
permutation conditions (\ref{CPT29})--(\ref{CPT32}).

As with the pseudoantiautomorphism
$\cA\rightarrow\overline{\widetilde{\cA}}$,
from the four products we have only two non-equivalent products.
Let us find squares of the non-equivalent matrices
$\sF=\cE_{d_1d_2\cdots d_g}$ ($g\equiv 0\pmod{2}$) and
$\sF=\cE_{c_1c_2\cdots c_s}$ ($s\equiv 1\pmod{2}$):
\begin{gather}
\sF^2=\left(\cE_{d_1d_2\cdots d_g}\right)^2=
\begin{cases}
+\sI, & \text{if $m+v\equiv 0,4\pmod{8}$},\\
-\sI, & \text{if $m+v\equiv 2,6\pmod{8}$};
\end{cases}\label{CPT36'}\\
\sF^2=\left(\cE_{c_1c_2\cdots c_s}\right)^2=
\begin{cases}
+\sI, & \text{if $u+l\equiv 3,7\pmod{8}$},\\
-\sI, & \text{if $u+l\equiv 1,5\pmod{8}$}.
\end{cases}\label{CPT36''}
\end{gather}

\section{The structure of $\sExt(\C_n)$}
As noted previously, the group $\sExt(\C_n)$ is 
a finite group\index{group!finite}
of order eight. This group contains as a subgroup the automorphism group
$\sAut_\pm(\C_n)$ (reflection group). Moreover, in the case of $\Pi\sim\sI$
(when the subalgebra $\cl_{p,q}$ has the ring 
$\K\simeq\R$, $p-q\equiv 0,2\pmod{8}$)
the group $\sExt(\C_n)$ is reduced to its subgroup
$\sAut_\pm(\C_n)$. The structure of the groups $\sAut_\pm(\C_n)$,
$\sAut_\pm(\cl_{p,q})$ is studied in detail (see Theorems \ref{taut} and
\ref{tautr}).

There are six finite groups of order eight (see the Table 3). One is cyclic
and two are direct group products of cyclic groups, hence these three are
Abelian. The remaining three groups are 
the quaternion group\index{group!quaternionic} $Q_4$ with
elements $\{\pm 1,\pm\bi,\pm\bj,\pm\bk\}$, 
the dihedral group\index{group!dihedral} $D_4$, and
the group $\overset{\ast}{\dZ}_4\otimes\dZ_2$. All these groups are
non--Abelian. As known, an important property of each finite group is its
order structure.\index{structure!order}
The order of a particular element $\alpha$ in the group
is the smallest integer $p$ for which $\alpha^p=1$. The Table 3
lists the number of distinct elements in each group which have order 2, 4,
or 8 (the identity 1 is the only element of order 1).
\begin{figure}
\begin{center}{\small
{\bf Table 3.} Finite groups of order 8.}
\end{center}
\begin{center}{\renewcommand{\arraystretch}{1.4}
\begin{tabular}{|l|c|ccc|}\hline
   &  & \multicolumn{3}{l}{Order structure}\vline\\ 
   & Type & 2 & 4 & 8 \\ \hline
1. $\dZ_2\otimes\dZ_2\otimes\dZ_2$ & Abelian & 7 & & \\ 
2. $\dZ_4\otimes\dZ_2$ & & 3 & 4 & \\ 
3. $\dZ_8$ & & 1 & 2 & 4\\ \hline
4. $D_4$ & Non--Abelian & 5 & 2 & \\
5. $Q_4$ & & 1 & 6 & \\
6. $\overset{\ast}{\dZ}_4\otimes\dZ_2$ & & 3 & 4 &\\ \hline
\end{tabular}
}
\end{center}
\end{figure}
Of course, $\dZ_8$ does not occur as a $G(p,q)$ 
(Salingaros group\index{group!Salingaros}), since
every element of $G(p,q)$ has order 1, 2, or 4. The groups
$\dZ_4\otimes\dZ_2$ and $\overset{\ast}{\dZ}_4\otimes\dZ_2$ have the same
order structure, but their signatures $(a,b,c,d,e,f,g)$ are different.
Moreover, the group $\overset{\ast}{\dZ}_4\otimes\dZ_2$ presents a first
example of the finite group of order 8 which has an important physical
meaning.\\[0.2cm]

{\it Example 1.}
Let us consider a Dirac algebra\index{algebra!Dirac} $\C_4$. In the algebra
$\C_4$ we can evolve four different 
real subalgebras\index{subalgebras!real} $\cl_{1,3}$,
$\cl_{3,1}$, $\cl_{4,0}$, $\cl_{0,4}$. 
Let us evolve the spacetime algebra\index{algebra!spacetime}
$\cl_{1,3}$. The algebra $\cl_{1,3}$ has the quaternionic division ring
$\K\simeq\BH$ ($p-q\equiv 6\pmod{8}$) and, therefore, admits the following
spinor representation (the well known $\gamma$-basis):
\begin{gather}
\gamma_0=\begin{pmatrix}
1 & 0 & 0 & 0\\
0 & 1 & 0 & 0\\
0 & 0 &-1 & 0\\
0 & 0 & 0 &-1
\end{pmatrix},\quad\gamma_1=\begin{pmatrix}
0 & 0 & 0 & 1\\
0 & 0 & 1 & 0\\
0 &-1 & 0 & 0\\
-1& 0 & 0 & 0
\end{pmatrix},\nonumber\\
\gamma_2=\begin{pmatrix}
0 & 0 & 0 &-i\\
0 & 0 & i & 0\\
0 & i & 0 & 0\\
-i& 0 & 0 & 0
\end{pmatrix},\quad\gamma_3=\begin{pmatrix}
0 & 0 & 1 & 0\\
0 & 0 & 0 &-1\\
-1& 0 & 0 & 0\\
0 & 1 & 0 & 0
\end{pmatrix}.\label{GammaB}
\end{gather}
The famous Dirac equation\index{equation!Dirac}
in the $\gamma$--basis looks like
\begin{equation}\label{Diraceq}
\left(i\gamma_0\frac{\partial}{\partial x_0}+
i\boldsymbol{\gamma}\frac{\partial}{\partial\bx}-m\right)\psi(x_0,\bx)=0.
\end{equation}\begin{sloppypar}\noindent
The invariance of the Dirac equation with respect to $P$--, $T$--, and
$C$--transformations leads to the following representation
(see, for example, \cite{BLP89} and also many other textbooks on quantum
field theory):\end{sloppypar}
\[
P\sim\gamma_0,\quad T\sim\gamma_1\gamma_3,\quad C\sim\gamma_2\gamma_0.
\]
Thus, we can form a finite group of order 8 ($CPT$--group)
\begin{multline}
\{1,\,P,\,T,\,PT,\,C,\,CP,\,CT,\,CPT\}\sim\\
\sim\left\{1,\,\gamma_0,\,\gamma_1\gamma_3,\,\gamma_0\gamma_1\gamma_3,\,
\gamma_2\gamma_0,\,\gamma_2,\,\gamma_2\gamma_0\gamma_1\gamma_3,\,
\gamma_2\gamma_1\gamma_3\right\}.
\label{DirG2}
\end{multline}
It is easy to verify that a multiplication table
of this group has a form
\begin{center}{\renewcommand{\arraystretch}{1.4}
\begin{tabular}{|c||c|c|c|c|c|c|c|c|}\hline
  & $1$ & $\gamma_0$ & $\gamma_{13}$ & $\gamma_{013}$ & $\gamma_{20}$ &
$\gamma_2$ & $\gamma_{2013}$ & $\gamma_{213}$\\ \hline\hline
$1$  & $1$ & $\gamma_0$ & $\gamma_{13}$ & $\gamma_{013}$ & $\gamma_2$ &
$\gamma_2$ & $\gamma_{2013}$ & $\gamma_{213}$\\ \hline
$\gamma_0$ & $\gamma_0$ & $1$ & $\gamma_{013}$ & $\gamma_{13}$ & $-\gamma_2$ &
$-\gamma_{20}$ & $-\gamma_{213}$ & $-\gamma_{2013}$\\ \hline
$\gamma_{13}$ & $\gamma_{13}$ & $\gamma_{013}$ & $-1$ & $-\gamma_0$ &
$\gamma_{2013}$ & $\gamma_{213}$ & $-\gamma_{20}$ & $-\gamma_2$\\ \hline
$\gamma_{013}$ & $\gamma_{013}$ & $\gamma_{13}$ & $-\gamma_0$ &
 $-1$ & $-\gamma_{213}$ & $-\gamma_{2013}$ & $\gamma_2$ & 
$\gamma_{20}$\\ \hline
$\gamma_{20}$ & $\gamma_{20}$ & $\gamma_2$ & $\gamma_{2013}$ &
$\gamma_{213}$ & $1$ & $\gamma_0$ & $\gamma_{13}$ &
$\gamma_{013}$\\ \hline
$\gamma_2$ & $\gamma_2$ & $\gamma_{20}$ & $\gamma_{213}$ & $\gamma_{2013}$ &
$-\gamma_0$ & $-1$ & $-\gamma_{013}$ & $-\gamma_{13}$\\ \hline
$\gamma_{2013}$ & $\gamma_{2013}$ & $\gamma_{213}$ & $-\gamma_{20}$ &
$\gamma_2$ & $\gamma_{13}$ & $\gamma_{013}$ & $-1$ & $-\gamma_0$\\ \hline
$\gamma_{213}$ & $\gamma_{213}$ & $\gamma_{2013}$ & $-\gamma_2$ &
$-\gamma_{20}$ & $-\gamma_{013}$ & $-\gamma_{13}$ & $\gamma_0$ & $1$\\ \hline
\end{tabular}\;\;$\sim$
}
\end{center}
\begin{center}{\renewcommand{\arraystretch}{1.4}
\begin{tabular}{|c||c|c|c|c|c|c|c|c|}\hline
     & $1$  & $P$  & $T$  & $PT$ & $C$  & $CP$ & $CT$ & $CPT$ \\ \hline\hline
$1$  & $1$  & $P$  & $T$  & $PT$ & $C$  & $CP$ & $CT$ & $CPT$ \\ \hline
$P$  & $P$  & $1$  & $PT$ & $T$  & $-CP$ & $-C$  & $-CPT$& $-CT$\\ \hline
$T$  & $T$  & $PT$ & $-1$  & $-P$  & $CT$ & $CPT$& $-C$  & $-CP$\\ \hline
$PT$ & $PT$ & $T$  & $-P$  & $-1$  & $-CPT$& $-CT$ & $CP$ & $C$\\ \hline
$C$  & $C$  & $CP$ & $CT$ & $CPT$& $1$  & $P$  & $T$  & $PT$\\ \hline
$CP$ & $CP$ & $C$  & $CPT$& $CT$ & $-P$  & $-1$  & $-PT$ & $-T$\\ \hline
$CT$ & $CT$ & $CPT$& $-C$  & $CP$ & $T$  & $PT$ & $-1$  & $-P$\\ \hline
$CPT$& $CPT$& $CT$ & $-CP$ & $-C$  & $-PT$ & $-T$  & $P$  & $1$\\ \hline
\end{tabular}.
}
\end{center}
\begin{sloppypar}
Hence it follows that the $CPT$--group (\ref{DirG2}) is a non--Abelian
finite group of the order structure (3,4). In more details, it is the group
$\overset{\ast}{\dZ}_4\otimes\dZ_2$ with the signature
$(+,-,-,+,-,-,+)$.\end{sloppypar}
\begin{theorem}\label{tautext}
Let $\C_n$ be the Clifford algebra over the field $\F=\C$ and let
$\sExt(\C_n)=\{\sI,\,\sW,\,\sE,\,\sC,\,\Pi,\,\sK,\,\sS,\,\sF\}$
be an extended automorphism group\index{group!automorphism!extended}
of the algebra $\C_n$, where
$\sW$, $\sE$, $\sC$, $\Pi$, $\sK$, $\sS$, $\sF$ are spinor representations
of the transformations $\cA\rightarrow\cA^\star$,
$\cA\rightarrow\widetilde{\cA}$, $\cA\rightarrow\widetilde{\cA^\star}$,
$\cA\rightarrow\overline{\cA}$, $\cA\rightarrow\overline{\cA^\star}$,
$\cA\rightarrow\overline{\widetilde{\cA}}$,
$\cA\rightarrow\overline{\widetilde{\cA^\star}}$. Then over the field $\F=\C$
in dependence on a division ring structure of the real subalgebras
$\cl_{p,q}\subset\C_n$ there exist following isomorphisms between
finite groups and groups $\sExt(\C_n)$:\\[0.2cm]
1) $\K\simeq\R$, types $p-q\equiv 0,2\pmod{8}$.\\
In this case the matrix $\Pi$ of the pseudoautomorphism
$\cA\rightarrow\overline{\cA}$ is proportional to the unit matrix
(identical transformation) and the extended automorphism group
$\sExt(\C_n)$ is reduced to the group of fundamental automorphisms,
$\sAut_\pm(\C_n)$.\\[0.2cm]
2) $\K\simeq\BH$, types $p-q\equiv 4,6\pmod{8}$.\\
In dependence on a spinbasis structure of the subalgebra $\cl_{p,q}$ there
exist the following group isomorphisms:
$\sExt_-(\C_n)\simeq\dZ_2\otimes\dZ_2\otimes\dZ_2$ with the
signature $(+,+,+,+,+,+,+)$ and $\sExt(\C_n)\simeq\dZ_4\otimes\dZ_2$
with $(+,+,+,-,-,-,-)$ for the type $p-q\equiv 4\pmod{8}$ if
$m,v,l,u\equiv 0\pmod{2}$, where $m$ and $l$ are quantities of
complex skewsymmetric and
symmetric matrices, and $u$ and $v$ are quantities of 
real skewsymmetric and symmetric
matrices of the spinbasis of $\cl_{p,q}$. Correspondingly, at
$m,v,l,u\equiv 0\pmod{2}$ there exist Abelian groups
$\sExt_-(\C_n)\simeq\dZ_4\otimes\dZ_2$ with $(+,-,-,d,e,f,g)$ for the
type $p-q\equiv 4\pmod{8}$ and $\sExt_-(\C_n)\simeq\dZ_4\otimes\dZ_2$
with $(-,+,-,d,e,f,g)$, $(-,-,+,d,e,f,g)$ for the type $p-q\equiv 6\pmod{8}$,
where among the symbols $d,e,f,g$ there are two pluses and two minuses.
\begin{sloppypar}
If $m,v,l,u\equiv 1\pmod{2}$ or if among $m,v,l,u$ there are both even and
odd numbers, then there exists the non--Abelian group
$\sExt_+(\C_n)\simeq Q_4$ with the signatures $(-,-,-,d,e,f,g)$ for
the type $p-q\equiv 6\pmod{8}$, where among $d,e,f,g$ there are one plus and
three minuses. And also there exist $\sExt_+(\C_n)\simeq D_4$ with
$(+,-,+,d,e,f,g)$, $(+,+,-,d,e,f,g)$ for the type $p-q\equiv 4\pmod{8}$ and
$\sExt_+(\C_n)\simeq D_4$ with $(-,+,+,d,e,f,g)$ for the type
$p-q\equiv 6\pmod{8}$, where among the symbols $d,e,f,g$ there are
three pluses and one minus. Besides, there exist the groups
$\sExt_+(\C_n)\simeq Q_4$ with the signatures $(+,-,-,-,-,-,-)$ for
$p-q\equiv 4\pmod{8}$ and $(-,+,-,-,-,-,-)$, $(-,-,+,-,-,-,-)$ for
$p-q\equiv 6\pmod{8}$. And also there exist the groups
$\sExt_+(\C_n)\simeq D_4$ with $(+,+,+,d,e,f,g)$, $(+,-,-,+,+,+,+)$ for
the type $p-q\equiv 4\pmod{8}$ and $(a,b,c,+,+,+,+)$ for the type
$p-q\equiv 6\pmod{8}$ (among the symbols $a,b,c$ there are one plus and two
minuses, and among $d,e,f,g$ there are two pluses and two minuses).
There exists the non--Abelian group $\sExt_+(\C_n)\simeq
\overset{\ast}{\dZ}_4\otimes\dZ_2$ with the signatures $(+,-,-,d,e,f,g)$ for
the type $p-q\equiv 4\pmod{8}$ and $(-,+,-,d,e,f,g)$, $(-,-,+,d,e,f,g)$
for the type $p-q\equiv 6\pmod{8}$, where among $d,e,f,g$ there are two
pluses and two minuses. And also there exist $\sExt_+(\C_n)\simeq
\overset{\ast}{\dZ}_4\otimes\dZ_2$ with $(+,-,+,d,e,f,g)$, $(+,+,-,d,e,f,g)$
for $p-q\equiv 4\pmod{8}$ and $(-,+,+,d,e,f,g)$ for $p-q\equiv 6\pmod{8}$,
where among $d,e,f,g$ there are one plus and three minuses. Finally, for
the type $p-q\equiv 6\pmod{8}$ there exists $\sExt_+(\C_n)\simeq
\overset{\ast}{\dZ}_4\otimes\dZ_2$ with $(-,-,-,d,e,f,g)$, where among the
symbols $d,e,f,g$ there are three pluses and one minus.
The full number of the different signatures $(a,b,c,d,e,f,g)$ is equal to 64.
\end{sloppypar}
\end{theorem}
\begin{proof}
First of all, it is necessary to define permutation relations between
the elements of the group $\sExt$. We start with the matrix of the
pseudoautomorphism $\cA\rightarrow\overline{\cA}$ (permutation relations
between the elements $\sW$, $\sE$ and $\sC$ are found in Theorem \ref{tautr}).
As known, for the types $p-q\equiv 4,6\pmod{8}$ the matrix $\Pi$
exists in the two different forms: 1) $\Pi=\cE_{\alpha_1\alpha_2\cdots
\alpha_a}$ is the product of all complex matrices of the spinbasis at
$a\equiv 0\pmod{2}$; 2) $\Pi=\cE_{\beta_1\beta_2\cdots\beta_b}$ is
the product of all real matrices of the spinbasis at $b\equiv 1\pmod{2}$.

Let us consider permutation relations of $\Pi$ with the matrix $\sK$ of the
pseudoautomorphism\index{pseudoautomorphism}
$\cA\rightarrow\overline{\cA^\star}$. The matrix $\sK$
also exists in the two different forms: $\sK=\cE_{\alpha_1\alpha_2\cdots
\alpha_a}$ at $a\equiv 1\pmod{2}$ and $\sK=\cE_{\beta_1\beta_2
\cdots\beta_b}$ at $b\equiv 0\pmod{2}$. In virtue of the definition
$\sK=\Pi\sW$, where $\sW=\cE_{12\cdots p+q}$ is the spinor
representation of the automorphism $\cA\rightarrow\cA^\star$, the matrix
$\Pi=\cE_{\alpha_1\alpha_2\cdots\alpha_a}$ corresponds to a
matrix $\sK=\cE_{\beta_1\beta_2\cdots\beta_b}$, since
$n=p+q$ is always even for the types $p-q\equiv 4,6\pmod{8}$. Correspondingly,
for the matrix $\Pi=\cE_{\beta_1\beta_2\cdots\beta_b}$ we obtain
$\sK=\cE_{\alpha_1\alpha_2\cdots\alpha_a}$, where
$a,b\equiv 1\pmod{2}$. It is easy to see that in both cases we have a
relation
\begin{equation}\label{CPT37}
\Pi\sK=(-1)^{ab}\sK\Pi,
\end{equation}
that is, at $a,b\equiv 0\pmod{2}$ the matrices $\Pi$ and $\sK$ always
commute and at $a,b\equiv 1\pmod{2}$ always anticommute.

Let us find now permutation relations of $\Pi$ with the matrix $\sS$ of
the pseudoantiautomorphism 
$\cA\rightarrow\overline{\widetilde{\cA}}$. As known, the matrix $\sS$
exists in the two non-equivalent forms: 
1) $\sS=\cE_{c_1c_2\cdots c_s}$ is the product of all complex
symmetric and real skewsymmetric matrices at $s\equiv 0\pmod{2}$;
2) $\sS=\cE_{d_1d_2\cdots d_g}$ is the product of all complex
skewsymmetric and real symmetric matrices at $g\equiv 1\pmod{2}$.
From $\sS=\Pi\sE$ it follows that 
$\Pi=\cE_{\alpha_1\alpha_2\cdots\alpha_a}$ corresponds to
$\sS=\cE_{c_1c_2\cdots c_s}$ if 
$\sE=\cE_{j_1j_2\cdots j_k}$ and to
$\sS=\cE_{d_1d_2\cdots d_g}$ if
$\sE=\cE_{i_1i_2\cdots i_{p+q-k}}$. In turn, the matrix
$\Pi=\cE_{\beta_1\beta_2\cdots\beta_b}$ corresponds to
$\sS=\cE_{d_1d_2\cdots d_g}$ if
$\sE=\cE_{j_1j_2\cdots j_k}$ and to
$\sS=\cE_{c_1c_2\cdots c_s}$ if
$\sE=\cE_{i_1i_2\cdots i_{p+q-k}}$. Thus, taking into account
that $\sS=\Pi\sE$, we obtain
\begin{eqnarray}
\Pi\sS&=&(-1)^{\frac{a(a-1)}{2}+\tau}\cE_{j_1j_2\cdots j_k},
\nonumber\\
\sS\Pi&=&(-1)^{\frac{a(a-1)}{2}+\tau-m+ak}\cE_{j_1j_2\cdots
j_k}\label{CPT38}
\end{eqnarray}
for the matrices $\Pi=\cE_{\alpha_1\alpha_2\cdots\alpha_a}$ and
$\sS=\cE_{c_1c_2\cdots c_s}$, where $m$ is the number of
complex skewsymmetric matrices of the spinbasis of $\cl_{p,q}$,
$p-q\equiv 4,6\pmod{8}$. Since a comparison $ak\equiv 0\pmod{2}$ holds
always, then the matrices $\Pi$ and $\sS$ commute at $m\equiv 0\pmod{2}$ and
anticommute at $m\equiv 1\pmod{2}$. Correspondingly,
\begin{eqnarray}
\Pi\sS&=&(-1)^{\frac{a(a-1)}{2}+\tau}\cE_{i_1i_2\cdots
i_{p+q-k}},\nonumber\\
\sS\Pi&=&(-1)^{\frac{a(a-1)}{2}+\tau-l+a(p+q-k)}\cE_{i_1i_2\cdots
i_{p+q-k}}\label{CPT39}
\end{eqnarray}
for the matrices $\Pi=\cE_{\alpha_1\alpha_2\cdots\alpha_a}$ and
$\sS=\cE_{d_1d_2\cdots d_g}$, where $l$ is the number of complex
symmetric matrices of the spinbasis. Since $a(p+q-k)\equiv 0\pmod{2}$
($a\equiv 0\pmod{2}$, $p+q-k\equiv 1\pmod{2}$), then in this case the
matrices $\Pi$ and $\sS$ commute at $l\equiv 0\pmod{2}$ and anticommute at
$l\equiv 1\pmod{2}$. Further, we have
\begin{eqnarray}
\Pi\sS&=&(-1)^{\frac{b(b-1)}{2}+\tau}\cE_{j_1j_2\cdots
j_k},\nonumber\\
\sS\Pi&=&(-1)^{\frac{b(b-1)}{2}+\tau-u+bk}\cE_{j_1j_2\cdots
j_k}\label{CPT40}
\end{eqnarray}
for the matrices $\Pi=\cE_{\beta_1\beta_2\cdots\beta_b}$ and
$\sS=\cE_{d_1d_2\cdots d_g}$, where $u$ is the number of real
skewsymmetric matrices of the spinbasis. Since $bk\equiv 0\pmod{2}$
($b\equiv 1\pmod{2}$, $k\equiv 0\pmod{2}$), then $\Pi$ and $\sS$ commute at
$u\equiv 0\pmod{2}$ and anticommute at $u\equiv 1\pmod{2}$. Finally,
\begin{eqnarray}
\Pi\sS&=&(-1)^{\frac{b(b-1)}{2}+\tau}\cE_{i_1i_2\cdots
i_{p+q-k}},\nonumber\\
\sS\Pi&=&(-1)^{\frac{b(b-1)}{2}+\tau-v+b(p+q-k)}\cE_{i_1i_2\cdots
i_{p+q-k}}\label{CPT41}
\end{eqnarray}
for the matrices $\Pi=\cE_{\beta_1\beta_2\cdots\beta_b}$ and
$\sS=\cE_{c_1c_2\cdots c_s}$, where $v$ is the number of real
symmetric matrices. Therefore, permutation conditions of the matrices
$\Pi$ and $\sS$ in this case have the form $b(p+q-k)\equiv v\pmod{2}$, that is,
$\Pi$ and $\sS$ commute at $v\equiv 1\pmod{2}$ and anticommute at
$v\equiv 0\pmod{2}$, since $b,p+q-k\equiv 1\pmod{2}$.

Now we find permutation conditions of $\Pi$ with the matrix $\sF$ of the
pseudoantiautomorphism
$\cA\rightarrow\overline{\widetilde{\cA^\star}}$. In turn, the matrix
$\sF$ exists also in the two non-equivalent forms: 
$\sF=\cE_{d_1d_2\cdots d_g}$ at $g\equiv 0\pmod{2}$ and
$\sF=\cE_{c_1c_2\cdots c_s}$ at $s\equiv 1\pmod{2}$.
From the definition $\sF=\Pi\sC$ it follows that
$\Pi=\cE_{\alpha_1\alpha_2\cdots\alpha_a}$ corresponds to
$\sF=\cE_{d_1d_2\cdots d_g}$ if
$\sC=\cE_{i_1i_2\cdots i_{p+q-k}}$ and to
$\sF=\cE_{c_1c_2\cdots c_s}$ if
$\sC=\cE_{j_1j_2\cdots j_k}$. The matrix
$\Pi=\cE_{\beta_1\beta_2\cdots\beta_b}$ corresponds to
$\sF=\cE_{c_1c_2\cdots c_s}$ if
$\sC=\cE_{i_1i_2\cdots i_{p+q-k}}$ and to
$\sF=\cE_{d_1d_2\cdots d_g}$ if
$\sC=\cE_{j_1j_2\cdots j_k}$.
Thus, taking into account that $\sF=\Pi\sC$, we obtain
\begin{eqnarray}
\Pi\sF&=&(-1)^{\frac{a(a-1)}{2}+\tau}\cE_{i_1i_2\cdots
i_{p+q-k}},\nonumber\\
\sF\Pi&=&(-1)^{\frac{a(a-1)}{2}+\tau-l+a(p+q-k)}\cE_{i_1i_2\cdots
i_{p+q-k}}\label{CPT42}
\end{eqnarray}
for the matrices $\Pi=\cE_{\alpha_1\alpha_2\cdots\alpha_a}$ and
$\sF=\cE_{d_1d_2\cdots d_g}$. It is easy to see that
$\Pi$ and $\sF$ commute at $l\equiv 0\pmod{2}$ and anticommute at
$l\equiv 1\pmod{2}$, since $a,p+q-k\equiv 0\pmod{2}$. Analogously,
\begin{eqnarray}
\Pi\sF&=&(-1)^{\frac{a(a-1)}{2}+\tau}\cE_{j_1j_2\cdots
j_k},\nonumber\\
\sF\Pi&=&(-1)^{\frac{a(a-1)}{2}+\tau-m+ak}\cE_{j_1j_2\cdots
j_k}\label{CPT43}
\end{eqnarray}
for $\Pi=\cE_{\alpha_1\alpha_2\cdots\alpha_a}$ and
$\sF=\cE_{c_1c_2\cdots c_s}$. Therefore, $\Pi$ and $\sF$
commute at $m\equiv 0\pmod{2}$ and anticommute at $m\equiv 1\pmod{2}$, since
$a\equiv 0\pmod{2}$, $k\equiv 1\pmod{2}$. Further, we have
\begin{eqnarray}
\Pi\sF&=&(-1)^{\frac{b(b-1)}{2}+\tau}\cE_{i_1i_2\cdots
i_{p+q-k}},\nonumber\\
\sF\Pi&=&(-1)^{\frac{b(b-1)}{2}+\tau-v+b(p+q-k)}\cE_{i_1i_2\cdots
i_{p+q-k}}\label{CPT44}
\end{eqnarray}
for the matrices $\Pi=\cE_{\beta_1\beta_1\cdots\beta_b}$ and
$\sF=\cE_{c_1c_2\cdots c_s}$. In this case, $\Pi$ and $\sF$
commute at $v\equiv 0\pmod{2}$ and anticommute at $v\equiv 1\pmod{2}$,
since $b\equiv 1\pmod{2}$, $p+q-k\equiv 0\pmod{2}$. Finally,
\begin{eqnarray}
\Pi\sF&=&(-1)^{\frac{b(b-1)}{2}+\tau}\cE_{j_1j_2\cdots j_k},
\nonumber\\
\sF\Pi&=&(-1)^{\frac{b(b-1)}{2}+\tau-u+bk}\cE_{j_1j_2\cdots
j_k}\label{CPT45}
\end{eqnarray}
for $\Pi=\cE_{\beta_1\beta_2\cdots\beta_b}$ and
$\sF=\cE_{d_1d_2\cdots d_g}$. Therefore, in this case
permutation conditions of the matrices $\Pi$ and $\sF$ have the form
$bk\equiv u\pmod{2}$, that is, $\Pi$ and $\sF$ commute at $u\equiv 1\pmod{2}$
and anticommute at $u\equiv 0\pmod{2}$, since $b,k\equiv 1\pmod{2}$.

Let us define now permutation conditions of $\Pi$ with the matrices
$\sW$, $\sE$ and $\sC$ of the transformations $\cA\rightarrow\cA^\star$,
$\cA\rightarrow\widetilde{\cA}$ and $\cA\rightarrow\widetilde{\cA^\star}$,
respectively. First of all, according to Theorem \ref{tpseudo} in the
case of subalgebras $\cl_{p,q}$ with the real ring $\K\simeq\R$
(types $p-q\equiv 0,2\pmod{8}$) the matrix $\Pi$ is proportional to the
unit matrix and, therefore, $\Pi$ commutes with $\sW$, $\sE$ and $\sC$.
In case of the ring $\K\simeq\BH$ (types $p-q\equiv 4,6\pmod{8}$) the
matrix $\Pi$ exists in the two forms: $\Pi_a$ at $a\equiv 0\pmod{2}$ and
$\Pi_b$ at $b\equiv 1\pmod{2}$. Since $a+b=p+q$, then the matrix $\sW$
can be represented by a product 
$\cE_{\alpha_1\alpha_2\cdots\alpha_a}\cE_{\beta_1\beta_2
\cdots\beta_b}$ and for 
$\Pi=\cE_{\alpha_1\alpha_2\cdots\alpha_a}$ we have
\begin{eqnarray}
\Pi\sW&=&(-1)^{\frac{a(a-1)}{2}}\sigma(\alpha_1)\sigma(\alpha_2)\cdots
\sigma(\alpha_a)\cE_{\beta_1\beta_2\cdots\beta_b},\nonumber\\
\sW\Pi&=&(-1)^{\frac{a(a-1)}{2}+ba}\sigma(\alpha_1)\sigma(\alpha_2)\cdots
\sigma(\alpha_a)\cE_{\beta_1\beta_2\cdots\alpha_b}.\label{CPT46}
\end{eqnarray}
Hence it follows that $\Pi$ and $\sW$ commute at $ab\equiv 0\pmod{2}$
and anticommute at $ab\equiv 1\pmod{2}$, but $a\equiv 0\pmod{2}$ and,
therefore, the matrices $\Pi$ and $\sW$ always commute in this case.
Taking $\Pi=\cE_{\beta_1\beta_2\cdots\beta_b}$, we find the
following conditions
\begin{eqnarray}
\Pi\sW&=&(-1)^{\frac{b(b-1)}{2}+ab}\sigma(\beta_1)\sigma(\beta_2)\cdots
\sigma(\beta_b)\cE_{\alpha_1\alpha_2\cdots\alpha_a},\nonumber\\
\sW\Pi&=&(-1)^{\frac{b(b-1)}{2}}\sigma(\beta_1)\sigma(\beta_2)\cdots
\sigma(\beta_b)\cE_{\alpha_1\alpha_2\cdots\alpha_a}.\label{CPT47}
\end{eqnarray}
Hence it follows that $ab\equiv 1\pmod{2}$, since in this case
$a,b\equiv 1\pmod{2}$ ($p+q=a+b$ is an even number). Therefore, at
$b\equiv 1\pmod{2}$ the matrices $\Pi$ and $\sW$ always anticommute.

Let us find now permutation conditions between $\Pi$ and the matrix
$\sE$ of the antiautomorphism $\cA\rightarrow\widetilde{\cA}$. As known
(see Theorem \ref{tautr}), the matrix $\sE$ exists in the two non-equivalent
forms. First of all, if $\Pi=\cE_{\alpha_1\alpha_2\cdots\alpha_a}$
and $\sE=\cE_{j_1j_2\cdots j_k}$, then it is obvious that
$\Pi$ and $\sE$ contain $m$ identical complex skewsymmetric matrices.
We can represent the matrices $\Pi$ and $\sE$ in the form of the following
products: $\Pi=\cE_{\alpha_1\alpha_2\cdots\alpha_l}\cE_{i_1
i_2\cdots i_m}$ and $\sE=\cE_{i_1i_2\cdots i_m}\cE_{j_1
j_2\cdots j_u}$, where $l$ and $u$ are the numbers of complex
symmetric and real skewsymmetric matrices, respectively. Therefore,
\begin{eqnarray}
\Pi\sE&=&(-1)^{\frac{m(m-1)}{2}}\sigma(i_1)\sigma(i_2)\cdots\sigma(i_m)
\cE_{\alpha_1\alpha_2\cdots\alpha_l}\cE_{j_1j_2\cdots
j_u},\nonumber\\
\sE\Pi&=&(-1)^{\frac{m(m-1)}{2}+m(u+l)}\sigma(i_1)\sigma(i_2)\cdots
\sigma(i_m)
\cE_{\alpha_1\alpha_2\cdots\alpha_l}\cE_{j_1j_2\cdots
j_u},\label{CPT48}
\end{eqnarray}
that is, $\Pi$ and $\sE$ commute at $m(u+l)\equiv 0\pmod{2}$ and
anticommute at $m(u+l)\equiv 1\pmod{2}$.

Analogously, if $\Pi=\cE_{\alpha_1\alpha_2\cdots\alpha_a}$ and
$\sE=\cE_{i_1i_2\cdots i_{p+q-k}}$, then it is easy to see that
in this case $\Pi$ and $\sE$ contain $l$ identical complex symmetric
matrices. Then $\Pi=\cE_{\alpha_1\alpha_2\cdots\alpha_m}
\cE_{i_1i_2\cdots i_l}$, $\sE=\cE_{i_1i_2\cdots i_l}
\cE_{i_1i_2\cdots i_v}$ ($v$ is the number of all real
symmetric matrices of the spinbasis) and
\begin{eqnarray}
\Pi\sE&=&(-1)^{\frac{l(l-1)}{2}}\sigma(i_1)\sigma(i_2)\cdots\sigma(i_l)
\cE_{\alpha_1\alpha_2\cdots\alpha_m}\cE_{i_1i_2\cdots
i_v},\nonumber\\
\sE\Pi&=&(-1)^{\frac{l(l-1)}{2}+l(m+v)}\sigma(i_1)\sigma(i_2)\cdots\sigma(i_l)
\cE_{\alpha_1\alpha_2\cdots\alpha_m}\cE_{i_1i_2\cdots
i_v}.\label{CPT49}
\end{eqnarray}
Therefore, in this case $\Pi$ and $\sE$ commute at $l(m+v)\equiv 0\pmod{2}$
and anticommute at $l(m+v)\equiv 1\pmod{2}$.
\begin{sloppypar}
In turn, the matrices $\Pi=\cE_{\beta_1\beta_2\cdots\beta_b}$
and $\sE=\cE_{j_1j_2\cdots j_k}$ contain $u$ identical real
skewsymmetric matrices. Therefore, $\Pi=\cE_{\beta_1\beta_2\cdots
\beta_v}\cE_{i_1i_2\cdots i_u}$, $\sE=\cE_{i_1i_2\cdots
i_u}\cE_{j_1j_2\cdots j_m}$ and\end{sloppypar}
\begin{eqnarray}
\Pi\sE&=&(-1)^{\frac{u(u-1)}{2}}\sigma(i_1)\sigma(i_2)\cdots\sigma(i_u)
\cE_{\beta_1\beta_2\cdots\beta_v}\cE_{j_1j_2\cdots
j_m},\nonumber\\
\sE\Pi&=&(-1)^{\frac{u(u-1)}{2}+u(m+v)}\sigma(i_1)\sigma(i_2)\cdots\sigma(i_u)
\cE_{\beta_1\beta_2\cdots\beta_v}\cE_{j_1j_2\cdots
j_m}.\label{CPT50}
\end{eqnarray}
Hence it follows that $\Pi$ and $\sE$ commute at $u(m+v)\equiv 0\pmod{2}$ and
anticommute at $u(m+v)\equiv 1\pmod{2}$.

Finally, the matrices $\Pi=\cE_{\beta_1\beta_2\cdots\beta_b}$ and
$\sE=\cE_{i_1i_2\cdots i_{p+q-k}}$ contain $v$ identical
real symmetric matrices. Therefore, in this case
$\Pi=\cE_{\beta_1\beta_2\cdots\beta_u}\cE_{i_1i_2\cdots
i_v}$, $\sE=\cE_{i_1i_2\cdots i_v}\cE_{i_1i_2\cdots
i_l}$ and
\begin{eqnarray}
\Pi\sE&=&(-1)^{\frac{v(v-1)}{2}}\sigma(i_1)\sigma(i_2)\cdots\sigma(i_v)
\cE_{\beta_1\beta_2\cdots\beta_u}\cE_{i_1i_2\cdots
i_l},\nonumber\\
\sE\Pi&=&(-1)^{\frac{v(v-1)}{2}+v(u+l)}\sigma(i_1)\sigma(i_2)\cdots\sigma(i_v)
\cE_{\beta_1\beta_2\cdots\beta_u}\cE_{i_1i_2\cdots
i_l},\label{CPT51}
\end{eqnarray}
that is, in this case $\Pi$ and $\sE$ commute at $v(u+l)\equiv 0\pmod{2}$ and
anticommute at $v(u+l)\equiv 1\pmod{2}$.

It is easy to see that permutation conditions of $\Pi$ with the matrix
$\sC$ of the antiautomorphism $\cA\rightarrow\widetilde{\cA^\star}$ are
analogous to the conditions (\ref{CPT48})--(\ref{CPT51}). Indeed,
at $\Pi=\cE_{\alpha_1\alpha_2\cdots\alpha_a}$ and
$\sC=\cE_{i_1i_2\cdots i_{p+q-k}}$ these conditions are analogous
to (\ref{CPT49}), that is, $l(m+v)\equiv 0,1\pmod{2}$. For the matrices
$\Pi=\cE_{\alpha_1\alpha_2\cdots\alpha_a}$ and
$\sC=\cE_{j_1j_2\cdots j_k}$ we obtain the condition (\ref{CPT48}),
that is, $m(u+l)\equiv 0,1\pmod{2}$. In turn, for
$\Pi=\cE_{\beta_1\beta_2\cdots\beta_b}$ and 
$\sC=\cE_{i_1i_2\cdots i_{p+q-k}}$ we have the condition
(\ref{CPT51}), that is, $v(u+l)\equiv 0,1\pmod{2}$. Finally, the matrices
$\Pi=\cE_{\beta_1\beta_2\cdots\beta_b}$ and
$\sC=\cE_{j_1j_2\cdots j_k}$ correspond to (\ref{CPT50}) with
$u(m+v)\equiv 0,1\pmod{2}$.

Let us consider now permutation conditions of the matrix $\sK$ of 
$\cA\rightarrow\overline{\cA^\star}$ with other elements
of the group $\sExt(\C_n)$. As it has been shown previously, the structure
of $\sK$ is analogous to the structure of $\Pi$, that is,
$\sK=\cE_{\alpha_1\alpha_2\cdots\alpha_a}$ at $a\equiv 1\pmod{2}$
and $\sK=\cE_{\beta_1\beta_2\cdots\beta_b}$ at $b\equiv 0\pmod{2}$.
Therefore, permutation conditions for $\sK$ are similar to the conditions
for $\Pi$. Permutation conditions between $\sK$ and $\Pi$ are defined by
the relation (\ref{CPT37}). Coming to the matrix $\sS$ of the
pseudoantiautomorphism $\cA\rightarrow\overline{\widetilde{\cA}}$ we see that
permutation conditions between $\sK$ and $\sS$ are analogous to
(\ref{CPT38})--(\ref{CPT41}). Namely, if $\sK=\cE_{\beta_1\beta_2
\cdots\beta_b}$ and $\sS=\cE_{c_1c_2\cdots c_s}$, then
from (\ref{CPT41}) it follows that $\sK$ and $\sS$ commute at 
$v\equiv 0\pmod{2}$ and anticommute at $v\equiv 1\pmod{2}$, since in this
case $b\equiv 0\pmod{2}$ and $p+q-k\equiv 0\pmod{2}$. Analogously, if
$\sK=\cE_{\alpha_1\alpha_2\cdots\alpha_b}$ and
$\sS=\cE_{d_1d_2\cdots d_g}$, then from (\ref{CPT39}) we find
that $\sK$ and $\sS$ commute at $l\equiv 0\pmod{2}$ and anticommute at
$l\equiv 1\pmod{2}$, since $a\equiv 1\pmod{2}$, $p+q-k\equiv 0\pmod{2}$ for
this case. Further, if $\sK=\cE_{\beta_1\beta_2\cdots\beta_b}$
and $\sS=\cE_{d_1d_2\cdots d_g}$, then from (\ref{CPT40}) we
find that in this case $\sK$ and $\sS$ commute at $u\equiv 0\pmod{2}$
and anticommute at $u\equiv 1\pmod{2}$, since $b\equiv 0\pmod{2}$,
$k\equiv 1\pmod{2}$. Finally, if $\sK=\cE_{\alpha_1\alpha_2\cdots
\alpha_a}$ and $\sS=\cE_{c_1c_2\cdots c_s}$, then from
(\ref{CPT38}) it follows that $\sK$ and $\sS$ commute at $m\equiv 1\pmod{2}$
and anticommute at $m\equiv 0\pmod{2}$, since in this case
$a,k\equiv 1\pmod{2}$.

In like manner we can find permutation conditions of $\sK$ with the matrix
$\sF$ of the pseudoantiautomorphism
$\cA\rightarrow\overline{\widetilde{\cA^\star}}$. Indeed, for
$\sK=\cE_{\beta_1\beta_2\cdots\beta_b}$ and
$\sF=\cE_{d_1d_2\cdots d_g}$ from (\ref{CPT45}) it follows
that $\sK$ and $\sF$ commute at $u\equiv 0\pmod{2}$ and anticommute at
$u\equiv 1\pmod{2}$, since in this case $b,k\equiv 0\pmod{2}$. Further, if
$\sK=\cE_{\alpha_1\alpha_2\cdots\alpha_a}$ and
$\sF=\cE_{c_1c_2\cdots c_s}$, then from (\ref{CPT43}) we find
that $\sK$ and $\sF$ commute at $m\equiv 0\pmod{2}$ and anticommute at
$m\equiv 1\pmod{2}$, since $a\equiv 1\pmod{2}$, $k\equiv 0\pmod{2}$.
In turn, for the matrices $\sK=\cE_{\beta_1\beta_2\cdots\beta_b}$
and $\sF=\cE_{c_1c_2\cdots c_s}$ from (\ref{CPT44}) we obtain that
$\sK$ and $\sF$ commute at $v\equiv 0\pmod{2}$ and anticommute at
$v\equiv 1\pmod{2}$, since $b\equiv 0\pmod{2}$, $p+q-k\equiv 1\pmod{2}$.
Finally, if $\sK=\cE_{\alpha_1\alpha_2\cdots\alpha_a}$ and
$\sF=\cE_{d_1d_2\cdots d_g}$, then from (\ref{CPT42}) it follows
that $\sK$ and $\sF$ commute at $l\equiv 1\pmod{2}$ and anticommute at
$l\equiv 0\pmod{2}$, since in this case $a,p+q-k\equiv 1\pmod{2}$.

It is easy to see that in virtue of similarity of the matrices $\sK$ and $\Pi$,
permutation conditions of $\sK$ with the elements of $\sAut_\pm(\cl_{p,q})$
are analogous to the conditions for $\Pi$. Indeed, if
$\sK=\cE_{\beta_1\beta_2\cdots\beta_b}$, then from (\ref{CPT47})
it follows that $\sK$ and $\sW$ always commute, since in this case
$a,b\equiv 0\pmod{2}$. In turn, if $\sK=\cE_{\alpha_1\alpha_2\cdots
\alpha_a}$, then from (\ref{CPT46}) we see that $\sK$ and $\sW$
always anticommute, since $a,b\equiv 1\pmod{2}$. Further, if
$\sK=\cE_{\beta_1\beta_2\cdots\beta_b}$ and
$\sE=\cE_{j_1j_2\cdots j_k}$, then from (\ref{CPT50}) it follows
that $\sK$ and $\sE$ commute at $u(m+v)\equiv 0\pmod{2}$ and anticommute at
$u(m+v)\equiv 1\pmod{2}$. For the matrices
$\sK=\cE_{\alpha_1\alpha_2\cdots\alpha_a}$ and
$\sE=\cE_{j_1j_2\cdots j_k}$ from (\ref{CPT48}) it follows
$m(u+l)\equiv 0,1\pmod{2}$. Correspondingly, for the matrices
$\sK=\cE_{\beta_1\beta_2\cdots\beta_b}$ and
$\sE=\cE_{i_1i_2\cdots i_{p+q-k}}$ from (\ref{CPT51}) we obtain
the condition $v(u+l)\equiv 0,1\pmod{2}$. 
For $\sK=\cE_{\alpha_1\alpha_2
\cdots\alpha_a}$ and $\sE=\cE_{i_1i_2\cdots i_{p+q-k}}$ 
from (\ref{CPT49}) we have $l(m+v)\equiv 0,1\pmod{2}$. In turn,
for $\sK=\cE_{\beta_1\beta_2\cdots\beta_b}$ and
$\sC=\cE_{i_1i_2\cdots i_{p+q-k}}$ from (\ref{CPT51}) we obtain
$v(u+l)\equiv 0,1\pmod{2}$, for $\sK=\cE_{\alpha_1\alpha_2\cdots
\alpha_a}$ and $\sC=\cE_{i_1i_2\cdots i_{p+q-k}}$ from
(\ref{CPT49}) we obtain $l(m+v)\equiv 0,1\pmod{2}$, for
$\sK=\cE_{\beta_1\beta_2\cdots\beta_b}$ and
$\sC=\cE_{j_1j_2\cdots j_k}$ from (\ref{CPT50}) it follows
$u(m+v)\equiv 0,1\pmod{2}$ and, finally, for
$\sK=\cE_{\alpha_1\alpha_2\cdots\alpha_a}$ and
$\sC=\cE_{j_1j_2\cdots j_k}$ from (\ref{CPT48}) we have
$m(u+l)\equiv 0,1\pmod{2}$. 

Let consider permutation conditions of the matrix $\sS$ of the
transformation $\cA\rightarrow\overline{\widetilde{\cA}}$ with other
elements of the group $\sExt(\C_n)$. Permutation conditions of $\sS$
with the matrices $\Pi$ and $\sK$ have been defined previously
(see (\ref{CPT38})--(\ref{CPT41})). Now we define permutation conditions of
$\sS$ with the matrix $\sF$ of the pseudoantiautomorphism
$\cA\rightarrow\overline{\widetilde{\cA^\star}}$ and elements of the
subgroup $\sAut_\pm(\cl_{p,q})$. So, let $\sS=\cE_{c_1c_2\cdots
c_s}$ and $\sF=\cE_{d_1d_2\cdots d_g}$, where
$s\equiv 0\pmod{2}$ and $g\equiv 0\pmod{2}$. As known, in this case
the product $\sS$ contains all complex symmetric matrices and all
real skewsymmetric matrices of the spinbasis. In turn, the product $\sF$
contains all complex skewsymmetric and real symmetric matrices. Therefore,
the product $\sS\sF$ does not contain identical matrices. Then
\begin{equation}\label{CPT52}
\sS\sF=(-1)^{sg}\sF\sS,
\end{equation}
that is, in this case $\sS$ and $\sF$ always commute, since
$s,g\equiv 0\pmod{2}$. If $\sS=\cE_{d_1d_2\cdots d_g}$ and
$\sF=\cE_{c_1c_2\cdots c_s}$, where $s,g\equiv 1\pmod{2}$, then
from (\ref{CPT52}) it follows that $\sS$ and $\sF$ always anticommute.

Further, if $\sS=\cE_{c_1c_2\cdots c_s}$, then permutation
conditions of $\sS$ with the matrix $\sW$ of the automorphism
$\cA\rightarrow\cA^\star$ have the form:
\begin{eqnarray}
\sS\sW&=&(-1)^{\frac{s(s-1)}{2}}\sigma(c_1)\sigma(c_2)\cdots\sigma(c_s)
\cE_{d_1d_2\cdots d_g},\nonumber\\
\sW\sS&=&(-1)^{\frac{s(s-1)}{2}+sg}\sigma(c_1)\sigma(c_2)\cdots\sigma(c_s)
\cE_{d_1d_2\cdots d_g},\label{CPT53}
\end{eqnarray}
that is, in this case $\sS$ always commutes with $\sW$, since
$s,g\equiv 0\pmod{2}$. In turn, the matrix $\sS=\cE_{d_1d_2\cdots
d_g}$ always anticommutes with $\sW$, since $s,g\equiv 1\pmod{2}$.

Let us define now permutation conditions between $\sS$ and $\sE$. If
$\sS=\cE_{c_1c_2\cdots c_s}$ and
$\sE=\cE_{j_1j_2\cdots j_k}$, then the product $\sS\sE$ contains
$u$ identical real skewsymmetric matrices.
Hence it follows that $\sS=\cE_{c_1c_2\cdots c_l}\cE_{i_1
i_2\cdots i_u}$ and $\sE=\cE_{i_1i_2\cdots i_u}
\cE_{j_1j_2\cdots j_m}$. Then
\begin{eqnarray}
\sS\sE&=&(-1)^{\frac{u(u-1)}{2}}\sigma(i_1)\sigma(i_2)\cdots\sigma(i_u)
\cE_{c_1c_2\cdots c_l}\cE_{j_1j_2\cdots j_m},\nonumber\\
\sE\sS&=&(-1)^{\frac{u(u-1)}{2}+u(l+m)}\sigma(i_1)\sigma(i_2)\cdots\sigma(i_u)
\cE_{c_1c_2\cdots c_l}\cE_{j_1j_2\cdots j_m},
\label{CPT54}
\end{eqnarray}
that is, $\sS$ and $\sE$ commute at $u(l+m)\equiv 0\pmod{2}$ and
anticommute at $u(l+m)\equiv 1\pmod{2}$.
\begin{sloppypar}
In turn, the products $\sS=\cE_{d_1d_2\cdots d_g}$ and
$\sE=\cE_{j_1j_2\cdots j_k}$ contain $m$ identical complex
skewsymmetric matrices. Therefore,
$\sS=\cE_{d_1d_2\cdots d_v}\cE_{i_1i_2\cdots i_m}$,
$\sE=\cE_{i_1i_2\cdots i_m}\cE_{j_1j_2\cdots j_u}$ and
\end{sloppypar}
\begin{eqnarray}
\sS\sE&=&(-1)^{\frac{m(m-1)}{2}}\sigma(i_1)\sigma(i_2)\cdots\sigma(i_m)
\cE_{d_1d_2\cdots d_v}\cE_{j_1j_2\cdots j_u},\nonumber\\
\sE\sS&=&(-1)^{\frac{m(m-1)}{2}+m(v+u)}\sigma(i_1)\sigma(i_2)\cdots\sigma(i_m)
\cE_{d_1d_2\cdots d_v}\cE_{j_1j_2\cdots j_u},
\label{CPT55}
\end{eqnarray}
that is, permutation conditions between $\sS$ and $\sE$ in this case
have the form $m(v+u)\equiv 0,1\pmod{2}$.
\begin{sloppypar}
Further, the products $\sS=\cE_{d_1d_2\cdots d_g}$ and
$\sE=\cE_{i_1i_2\cdots i_{p+q-k}}$ contain $v$ identical real
symmetric matrices. Then $\sS=\cE_{d_1d_2\cdots d_m}\cE_{i_1
i_2\cdots i_v}$,
$\sE=\cE_{i_1i_2\cdots i_v}\cE_{i_1i_2\cdots i_l}$ and
\end{sloppypar}
\begin{eqnarray}
\sS\sE&=&(-1)^{\frac{v(v-1)}{2}}\sigma(i_1)\sigma(i_2)\cdots\sigma(i_v)
\cE_{d_1d_2\cdots d_m}\cE_{i_1i_2\cdots i_l},\nonumber\\
\sE\sS&=&(-1)^{\frac{v(v-1)}{2}+v(m+l)}\sigma(i_1)\sigma(i_2)\cdots\sigma(i_v)
\cE_{d_1d_2\cdots d_m}\cE_{i_1i_2\cdots i_l},
\label{CPT56}
\end{eqnarray}
that is, permutation conditions between $\sS$ and $\sE$ are
$v(m+l)\equiv 0,1\pmod{2}$.
\begin{sloppypar}
Finally, the products $\sS=\cE_{c_1c_2\cdots c_s}$ and
$\sE=\cE_{i_1i_2\cdots i_{p+q-k}}$ contain $l$ identical complex
symmetric matrices and, therefore,
$\sS=\cE_{c_1c_2\cdots c_u}\cE_{i_1i_2\cdots i_l}$,
$\sE=\cE_{i_1i_2\cdots i_l}\cE_{i_1i_2\cdots i_v}$.
Then\end{sloppypar}
\begin{eqnarray}
\sS\sE&=&(-1)^{\frac{l(l-1)}{2}}\sigma(i_1)\sigma(i_2)\cdots\sigma(i_l)
\cE_{c_1c_2\cdots c_u}\cE_{i_1i_2\cdots i_v},\nonumber\\
\sE\sS&=&(-1)^{\frac{l(l-1)}{2}+l(u+v)}\sigma(i_1)\sigma(i_2)\cdots\sigma(i_l)
\cE_{c_1c_2\cdots c_u}\cE_{i_1i_2\cdots i_v}
\label{CPT57}
\end{eqnarray}
and permutation conditions for $\sS$ and $\sE$ in this case have the form
$l(u+v)\equiv 0,1\pmod{2}$.

It is easy to see that permutation conditions between $\sS$ and $\sC$ are
analogous to (\ref{CPT54})--(\ref{CPT57}). Indeed, if
$\sS=\cE_{c_1c_2\cdots c_s}$ and
$\sC=\cE_{i_1i_2\cdots i_{p+q-k}}$, then from (\ref{CPT57})
it follows the comparison $l(u+v)\equiv 0,1\pmod{2}$. In turn,
if $\sS=\cE_{d_1d_2\cdots d_g}$ and 
$\sC=\cE_{i_1i_2\cdots i_{p+q-k}}$, then from (\ref{CPT56})
we obtain $v(m+l)\equiv 0,1\pmod{2}$. Further, for
$\sS=\cE_{d_1d_2\cdots d_g}$ and
$\sC=\cE_{j_1j_2\cdots j_k}$ from (\ref{CPT55}) it follows that
$m(v+u)\equiv 0,1\pmod{2}$. Analogously, for $\sS=\cE_{c_1c_2\cdots
c_s}$ and $\sC=\cE_{j_1j_2\cdots j_k}$ from (\ref{CPT54})
we have $u(l+m)\equiv 0,1\pmod{2}$.

Finally, let us consider permutation conditions of the matrix $\sF$ of
$\cA\rightarrow\overline{\widetilde{\cA^\star}}$ with other elements
of the group $\sExt(\C_n)$. Permutation conditions of $\sF$ with the
matrices $\Pi$, $\sK$ and $\sS$ have been found previously (see
(\ref{CPT42})--(\ref{CPT45}), (\ref{CPT52})). Now we define permutation
conditions between $\sF$ and the elements of the subgroups
$\sAut_\pm(\cl_{p,q})$. It is easy to see that permutation conditions between
$\sF$ and $\sW$ are equivalent to (\ref{CPT53}), that is, 
$\sF=\cE_{d_1d_2\cdots d_g}$ always commute with $\sW$,
since $s,g\equiv 0\pmod{2}$, and $\sF=\cE_{c_1c_2\cdots c_s}$
always anticommute with $\sW$, since in this case $s,g\equiv 1\pmod{2}$.
In turn, permutation conditions between $\sF$ and $\sE$ are equivalent to
(\ref{CPT54})--(\ref{CPT57}). Indeed, if
$\sF=\cE_{d_1d_2\cdots d_g}$ and 
$\sE=\cE_{j_1j_2\cdots j_k}$, then from (\ref{CPT55}) it follows
the comparison $m(v+u)\equiv 0,1\pmod{2}$. For
$\sF=\cE_{c_1c_2\cdots c_s}$ and
$\sE=\cE_{j_1j_2\cdots j_k}$ from (\ref{CPT54}) we have
$u(l+m)\equiv 0,1\pmod{2}$. Analogously, for
$\sF=\cE_{c_1c_2\cdots c_s}$ and
$\sE=\cE_{i_1i_2\cdots i_{p+q-k}}$ from (\ref{CPT57}) we obtain
$l(u+v)\equiv 0,1\pmod{2}$, and for
$\sF=\cE_{d_1d_2\cdots d_g}$, 
$\sE=\cE_{i_1i_2\cdots i_{p+q-k}}$ from (\ref{CPT56}) it follows
$v(m+l)\equiv 0,1\pmod{2}$. It is easy to see that permutation conditions
between $\sF$ and $\sC$ are equivalent to (\ref{CPT54})--(\ref{CPT57}).
Namely, for $\sF=\cE_{d_1d_2\cdots d_g}$, 
$\sC=\cE_{i_1i_2\cdots i_{p+q-k}}$ from (\ref{CPT56}) we obtain
$v(m+l)\equiv 0,1\pmod{2}$, for $\sF=\cE_{c_1c_2\cdots c_s}$,
$\sC=\cE_{i_1i_2\cdots i_{p+q-k}}$ from (\ref{CPT57}) it follows
$l(u+v)\equiv 0,1\pmod{2}$. Correspondingly, for
$\sF=\cE_{c_1c_2\cdots c_s}$, 
$\sC=\cE_{j_1j_2\cdots j_k}$ from (\ref{CPT54}) we find
$u(l+m)\equiv 0,1\pmod{2}$, and for
$\sF=\cE_{d_1d_2\cdots d_g}$,
$\sC=\cE_{j_1j_2\cdots j_k}$ from (\ref{CPT55}) we see that
$m(v+u)\equiv 0,1\pmod{2}$.

Now, we are in a position to define a detailed classification for extended
automorphism groups $\sExt(\C_n)$. First of all, since for the subalgebras
$\cl_{p,q}$ over the ring $\K\simeq\R$ the group $\sExt(\C_n)$ is reduced
to $\sAut_\pm(\C_n)$, then all essentially different groups $\sExt(\C_n)$
correspond to the subalgebras $\cl_{p,q}$ with the quaternionic ring 
$\K\simeq\BH$, $p-q\equiv 4,6\pmod{8}$. Let us classify the groups
$\sExt(\C_n)$ with respect to their subgroups $\sAut_\pm(\cl_{p,q})$.
Taking into account the structure of $\sAut_\pm(\cl_{p,q})$ at
$p-q\equiv 4,6\pmod{8}$ (see Theorem \ref{tautr}) we obtain for the group
$\sExt(\C_n)=\left\{\sI,\sW,\sE,\sC,\Pi,\sK,\sS,\sF\right\}$ the following
possible realizations:
\begin{gather}
\sExt^1=\left\{\sI,\cE_{12\cdots p+q},\cE_{j_1j_2\cdots j_k},
\cE_{i_1i_2\cdots i_{p+q-k}},\cE_{\alpha_1\alpha_2\cdots\alpha_a},
\cE_{\beta_1\beta_2\cdots\beta_b},\cE_{c_1c_2\cdots c_s},
\cE_{d_1d_2\cdots d_g}\right\},\nonumber\\
\sExt^2=\left\{\sI,\cE_{12\cdots p+q},\cE_{j_1j_2\cdots j_k},
\cE_{i_1i_2\cdots i_{p+q-k}},\cE_{\beta_1\beta_2\cdots\beta_b},
\cE_{\alpha_1\alpha_2\cdots\alpha_a},\cE_{d_1d_2\cdots d_g},
\cE_{c_1c_2\cdots c_s}\right\},\nonumber\\
\sExt^3=\left\{\sI,\cE_{12\cdots p+q},\cE_{i_1i_2\cdots i_{p+q-k}},
\cE_{j_1j_2\cdots j_k}, \cE_{\alpha_1\alpha_2\cdots\alpha_a},
\cE_{\beta_1\beta_2\cdots\beta_b},\cE_{d_1d_2\cdots d_g},
\cE_{c_1c_2\cdots c_s}\right\},\nonumber\\
\sExt^4=\left\{\sI,\cE_{12\cdots p+q},\cE_{i_1i_2\cdots i_{p+q-k}},
\cE_{j_1j_2\cdots j_k}, \cE_{\beta_1\beta_2\cdots\beta_b},
\cE_{\alpha_1\alpha_2\cdots\alpha_a},\cE_{c_1c_2\cdots c_s},
\cE_{d_1d_2\cdots d_g}\right\}.\nonumber
\end{gather}
The groups $\sExt^1(\C_n)$ and $\sExt^2(\C_n)$ have Abelian subgroups
$\sAut_-(\cl_{p,q})$ ($\dZ_2\otimes\dZ_2$ or $\dZ_4$). In turn, the groups
$\sExt^3(\C_n)$ and $\sExt^4(\C_n)$ have only non-Abelian subgroups
$\sAut_+(\cl_{p,q})$ ($Q_4/\dZ_2$ or $D_4/\dZ_2$).

Let us start with the group $\sExt^1(\C_n)$. All the elements of
$\sExt^1(\C_n)$ are even products, that is, $p+q\equiv 0\pmod{2}$,
$k\equiv 0\pmod{2}$, $a\equiv 0\pmod{2}$, $b\equiv 0\pmod{2}$,
$s\equiv 0\pmod{2}$ and $g\equiv 0\pmod{2}$. At this point, the elements
$\sI$, $\sW$, $\sE$, $\sC$ form Abelian subgroups $\dZ_2\otimes\dZ_2$ or
$\dZ_4$ (see Theorem \ref{tautr}). In virtue of (\ref{CPT37}) the element
$\Pi$ commutes with $\sK$, and from (\ref{CPT38}) it follows that $\Pi$
commutes with $\sS$ at $m\equiv 0\pmod{2}$. From (\ref{CPT42}) we see that
$\Pi$ commutes with $\sF$ at $l\equiv 0\pmod{2}$, and from (\ref{CPT46}) it
follows that $\Pi$ commutes with $\sW$. The conditions (\ref{CPT48})
show that $\Pi$ commutes with $\sE$ at $m(u+l)\equiv 0\pmod{2}$ and
commutes with $\sC$ at $l(m+v)\equiv 0\pmod{2}$. Further, the element
$\sK$ commutes with $\sS$ at $v\equiv 0\pmod{2}$ and commutes with $\sF$ at
$u\equiv 0\pmod{2}$. And also $\sK\in\sExt^1$ always commutes with $\sW$
and commutes correspondingly with $\sE$ and $\sC$ at $u(m+v)\equiv 0\pmod{2}$
and $v(u+l)\equiv 0\pmod{2}$. From (\ref{CPT52}) and (\ref{CPT53})
it follows that the element $\sS$ always commutes with $\sF$ and $\sW$.
The conditions (\ref{CPT54}) and (\ref{CPT57}) show that $\sS$ commutes
with $\sE$ and $\sC$ correspondingly at $u(l+m)\equiv 0\pmod{2}$ and
$l(u+v)\equiv 0\pmod{2}$. In virtue of (\ref{CPT53}) the element $\sF$
always commutes with $\sW$. Finally, from (\ref{CPT55}) and (\ref{CPT56})
it follows that $\sF$ commutes with $\sE$ and $\sC$ correspondingly at
$m(v+u)\equiv 0\pmod{2}$ and $v(m+l)\equiv 0\pmod{2}$. Thus,
the group $\sExt^1(\C_n)$ is Abelian at $m,l,u,v\equiv 0\pmod{2}$.
In case of the subgroup $\sAut_-(\cl_{p,q})\simeq\dZ_2\otimes\dZ_2$
($p-q\equiv 4\pmod{8}$) we obtain an Abelian group
$\sExt^1_-(\C_n)\simeq\dZ_2\otimes\dZ_2\otimes\dZ_2$ with the signature
$(+,+,+,+,+,+,+)$ for the elements $\Pi$, $\sK$, $\sS$ and $\sF$ with
positive squares ($m-l\equiv 0,4\pmod{8}$, $v-u\equiv 0,4\pmod{8}$,
$u+l\equiv 0,4\pmod{8}$ and $m+v\equiv 0,4\pmod{8}$).
It is easy to see that for the type $p-q\equiv 4\pmod{8}$ there exists also
$\sExt^1_-(\C_n)\simeq\dZ_4\otimes\dZ_2$ with the signature
$(+,+,+,-,-,-,-)$ and the subgroup $\dZ_2\otimes\dZ_2$, where
$m-l\equiv 2,6\pmod{8}$, $v-u\equiv 2,6\pmod{8}$, $u+l\equiv 2,6\pmod{8}$ and
$m+v\equiv 2,6\pmod{8}$. Further, for the type $p-q\equiv 4\pmod{8}$
there exist Abelian groups $\sExt^1_-(\C_n)\simeq\dZ_4\otimes\dZ_2$ with
the signatures $(+,-,-,d,e,f,g)$ and subgroups $\dZ_4$, where among the
symbols $d$, $e$, $f$, $g$ there are two pluses and two minuses.
Correspondingly, at $m,v,l,u\equiv 0\pmod{2}$ for the type
$p-q\equiv 6\pmod{8}$ there exist Abelian groups 
$\sExt^1_-(\C_n)\simeq\dZ_4\otimes\dZ_2$ with the signatures
$(-,+,-,d,e,f,g)$ and $(-,-,+,d,e,f,g)$ if $m-u\equiv 0,1,4,5\pmod{8}$,
$v-l\equiv 2,3,6,7\pmod{8}$ and $m-u\equiv 2,3,6,7\pmod{8}$,
$v-l\equiv 0,1,4,5\pmod{8}$.

It is easy to see that from the comparison $k,a,b,s,g\equiv 0\pmod{2}$
it follows that the numbers $m$, $v$, $l$, $u$ are simultaneously even or
odd. The case $m,v,l,u\equiv 0\pmod{2}$, considered previously, leads
to the Abelian groups $\sExt^1_-$. In contrast to this, the case
$m,v,l,u\equiv 1\pmod{2}$ leads to non-Abelian groups $\sExt^1_+$
with Abelian subgroups $\dZ_4$ and $\dZ_2\otimes\dZ_2$ (it follows from
(\ref{CPT37})--(\ref{CPT38})). Namely, in this case we have the group
$\sExt^1_+\simeq\overset{\ast}{\dZ}_4\otimes\dZ_2$ with the subgroup $\dZ_4$
for $p-q\equiv 4,6\pmod{8}$ (it should be noted that signatures of
the groups $\dZ_4\otimes\dZ_2$ and $\overset{\ast}{\dZ}_4\otimes\dZ_2$
do not coincide), the group $\sExt^1_+\simeq Q_4$ with the subgroup
$\dZ_4$ for $p-q\equiv 4,6\pmod{8}$ and $\sExt^1_+\simeq D_4$ with
$\dZ_2\otimes\dZ_2$ for the type $p-q\equiv 4\pmod{8}$.

Let us consider now the group $\sExt^2(\C_n)$. In this case among the
elements of $\sExt^2$ there are both even and odd elements:
$k\equiv 0\pmod{2}$, $b\equiv 1\pmod{2}$, $a\equiv 1\pmod{2}$,
$g\equiv 1\pmod{2}$ and $s\equiv 1\pmod{2}$. At this point, the elements
$\sI$, $\sW$, $\sE$, $\sC$ form Abelian subgroups $\dZ_2\otimes\dZ_2$
and $\dZ_4$. In virtue of (\ref{CPT37}) the element $\Pi$ always
anticommutes with $\sK$, and from (\ref{CPT52}) it follows that the
elements $\sS$ and $\sF$ always anticommute. Therefore, all the groups
$\sExt^2$ are non-Abelian. Among these groups there are the following
isomorphisms: $\sExt^2_+\simeq\overset{\ast}{\dZ}_4\otimes\dZ_2$ with
the signatures $(+,-,-,d,e,f,g)$ for the type $p-q\equiv 4\pmod{8}$ and
$(-,+,-,d,e,f,g)$, $(-,-,+,d,e,f,g)$ for $p-q\equiv 6\pmod{8}$, where among
the symbols $d$, $e$, $f$, $g$ there are two pluses and two minuses;
$\sExt^2_+\simeq Q_4$ with $(+,-,-,-,-,-,-)$ for $p-q\equiv 4\pmod{8}$ and
$(-,+,-,-,-,-,-)$, $(-,-,+,-,-,-,-)$ for $p-q\equiv 6\pmod{8}$;
$\sExt^2_+\simeq D_4$ with $(a,b,c,+,+,+,+)$ and $(+,+,+,d,e,f,g)$, where
among $a$, $b$, $c$ there are two minuses and one plus, and among
$d$, $e$, $f$, $g$ there are two pluses and two minuses. For all the
groups $\sExt^2$ among the numbers $m$, $v$, $u$, $l$ there are both even and
odd numbers.
\begin{sloppypar}
Let consider the group $\sExt^3(\C_n)$. First of all, the groups
$\sExt^3$ contain non-Abelian subgroups $\sAut_+(\cl_{p,q})$ (the elements
$\sE$ and $\sC$ are odd). Therefore, all the groups $\sExt^3$ are
non-Abelian. Among these groups there are the following isomorphisms:
$\sExt^3_+\simeq D_4$ with $(+,-,+,d,e,f,g)$ and $(+,+,-,d,e,f,g)$ for the
type $p-q\equiv 4\pmod{8}$, where among $d$, $e$, $f$, $g$ there are
three pluses and one minus; $\sExt^3_+\simeq Q_4$ with $(-,-,-,d,e,f,g)$,
where among $d$, $e$, $f$, $g$ there are one plus and three minuses;
$\sExt^3_+\simeq D_4$ with $(-,+,+,d,e,f,g)$, where among $d$, $e$, $f$, $g$
there are three pluses and one minus (the type $p-q\equiv 6\pmod{8}$).
Besides, there exist the groups $\sExt^3_+\simeq\overset{\ast}{\dZ}_4
\otimes\dZ_2$ with the signatures $(+,-,+,d,e,f,g)$, $(+,+,-,d,e,f,g)$
for the type $p-q\equiv 4\pmod{8}$ and $(-,-,-,d,e,f,g)$,
$(-,+,+,d,e,f,g)$ for $p-q\equiv 6\pmod{8}$, where among $d$, $e$, $f$, $g$
there are one plus and three minuses.
\end{sloppypar}
Finally, let us consider the group $\sExt^4(\C_n)$. These groups contain
non--Abelian subgroups $\sAut_+(\cl_{p,q})$ and, therefore, all
$\sExt^4$ are non--Abelian. The isomorphism structure of $\sExt^4$ is
similar to $\sExt^3$.

It is easy to see that a full number of all possible signatures
$(a,b,c,d,e,f,g)$ is equal to $2^7=128$. At this point, we have eight 
signature types: (seven `$+$'), (one `$-$', six `$+$'),
(two `$-$', five `$+$'), (three `$-$', four `$+$'),
(four `$-$', three `$+$'), (five `$-$', two `$+$'),
(six `$-$', one `$+$'), (seven `$-$'). However, only four types from
enumerated above correspond to finite groups of order 8:
(seven `$+$') $\rightarrow\dZ_2\otimes\dZ_2\otimes\dZ_2$,
(two `$-$', five `$+$') $\rightarrow D_4$,
(four `$-$', three `$+$') $\rightarrow\dZ_4\otimes\dZ_2$
($\overset{\ast}{\dZ}_4\otimes\dZ_2$) and
(six `$-$', one `$+$') $\rightarrow Q_4$. Therefore, for the group
$\sExt$ there exist 64 different realizations.
\end{proof}
{\it Example 2.}
Let us study an extended automorphism group of the Dirac
algebra\index{algebra!Dirac} $\C_4$. 
We evolve in $\C_4$ the real subalgebra with the
quaternionic ring. Let it be the spacetime algebra $\cl_{p,q}$ with a
spinbasis defined by the matrices (\ref{GammaB}). We define now elements
of the group $\sExt(\C_4)$. First of all, the matrix of the automorphism
$\cA\rightarrow\cA^\star$ has a form: $\sW=\gamma_0\gamma_1\gamma_2\gamma_3$.
Further, since
\[
\gamma^{\sT}_0=\gamma_0,\quad\gamma^{\sT}_1=-\gamma_1,\quad
\gamma^{\sT}_2=-\gamma_2,\quad\gamma^{\sT}_3=-\gamma_3,
\]
then in accordance with Theorem \ref{tautr} the matrix $\sE$ of the
antiautomorphism $\cA\rightarrow\widetilde{\cA}$ is an even product of
skewsymmetric matrices of the spinbasis (\ref{GammaB}), that is,
$\sE=\gamma_1\gamma_3$. From the definition $\sC=\sE\sW$ we find that
the matrix of the antiautomorphism $\cA\rightarrow\widetilde{\cA^\star}$
has a form $\sC=\gamma_0\gamma_2$. $\gamma$--basis contains three
real matrices $\gamma_0$, $\gamma_1$ and $\gamma_3$, therefore, for the
matrix of the pseudoautomorphism $\cA\rightarrow\overline{\cA}$ we obtain
$\Pi=\gamma_0\gamma_1\gamma_3$ (see Theorem \ref{tpseudo}). Further,
in accordance with $\sK=\Pi\sW$ for the matrix of the pseudoautomorphism
$\cA\rightarrow\overline{\cA^\star}$ we have $\sK=\gamma_2$. Finally,
for the pseudoantiautomorphisms 
$\cA\rightarrow\overline{\widetilde{\cA}}$,
$\cA\rightarrow\overline{\widetilde{\cA^\star}}$ from the definitions
$\sS=\Pi\sE$, $\sF=\Pi\sC$ it follows $\sS=\gamma_0$,
$\sF=\gamma_1\gamma_2\gamma_3$. Thus, we come to the following extended
automorphism group:\index{group!automorphism!extended}
\begin{multline}
\sExt(\C_4)\simeq\{\sI,\,\sW,\,\sE,\,\sC,\,\Pi,\,\sK,\,\sS,\,\sF\}\simeq\\
\{\sI,\,\gamma_0\gamma_1\gamma_2\gamma_3,\,\gamma_1\gamma_3,\,
\gamma_0\gamma_2,\,\gamma_0\gamma_1\gamma_3,\,\gamma_2,\,\gamma_0,\,
\gamma_1\gamma_2\gamma_3\}.\label{Dirac3}
\end{multline}
The multiplication table of this group has a form:
\begin{center}{\renewcommand{\arraystretch}{1.4}
\begin{tabular}{|c||c|c|c|c|c|c|c|c|}\hline
  & $\sI$ & $\gamma_{0123}$ & $\gamma_{13}$ & $\gamma_{02}$ & $\gamma_{013}$ &
$\gamma_{2}$ & $\gamma_{0}$ & $\gamma_{123}$\\ \hline\hline
$\sI$  & $\sI$ & $\gamma_{0123}$ & $\gamma_{13}$ & $\gamma_{02}$ & $\gamma_{013}$ &
$\gamma_{2}$ & $\gamma_{0}$ & $\gamma_{123}$\\ \hline
$\gamma_{0123}$ & $\gamma_{0123}$ & $-\sI$ & $\gamma_{02}$ & $-\gamma_{012}$ 
& $-\gamma_2$ &
$\gamma_{013}$ & $-\gamma_{123}$ & $\gamma_{0}$\\ \hline
$\gamma_{13}$ & $\gamma_{13}$ & $\gamma_{02}$ & $-\sI$ & $-\gamma_{0123}$ &
$-\gamma_{0}$ & $-\gamma_{123}$ & $\gamma_{013}$ & $\gamma_2$\\ \hline
$\gamma_{02}$ & $\gamma_{02}$ & $-\gamma_{13}$ & $-\gamma_{0123}$ &
$\sI$ & $\gamma_{123}$ & $-\gamma_{0}$ & $-\gamma_2$ & 
$\gamma_{013}$\\ \hline
$\gamma_{013}$ & $\gamma_{013}$ & $\gamma_2$ & $-\gamma_{0}$ &
$-\gamma_{123}$ & $-\sI$ & $-\gamma_{0123}$ & $\gamma_{13}$ &
$\gamma_{02}$\\ \hline
$\gamma_2$ & $\gamma_2$ & $-\gamma_{013}$ & $-\gamma_{123}$ & $\gamma_{0}$ &
$\gamma_{0123}$ & $-\sI$ & $-\gamma_{02}$ & $\gamma_{13}$\\ \hline
$\gamma_{0}$ & $\gamma_{0}$ & $\gamma_{123}$ & $\gamma_{013}$ &
$\gamma_2$ & $\gamma_{13}$ & $\gamma_{02}$ & $\sI$ & $\gamma_{0123}$\\ \hline
$\gamma_{123}$ & $\gamma_{123}$ & $-\gamma_{0}$ & $\gamma_2$ &
$-\gamma_{013}$ & $-\gamma_{02}$ & $\gamma_{13}$ & $-\gamma_{0123}$ & 
$\sI$\\ \hline
\end{tabular}\;\;$\sim$
}
\end{center}
\begin{center}{\renewcommand{\arraystretch}{1.4}
\begin{tabular}{|c||c|c|c|c|c|c|c|c|}\hline
     & $\sI$  & $\sW$  & $\sE$  & $\sC$ & $\Pi$  & $\sK$ & $\sS$ & $\sF$ \\ \hline\hline
$\sI$  & $\sI$  & $\sW$  & $\sE$  & $\sC$ & $\Pi$  & $\sK$ & $\sS$ & $\sF$ \\ \hline
$\sW$  & $\sW$  & $-\sI$  & $\sC$ & $-\Pi$  & $-\sK$ & $\Pi$  & $-\sF$& $\sS$\\ \hline
$\sE$  & $\sE$  & $\sC$ & $-\sI$  & $-\sW$  & $-\sS$ & $-\sF$& $\Pi$  & $\sK$\\ \hline
$\sC$ & $\sC$ & $-\sE$  & $-\sW$  & $\sI$  & $\sF$& $-\sS$ & $-\sK$ & $\Pi$\\ \hline
$\Pi$  & $\Pi$  & $\sK$ & $-\sS$ & $-\sF$& $-\sI$  & $-\sW$  & $\sE$  & $\sC$\\ \hline
$\sK$ & $\sK$ & $-\Pi$  & $-\sF$& $\sS$ & $\sW$  & $-\sI$  & $-\sC$ & $\sE$\\ \hline
$\sS$ & $\sS$ & $\sF$& $\Pi$  & $\sK$ & $\sE$  & $\sC$ & $\sI$  & $\sW$\\ \hline
$\sF$& $\sF$& $-\sS$ & $\sK$ & $-\Pi$  & $-\sC$ & $\sE$  & $-\sW$  & $\sI$\\ \hline
\end{tabular}.
}
\end{center}
As follows from this table, the group $\sExt(\C_4)$ is non-Abelian.
$\sExt_+(\C_4)$ contains Abelian group of spacetime reflections
$\sAut_-(\cl_{1,3})\simeq\dZ_4$ as a subgroup. It is easy to see that the group
(\ref{Dirac3}) is a group of the form $\sExt^2_+$ with order structure
$(3,4)$. More precisely, the group (\ref{Dirac3}) is a finite group
$\overset{\ast}{\dZ}_4\otimes\dZ_2$ with the signature
$(-,-,+,-,-,+,+)$.

Coming back to example 1 we see that the groups (\ref{DirG2}) and
(\ref{Dirac3}) are isomorphic:
\[
\{1,\,P,\,T,\,PT,\,C,\,CP,\,CT,\,CPT\}\simeq
\{\sI,\,\sW,\,\sE,\,\sC,\,\Pi,\,\sK,\,\sS,\,\sF\}\simeq
\overset{\ast}{\dZ}_4\otimes\dZ_2.
\]
Moreover, the subgroups of spacetime reflections of these groups are
also isomorphic:
\[
\{1,\,P,\,T,\,PT\}\simeq\{\sI,\,\sW,\sE,\,\sC\}\simeq\dZ_4.
\]
Thus, we come to the following result: the finite group (\ref{DirG2}),
derived from the analysis of invariance properties of the Dirac equation
with respect to discrete transformations $C$, $P$ and $T$, is isomorphic
to an extended automorphism group of the Dirac algebra $\C_4$.
This result allows us to study discrete symmetries and their group
structure for physical fields of any spin (without handling to analysis
of relativistic wave equations).

%
\section{Clifford-Lipschitz groups}
The Lipschitz group $\Lip_{p,q}$, also called the Clifford group, introduced
by Lipschitz in 1886 \cite{Lips}, may be defined as the subgroup of
invertible elements $s$ of the algebra $\cl_{p,q}$:
\[
\Lip_{p,q}=\left\{s\in\cl^+_{p,q}\cup\cl^-_{p,q}\;|\;\forall \bx\in\R^{p,q},\;
s\bx s^{-1}\in\R^{p,q}\right\}.
\]
The set $\Lip^+_{p,q}=\Lip_{p,q}\cap\cl^+_{p,q}$ is called {\it special
Lipschitz group}\index{group!special Lipschitz} \cite{Che55}.

Let $N:\;\cl_{p,q}\rightarrow\cl_{p,q},\;N(\bx)=\bx\widetilde{\bx}$.
If $\bx\in\R^{p,q}$, then $N(\bx)=\bx(-\bx)=-\bx^2=-Q(\bx)$. Further, the
group $\Lip_{p,q}$ has a subgroup
\begin{equation}\label{Pin}
\pin(p,q)=\left\{s\in\Lip_{p,q}\;|\;N(s)=\pm 1\right\}.
\end{equation}
Analogously, {\it a spinor group}\index{group!spinor}
$\spin(p,q)$ is defined by the set
\begin{equation}\label{Spin}
\spin(p,q)=\left\{s\in\Lip^+_{p,q}\;|\;N(s)=\pm 1\right\}.
\end{equation}
It is obvious that $\spin(p,q)=\pin(p,q)\cap\cl^+_{p,q}$.
The group $\spin(p,q)$ contains a subgroup
\begin{equation}\label{Spin+}
\spin_+(p,q)=\left\{s\in\spin(p,q)\;|\;N(s)=1\right\}.
\end{equation}
The groups $O(p,q),\,SO(p,q)$ and $SO_+(p,q)$ are
isomorphic, respectively, 
to the following quotient groups\index{group!quotient}
\begin{eqnarray}
O(p,q)&\simeq&\pin(p,q)/\dZ_2,\nonumber\\
SO(p,q)&\simeq&\spin(p,q)/\dZ_2,\nonumber\\
SO_+(p,q)&\simeq&\spin_+(p,q)/\dZ_2,\nonumber
\end{eqnarray}
\begin{sloppypar}\noindent
where the kernel\index{kernel}
$\dZ_2=\{1,-1\}$. Thus, the groups $\pin(p,q)$, $\spin(p,q)$
and $\spin_+(p,q)$ are the universal coverings of the groups $O(p,q),\,SO(p,q)$
and $SO_+(p,q)$, respectively.\end{sloppypar}

Further, since $\cl^+_{p,q}\simeq\cl^+_{q,p}$, then
\[
\spin(p,q)\simeq\spin(q,p).
\]
In contrast with this, the groups $\pin(p,q)$ and $\pin(q,p)$ are 
non--isomorphic. Let us denote $\spin(n)=\spin(n,0)\simeq\spin(0,n)$.
\begin{theorem}[{\rm\cite{Cor84}}]\label{t3}
The spinor groups
\[
\spin(2),\;\;\spin(3),\;\;\spin(4),\;\;\spin(5),\;\;\spin(6)
\]
are isomorphic to the unitary groups
\[
U(1),\;\;Sp(1)\sim SU(2),\;\;SU(2)\times SU(2),\;\;Sp(2),\;\;SU(4).
\]
\end{theorem} 
In case of the types $p-q\equiv 1,5\pmod{8}$
the algebra $\cl_{p,q}$ is isomorphic to a direct
sum of two mutually annihilating 
simple ideals\index{ideal!mutually annihilating} $\frac{1}{2}(1\pm\omega)
\cl_{p,q}$: $\cl_{p,q}\simeq\frac{1}{2}(1+\omega)\cl_{p,q}\oplus\frac{1}{2}
(1-\omega)\cl_{p,q}$, where $\omega=\e_{12\ldots p+q},\,p-q\equiv 1,5
\pmod{8}$. At this point, the each ideal is isomorpic to $\cl_{p,q-1}$ or
$\cl_{q,p-1}$. Therefore, for the Clifford--Lipschitz groups we have the
following isomorphisms
\begin{eqnarray}
\pin(p,q)&\simeq&\pin(p,q-1)\bigcup\pin(p,q-1)\nonumber\\
&\simeq&\pin(q,p-1)\bigcup\pin(q,p-1).\label{Pinodd1}
\end{eqnarray}
Or, since $\cl_{p,q-1}\simeq\cl^+_{p,q}\subset\cl_{p,q}$, then 
according to (\ref{Spin+})
\[
\pin(p,q)\simeq\spin(p,q)\bigcup\spin(p,q)
\]
if $p-q\equiv 1,5\pmod{8}$.

Further, in the case of $p-q\equiv 3,7\pmod{8}$ 
the $\cl_{p,q}$ is isomorphic to a complex algebra $\C_{p+q-1}$. Therefore,
for the $\pin$ groups we obtain
\begin{eqnarray}
\pin(p,q)&\simeq&\pin(p,q-1)\bigcup\e_{12\ldots p+q}\pin(p,q-1)\nonumber\\
&\simeq&\pin(q,p-1)\bigcup\e_{12\ldots p+q}\pin(q,p-1)\label{e13}
\end{eqnarray}
if $p-q\equiv 1,5\pmod{8}$ and correspondingly
\begin{equation}\label{e14}
\pin(p,q)\simeq\spin(p,q)\bigcup\e_{12\ldots p+q}\spin(p,q).
\end{equation}
In case of $p-q\equiv 3,7\pmod{8}$ we have isomorphisms which are analoguos
to (\ref{e13})-(\ref{e14}), since $\omega\cl_{p,q}\sim\cl_{p,q}$.
Generalizing we obtain the following
\begin{theorem}\label{t4}
Let $\pin(p,q)$ and $\spin(p,q)$ be the Clifford-Lipschitz groups of the
invertible elements of the algebras $\cl_{p,q}$ with odd dimensionality,
$p-q\equiv 1,3,5,7\pmod{8}$. Then
\begin{eqnarray}
\pin(p,q)&\simeq&\pin(p,q-1)\bigcup\omega\pin(p,q-1)\nonumber\\
&\simeq&\pin(q,p-1)\bigcup\omega\pin(q,p-1)\nonumber
\end{eqnarray}
and
\[
\pin(p,q)\simeq\spin(p,q)\bigcup\omega\spin(p,q),
\]
where $\omega=\e_{12\ldots p+q}$ is a volume element of $\cl_{p,q}$.
\end{theorem}
In case of low dimensionalities from Theorem \ref{t3} and Theorem
\ref{t4} it immediately follows
\begin{theorem}\label{t5}
For $p+q\leq 5$ and $p-q\equiv 3,5\pmod{8}$,
\begin{eqnarray}
\pin(3,0)&\simeq&SU(2)\cup iSU(2),\nonumber\\
\pin(0,3)&\simeq&SU(2)\cup eSU(2),\nonumber\\
\pin(5,0)&\simeq&Sp(2)\cup eSp(2),\nonumber\\
\pin(0,5)&\simeq&Sp(2)\cup iSp(2).\nonumber
\end{eqnarray}
\end{theorem}
\begin{proof}\begin{sloppypar}\noindent
Indeed, in accordance with Theorem \ref{t4} $\pin(3,0)\simeq\spin(3)
\cup\e_{123}\spin(3)$. Further, from Theorem \ref{t3} we have
$\spin(3)\simeq SU(2)$, and a square of the element $\omega=\e_{123}$ is
equal to $-1$, therefore, $\omega\sim i$. Thus, $\pin(3,0)\simeq SU(2)\cup
iSU(2)$. For the group $\pin(0,3)$ a square of $\omega$ is equal to $+1$,
therefore, $\pin(0,3)\simeq SU(2)\cup eSU(2)$, $e$ is a double unit.
As expected, $\pin(3,0)\not\simeq\pin(0,3)$. The isomorphisms for the
groups $\pin(5,0)$ and $\pin(0,5)$ are analogously proved.\end{sloppypar}
\end{proof}

In turn, over the field $\F=\C$ there exists a complex Clifford--Lipschitz
group\index{group!complex Clifford-Lipschitz}
\[
\Lip_n=\left\{s\in\C^+_n\cup\C^-_n\;|\;\forall\bx\in\C^n,\;s\bx s^{-1}\in
\C^n\right\}.
\]
The group $\Lip_n$ has a subgroup
\begin{equation}\label{CL}
\pin(n,\C)=\left\{s\in\Lip_n\;|\; N(s)=\pm 1\right\}.
\end{equation}
$\pin(n,\C)$ is an universal covering of 
the complex orthogonal group\index{group!complex orthogonal}
$O(n,\C)$. When $n\equiv 1\pmod{2}$ we have
\begin{equation}\label{PinoddC}
\pin(n,\C)\simeq\pin(n-1,\C)\bigcup\e_{12\cdots n}\pin(n-1,\C).
\end{equation}
\subsection{$PT$-structures}
On the other hand, there exists a more detailed version of the $\pin$--group
(\ref{Pin}) proposed by D\c{a}browski in 1988 \cite{Dab88}. In general,
there are eight double coverings of the orthogonal group 
$O(p,q)$ \cite{Dab88,BD89}:
\[
\rho^{a,b,c}:\;\;\pin^{a,b,c}(p,q)\longrightarrow O(p,q),
\]\begin{sloppypar}\noindent
where $a,b,c\in\{+,-\}$. As known, the group $O(p,q)$ consists of four
connected components: identity connected component $O_0(p,q)$, and three
components corresponding to space inversion $P$, time reversal
$T$, and the combination of these two $PT$, i.e., $O(p,q)=(O_0(p,q))\cup
P(Q_0(p,q))\cup T(O_0(p,q))\cup PT(O_0(p,q))$. Further, since the
four--element group (reflection group) $\{1,\,P,\,T,\,PT\}$ is isomorphic to
the finite group $\dZ_2\otimes\dZ_2$ 
(Gauss--Klein viergruppe\index{group!Gauss-Klein} \cite{Sal81a,Sal84}), then
$O(p,q)$ may be represented by 
a semidirect product\index{product!semidirect} $O(p,q)\simeq O_0(p,q)
\odot(\dZ_2\otimes\dZ_2)$. The signs of $a,b,c$ correspond to the signs of the
squares of the elements in $\pin^{a,b,c}(p,q)$ which cover space inversion
$P$, time reversal $T$ and a combination of these two
$PT$ ($a=-P^2,\,b=T^2,\,c=-(PT)^2$ in D\c{a}browski's notation \cite{Dab88} and
$a=P^2,\,b=T^2,\,c=(PT)^2$ in Chamblin's notation \cite{Ch94} which we will
use below).
An explicit form of the group $\pin^{a,b,c}(p,q)$ is given by the following
semidirect product\end{sloppypar}
\begin{equation}\label{Pinabc}
\pin^{a,b,c}(p,q)\simeq\frac{(\spin_+(p,q)\odot C^{a,b,c})}{\dZ_2},
\end{equation}
where $C^{a,b,c}$ are the four double coverings of
$\dZ_2\otimes\dZ_2$. 
All the eight universal coverings of the orthogonal group
$O(p,q)$ are given in the Table 4.
\begin{figure}
\begin{center}{\small {\bf Table 4.} $PT$-structures.}\end{center}
\begin{center}
{\renewcommand{\arraystretch}{1.4}
\begin{tabular}{|c|l|l|}\hline
$a$ $b$ $c$ & $C^{a,b,c}$ & Remark \\ \hline
$+$ $+$ $+$ & $\dZ_2\otimes\dZ_2\otimes\dZ_2$ & $PT=TP$\\
$+$ $-$ $-$ & $\dZ_2\otimes\dZ_4$ & $PT=TP$\\
$-$ $+$ $-$ & $\dZ_2\otimes\dZ_4$ & $PT=TP$\\
$-$ $-$ $+$ & $\dZ_2\otimes\dZ_4$ & $PT=TP$\\ \hline
$-$ $-$ $-$ & $Q_4$ & $PT=-TP$\\
$-$ $+$ $+$ & $D_4$ & $PT=-TP$\\
$+$ $-$ $+$ & $D_4$ & $PT=-TP$\\
$+$ $+$ $-$ & $D_4$ & $PT=-TP$\\ \hline
\end{tabular}
}
\end{center}
\end{figure}
Here $\dZ_4$, $Q_4$, and $D_4$ are complex\index{group!complex}, 
quaternion\index{group!quaternionic}, and
dihedral groups\index{group!dihedral}, respectively.
According to \cite{Dab88} the group $\pin^{a,b,c}(p,q)$ satisfying the
condition
$PT=-TP$ is called {\it Cliffordian}\index{group!Cliffordina}, 
and respectively {\it non--Cliffordian}\index{group!non-Cliffordian} 
when $PT=TP$. Taking into account the Proposition \ref{prop1} and
Theorems \ref{taut} and \ref{tautr}, we come to the following
\begin{theorem}[{\rm\cite{Var99}}]\label{t10}
Let $\pin^{a,b,c}(p,q)$ be an universal covering of 
the complex orthogonal group\index{group!complex orthogonal}
$O(n,\C)$ of the space $\C^n$ associated with the complex algebra
$\C_n$.
Squares of the
symbols $a,b,c\in
\{-,+\}$ are correspond to squares of the elements of the finite group 
$\sAut=\{\sI,\sW,\sE,\sC\}:\;a=\sW^2,\,b=\sE^2,\,c=\sC^2$, where $\sW,\sE$
 and $\sC$
are correspondingly the matrices of the fundamental automorphisms $\cA\rightarrow
\cA^\star,\,\cA\rightarrow\widetilde{\cA}$ and $\cA\rightarrow
\widetilde{\cA^\star}$ of $\C_{n}$. Then over the field
$\F=\C$ for the algebra $\C_n$ there exist 
two non--isomorphic universal coverings of the group
$O(n,\C)$:\\
1) Non--Cliffordian groups\index{group!non-Cliffordian}
\[
\pin^{+,+,+}(n,\C)\simeq\frac{(\spin_+(n,\C)\odot\dZ_2\otimes\dZ_2\otimes\dZ_2)}
{\dZ_2},
\]
if $n\equiv 0\pmod{4}$ and
\[
\pin^{+,+,+}(n,\C)\simeq\pin^{+,+,+}(n-1,\C)\bigcup\e_{12\ldots n}
\pin^{+,+,+}(n-1,\C),
\]
if $n\equiv 1\pmod{4}$.\\
2) Cliffordian groups\index{group!Cliffordian}
\[
\pin^{-,-,-}(n,\C)\simeq\frac{(\spin_+(n,\C)\odot Q_4)}{\dZ_2},
\]
if $n\equiv 2\pmod{4}$ and
\[
\pin^{-,-,-}(n,\C)\simeq\pin^{-,-,-}(n-1,\C)\bigcup\e_{12\ldots n}
\pin^{-,-,-}(n-1,\C),
\]
if $n\equiv 3\pmod{4}$.
\end{theorem}
\begin{theorem}[{\rm\cite{Var00}}]\label{tgroupr}\begin{sloppypar}\noindent
Let $\pin^{a,b,c}(p,q)$ be an universal covering of the orthogonal group
$O(p,q)$ of the real space $\R^{p,q}$ associated with the algebra
$\cl_{p,q}$.
The squares of symbols $a,b,c\in
\{-,+\}$ correspond to the squares of the elements of a finite group
$\sAut(\cl_{p,q})=\{\sI,\sW,\sE,\sC\}:\;a=\sW^2,\,b=\sE^2,\,c=\sC^2$, 
where $\sW,\sE$ and $\sC$
are the matrices of the 
fundamental automorphisms\index{automorphism!fundamental} $\cA\rightarrow
\cA^\star,\,\cA\rightarrow\widetilde{\cA}$ and $\cA\rightarrow
\widetilde{\cA^\star}$ of the algebra $\cl_{p,q}$, respectively.
Then over the field $\F=\R$ 
in dependence on a division ring structure of the algebra $\cl_{p,q}$,
there exist eight universal coverings of the orthogonal group $O(p,q)$:\\[0.2cm]
1) A non--Cliffordian group\index{group!non-Cliffordian}\end{sloppypar}
\[
\pin^{+,+,+}(p,q)\simeq\frac{(\spin_0(p,q)\odot\dZ_2\otimes\dZ_2\otimes\dZ_2)}
{\dZ_2},
\]
exists if $\K\simeq\R$ and the numbers $p$ and $q$ form the type 
$p-q\equiv 0\pmod{8}$ and $p,q\equiv 0\pmod{4}$, and also if
$p-q\equiv 4\pmod{8}$ and $\K\simeq\BH$. The algebras $\cl_{p,q}$ with the
rings $\K\simeq\R\oplus\R,\,\K\simeq\BH\oplus\BH$ ($p-q\equiv 1,5\pmod{8}$)
admit the group $\pin^{+,+,+}(p,q)$ if in the direct sums there are
addendums of the type
$p-q\equiv 0\pmod{8}$ or $p-q\equiv 4\pmod{8}$. The types $p-q\equiv 3,7
\pmod{8}$, $\K\simeq\C$ admit a non--Cliffordian group $\pin^{+,+,+}(p+q-1,
\C)$ if $p\equiv 0\pmod{2}$ and $q\equiv 1\pmod{2}$. Further, 
non--Cliffordian groups
\[
\pin^{a,b,c}(p,q)\simeq\frac{(\spin_0(p,q)\odot(\dZ_2\otimes\dZ_4)}{\dZ_2},
\]
with $(a,b,c)=(+,-,-)$ exist if $p-q\equiv 0\pmod{8}$, 
$p,q\equiv 2\pmod{4}$ and $\K\simeq\R$, and also if
$p-q\equiv 4\pmod{8}$ and $\K\simeq\BH$. Non--Cliffordian
groups with the signatures
$(a,b,c)=(-,+,-)$ and $(a,b,c)=(-,-,+)$ exist over the ring
$\K\simeq\R$ ($p-q\equiv 2\pmod{8}$) if $p\equiv
2\pmod{4},\,q\equiv 0\pmod{4}$ and $p\equiv 0\pmod{4},\,q\equiv 2\pmod{4}$,
respectively,
and also these groups exist over the ring $\K\simeq\BH$ if
$p-q\equiv 6\pmod{8}$. 
The algebras $\cl_{p,q}$ with the rings
$\K\simeq\R\oplus\R,\,\K\simeq\BH\oplus\BH$ ($p-q\equiv 1,5\pmod{8}$)
admit the group $\pin^{+,-,-}(p,q)$ if in the direct sums there are addendums
of the type $p-q\equiv 0\pmod{8}$ or $p-q\equiv 4\pmod{8}$, and also admit the
groups $\pin^{-,+,-}(p,q)$ and $\pin^{-,-,+}(p,q)$ if in the direct sums
there are addendums of the type $p-q\equiv 2\pmod{8}$ 
or $p-q\equiv 6\pmod{8}$.\\[0.2cm]
2) A Cliffordian group\index{group!Cliffordian}
\[
\pin^{-,-,-}(p,q)\simeq\frac{(\spin_0(p,q)\odot Q_4)}{\dZ_2},
\]
exists if $\K\simeq\R$ ($p-q\equiv 2\pmod{8}$) and $p\equiv 3\pmod{4},\,
q\equiv 1\pmod{4}$, and also if $p-q\equiv 6\pmod{8}$ and $\K\simeq\BH$.
The algebras $\cl_{p,q}$ with the rings 
$\K\simeq\R\oplus\R,\,\K\simeq\BH\oplus\BH$ ($p-q\equiv 1,5\pmod{8}$)
admit the group $\pin^{-,-,-}(p,q)$ if in the direct sums there are 
addendums of the type
$p-q\equiv 2\pmod{8}$ or $p-q\equiv 6\pmod{8}$. The types $p-q\equiv 3,7
\pmod{8}$, $\K\simeq\C$ admit a Cliffordian group $\pin^{-,-,-}(p+q-1,\C)$,
if $p\equiv 1\pmod{2}$ and $q\equiv 0\pmod{2}$. Further, Cliffordian groups
\[
\pin^{a,b,c}(p,q)\simeq\frac{(\spin_0(p,q)\odot D_4)}{\dZ_2},
\]
with $(a,b,c)=(-,+,+)$ exist if $\K\simeq\R$ ($p-q\equiv 2\pmod{8}$)
and $p\equiv 1\pmod{4},\,q\equiv 3\pmod{4}$,
and also if $p-q\equiv 6\pmod{8}$ and $\K\simeq\BH$. Cliffordian groups with
the signatures
$(a,b,c)=(+,-,+)$ and $(a,b,c)=(+,+,-)$ exist over the ring
$\K\simeq\R$ ($p-q\equiv 0\pmod{8}$) if
$p,q\equiv 3\pmod{4}$ and $p,q\equiv 1\pmod{4}$, respectively,
and also these groups
exist over the ring $\K\simeq\BH$ if $p-q\equiv 4\pmod{8}$.
The algebras $\cl_{p,q}$ with the rings 
$\K\simeq\R\oplus\R,\,\K\simeq\BH\oplus\BH$ ($p-q\equiv 1,5\pmod{8}$)
admit the group $\pin^{-,+,+}(p,q)$ if in the direct sums there are addendums
of the type $p-q\equiv 2\pmod{8}$ or $p-q\equiv 6\pmod{8}$, and also admit the
groups $\pin^{+,-,+}(p,q)$ and $\pin^{+,+,-}(p,q)$ if in the direct sums there
are addendums of the type $p-q\equiv 0\pmod{8}$ or $p-q\equiv 4\pmod{8}$.
\end{theorem}

\subsection{$CPT$-structures}\begin{sloppypar}\noindent
As it has been shown previously, there exist 64 different signatures
$(a,b,c,d,e,f,g)$ for the extended automorphism group $\sExt(\C_n)$ of the
complex Clifford algebra $\C_n$. At this point, the group of fundamental
automorphisms, $\sAut_\pm(\cl_{p,q})$, which has 8 different signatures
$(a,b,c)$, is defined as a subgroup of $\sExt(\C_n)$. As known,
the Clifford--Lipschitz group\index{group!Clifford-Lipschitz}
$\pin(n,\C)$ (an universal covering of the
complex orthogonal group $O(n,\C)$) is completely constructed within
the algebra $\C_n$ (see definition (\ref{CL})). If we take into account
spacetime reflections, then according to \cite{Shi58,Shi60,Dab88} there
exist 8 types of universal covering ($PT$--structures) described by the
group $\pin^{a,b,c}(n,\C)$. As it shown in \cite{Var99,Var00}, the
group $\pin^{a,b,c}(n,\C)$ (correspondingly $\pin^{a,b,c}(p,q)$ over the
field $\F=\R$) is completely defined within the algebra $\C_n$
(correspondingly $\cl_{p,q}$) by means of identification of the reflection
subgroup $\{1,P,T,PT\}$ with the automorphism group
$\{\Id,\star,\widetilde{\phantom{cc}},\widetilde{\star}\}$ of $\C_n$
(correspondingly $\cl_{p,q}$). In turn, the pseudoautomorphism
$\cA\rightarrow\overline{\cA}$ of $\C_n$ and the extended automorphism 
group $\{\Id,\star,\widetilde{\phantom{cc}},\widetilde{\star},
\overline{\phantom{cc}},\overline{\star},\overline{\widetilde{\phantom{cc}}},
\overline{\widetilde{\star}}\}\simeq\dZ_2\otimes\dZ_2\otimes\dZ_2$
allows us to give a further generalization of the Clifford--Lipschitz
group (\ref{CL}) (correspondingly (\ref{Pin})). We claim that there
exist 64 types of the universal covering ($CPT$--structures) for the complex
orthogonal group $O(n,\C)$:\end{sloppypar}
\begin{equation}\label{CCL}
\pin^{a,b,c,d,e,f,g}(n,\C)\simeq
\frac{(\spin_+(n,\C)\odot C^{a,b,c,d,e,f,g})}{\dZ_2},
\end{equation}
where $C^{a,b,c,d,e,f,g}$ are five double coverings of the group
$\dZ_2\otimes\dZ_2\otimes\dZ_2$. All the posiible double coverings
$C^{a,b,c,d,e,f,g}$ are given in the Table 5.
\begin{figure}
\begin{center}{\small {\bf Table 5.} $CPT$-structures.}\end{center}
\begin{center}{\renewcommand{\arraystretch}{1.4}
\begin{tabular}{|l|l|l|}\hline
$a\;b\;c\;d\;e\;f\;g$ & $C^{a,b,c,d,e,f,g}$ & Type \\ \hline
$+\;+\;+\;+\;+\;+\;+$ & $\dZ_2\otimes\dZ_2\otimes\dZ_2\otimes\dZ_2$ & Abelian \\ 
three `$+$' and four `$-$' & $\dZ_4\otimes\dZ_2\otimes\dZ_2$ & \\ \hline
one `$+$' and six `$-$' & $Q_4\otimes\dZ_2$ & Non--Abelian \\
five `$+$' and two `$-$' & $D_4\otimes\dZ_2$ &   \\
three `$+$' and four `$-$' & $\overset{\ast}{\dZ}_4\otimes\dZ_2\otimes\dZ_2$
& \\ \hline
\end{tabular}
}
\end{center}
\end{figure}
The group (\ref{CCL}) with non--Abelian $C^{a,b,c,d,e,f,g}$ we will call
{\it Cliffordian}\index{group!Cliffordian}
and respectively {\it non--Cliffordian}\index{group!non-Cliffordian} when
$C^{a,b,c,d,e,f,g}$ is Abelian.

Analogously, over the field $\F=\R$ there exist 64 universal coverings
of the real orthogonal group $O(p,q)$:
\[
\rho^{a,b,c,d,e,f,g}:\;\pin^{a,b,c,d,e,f,g}\longrightarrow O(p,q),
\]
where
\begin{equation}\label{RCL}
\pin^{a,b,c,d,e,f,g}(p,q)\simeq
\frac{(\spin_+(p,q)\odot C^{a,b,c,d,e,f,g})}{\dZ_2}.
\end{equation}
It is easy to see that in case of the algebra $\cl_{p,q}$ (or subalgebra
$\cl_{p,q}\subset\C_n$) with the real division ring $\K\simeq\R$,
$p-q\equiv 0,2\pmod{8}$, $CPT$--structures, defined by the groups
(\ref{CCL}) and (\ref{RCL}), are reduced to the eight
Shirokov--D\c{a}browski $PT$--structures.

Further, using the well-known isomorphism (\ref{PinoddC}) we obtain for
the group $O(n,\C)$ with odd dimensionality the following universal
covering:
\[
\pin^{a,b,c,d,e,f,g}(n,\C)\simeq\pin^{a,b,c,d,e,f,g}(n-1,\C)\bigcup
\e_{12\cdots n}\pin^{a,b,c,d,e,f,g}(n-1,\C).
\]
Correspondingly, in virtue of $\cl_{p,q}\simeq\C_{n-1}$ 
($p-q\equiv 3,7\pmod{8}$, $n=p+q$)
and (\ref{Pinodd1})
for the group $O(p,q)$ with odd dimensionality we have
\[
\pin^{a,b,c,d,e,f,g}(p,q)\simeq\pin^{a,b,c,d,e,f,g}(n-1,\C)
\]
for $p-q\equiv 3,7\pmod{8}$ and
\begin{eqnarray}
\pin^{a,b,c,d,e,f,g}(p,q)&\simeq&\pin^{a,b,c,d,e,f,g}(p,q-1)\bigcup
\e_{12\cdots n}\pin^{a,b,c,d,e,f,g}(p,q-1),\nonumber\\
\pin^{a,b,c,d,e,f,g}(p,q)&\simeq&\pin^{a,b,c,d,e,f,g}(q,p-1)\bigcup
\e_{12\cdots n}\pin^{a,b,c,d,e,f,g}(q,p-1)\nonumber
\end{eqnarray}
for the types $p-q\equiv 1,5\pmod{8}$. Hence it immediately follows
\begin{theorem}
Let $\pin^{a,b,c,d,e,f,g}(n,\C)$ be an universal covering of the complex
orthogonal group $O(n,\C)$ of the space $\C^n$ associated with the complex
algebra $\C_n$. Squares of the symbols $a,b,c,d,e,f,g\in\{-,+\}$
correspond to squares of the elements of the finite group
$\sExt=\{\sI,\sW,\sE,\sC,\Pi,\sK,\sS,\sF\}$: $a=\sW^2$, $b=\sE^2$,
$c=\sC^2$, $d=\Pi^2$, $e=\sK^2$, $f=\sS^2$, $g=\sF^2$, where
$\sW$, $\sE$, $\sC$, $\Pi$, $\sK$, $\sS$, $\sF$ 
are spinor representations of the
automorphisms $\cA\rightarrow\cA^\star$, $\cA\rightarrow\widetilde{\cA}$,
$\cA\rightarrow\widetilde{\cA^\star}$, $\cA\rightarrow\overline{\cA}$,
$\cA\rightarrow\overline{\cA^\star}$, $\cA\rightarrow
\overline{\widetilde{\cA}}$, 
$\cA\rightarrow\overline{\widetilde{\cA^\star}}$. Then over the field
$\F=\C$ in dependence on a division ring structure $\K=f\cl_{p,q}f$ of
the real subalgebras $\cl_{p,q}\subset\C_n$, there exist the following
universal coverings ($CPT$--structures) of the group $O(n,\C)$:\\[0.2cm]
1) $\K\simeq\R$, $p-q\equiv 0,2\pmod{8}$.\\
In this case $CPT$--structures are reduced to the eight
Shirokov--D\c{a}browski $PT$--structures
\[
\pin^{a,b,c}(n,\C)\simeq\frac{(\spin_+(n,\C)\odot C^{a,b,c})}{\dZ_2},
\]
where $C^{a,b,c}$ are double coverings of the group
$\{1,P,T,PT\}\simeq\{\sI,\sW,\sE,\sC\}\simeq\dZ_2\otimes\dZ_2$.\\[0.2cm]
2) $\K\simeq\BH$, $p-q\equiv 4,6\pmod{8}$.\\
In this case we have 64 universal coverings:\\
a) Non--Cliffordian group\index{group!non-Cliffordian}
\[
\pin^{+,+,+,+,+,+,+}(n,\C)\simeq
\frac{(\spin_+(n,\C)\odot\dZ_2\otimes\dZ_2\otimes\dZ_2\otimes\dZ_2)}{\dZ_2}
\]\begin{sloppypar}\noindent
exists if the subalgebra $\cl_{p,q}$ admits the type $p-q\equiv 4\pmod{8}$.
Non--Cliffordian groups\end{sloppypar}
\[
\pin^{a,b,c,d,e,f,g}(n,\C)\simeq
\frac{(\spin_+(n,\C)\odot\dZ_4\otimes\dZ_2\otimes\dZ_2)}{\dZ_2}
\]
exist with the signature $(+,+,+,-,-,-,-)$ when $\cl_{p,q}$ has the type
$p-q\equiv 4\pmod{8}$, and also these groups with the
signatures $(a,b,c,d,e,f,g)$ exist when
$p-q\equiv 6\pmod{8}$, where among the symbols $a$, $b$, $c$ there are
two minuses and one plus, and among $d$, $e$, $f$, $g$ -- two pluses and
two minuses.\\
b) Cliffordian groups\index{group!Cliffordian}
\[
\pin^{a,b,c,d,e,f,g}(n,\C)\simeq
\frac{(\spin_+(n,\C)\odot Q_4\otimes\dZ_2)}{\dZ_2}
\]
exist with the signature $(+,-,-,-,-,-,-)$ when $p-q\equiv 4\pmod{8}$ and with
the signatures $(-,+,-,-,-,-,-)$, $(-,-,+,-,-,-,-)$ when
$p-q\equiv 6\pmod{8}$. And also these groups exist with the signature
$(-,-,-,d,e,f,g)$ if $p-q\equiv 6\pmod{8}$, where among the symbols
$d$, $e$, $f$, $g$ there are one plus and three minuses.
Cliffordian groups
\[
\pin^{a,b,c,d,e,f,g}(n,\C)\simeq
\frac{(\spin_+(n,\C)\odot D_4\otimes\dZ_2)}{\dZ_2}
\]\begin{sloppypar}\noindent
exist with the signatures $(+,-,-,+,+,+,+)$ and $(+,+,+,d,e,f,g)$ when
$\cl_{p,q}$ has the type $p-q\equiv 4\pmod{8}$ and among $d$, $e$, $f$, $g$
there are two minuses and two pluses, and also these groups exist with
$(+,-,+,d,e,f,g)$ and $(+,+,-,d,e,f,g)$, where among $d$, $e$, $f$, $g$
there are three pluses and one minus. Cliffordian groups of this type
exist also with the signatures $(a,b,c,+,+,+,+)$ and $(-,+,+,d,e,f,g)$ when
$p-q\equiv 6\pmod{8}$, where among $a$, $b$, $c$ there are two minuses and
one plus, and among $d$, $e$, $f$, $g$ there are three pluses and one minus.
Cliffordian groups\end{sloppypar}
\[
\pin^{a,b,c,d,e,f,g}(n,\C)\simeq
\frac{(\spin_+(n,\C)\odot\overset{\ast}{\dZ}_4\otimes\dZ_2\otimes\dZ_2)}
{\dZ_2}
\]
exist with the signatures $(+,-,-,d,e,f,g)$ when $p-q\equiv 4\pmod{8}$ and
among the symbols $d$, $e$, $f$, $g$ there are two pluses and two minuses,
and also these groups exist with $(+,-,+,d,e,f,g)$ and $(+,+,-,d,e,f,g)$,
where $p-q\equiv 4\pmod{8}$ and among $d$, $e$, $f$, $g$ there are one plus
and three minuses. Cliffordian groups of this type exist also with
$(-,+,-,d,e,f,g)$ and $(-,-,+,d,e,f,g)$ if $p-q\equiv 6\pmod{8}$, where
among $d$, $e$, $f$, $g$ there are two pluses and two minuses. And also
these groups exist with $(-,-,-,d,e,f,g)$ if $p-q\equiv 6\pmod{8}$, where
among $d$, $e$, $f$, $g$ there are three pluses and one minus.
\end{theorem}

\section{Quotient representations of the\protect\newline
Clifford-Lipschitz groups}
\index{representation!quotient}
In accordance with (\ref{Simple}) and (\ref{Semi-Simple}), the map
$\gamma$ gives the 
left-regular spinor representation\index{representation!spinor!left-regular}
$\fR$ of $\cl(Q)$ in
$\dS$ and $\dS\oplus\dS$, respectively. The representation $\fR$ is
faithful\index{representation!faithful}
if its kernel is zero, that is, $\fR(a)x=0$,
$\forall x\in\dS\Rightarrow a=0$. If the representation $\fR$ has only
two invariant subspaces\index{subspace!invariant}
$\dS$ and $\{0\}$, then $\fR$ is said to be simple
or irreducible. On the contrary case, $\fR$ is said to be semi-simple,
that is, it is a direct sum of simple modules, and in this case
$\dS$ is a direct sum of subspaces which are globally invariant under
$\fR(a)$, $\forall a\in\cl(Q)$. In virtue of the definition (\ref{Pin}),
the representation $\fR$ of $\cl_{p,q}$ induces a representation of
$\pin(p,q)$ which we will denote by the same symbol $\fR$, and also
induces a representation of the group $\spin(p,q)$ which we will denote
by $\fR^+$. Analogously, in virtue of (\ref{CL}), the representation
$\fC$ of the complex algebra $\C_n$ induces a representation of
$\pin(n,\C)$ which we will denote also by the same symbol $\fC$.

Further, over the field $\F=\R$ at $p+q\equiv 0\pmod{2}$ there are four
types of real algebras $\cl_{p,q}$: two types $p-q\equiv 0,2\pmod{8}$ with
a real division ring $\K\simeq\R$ and two types $p-q\equiv 4,6\pmod{8}$ with a
quaternionic division ring $\K\simeq\BH$. Thus, in this case
the representation $\fR$ of $\pin(p,q)$ is divided into the following
four classes:
\begin{eqnarray}
\fR^0_m&\leftrightarrow&\cl_{p,q},\;p-q\equiv 0\pmod{8},\;\K\simeq\R,
\nonumber\\
\fR^2_m&\leftrightarrow&\cl_{p,q},\;p-q\equiv 2\pmod{8},\;\K\simeq\R,
\nonumber\\
\fH^4_m&\leftrightarrow&\cl_{p,q},\;p-q\equiv 4\pmod{8},\;\K\simeq\BH,
\nonumber\\
\fH^6_m&\leftrightarrow&\cl_{p,q},\;p-q\equiv 6\pmod{8},\;\K\simeq\BH,
\label{Ident}
\end{eqnarray}
here $m=\frac{p+q}{2}$.
We will call the representations $\fH^4_m$ and $\fH^6_m$ as
{\it quaternionic representations}\index{representations!quaternionic}
of the group $\pin(p,q)$. 
Using the correspondence (\ref{Ident}), we can define the Periodic Table
for the real representations of $\pin(p,q)$ (see the Table 6).
\begin{figure}
\begin{center}{\small {\bf Table 6.} Real representations of $\pin(p,q)$.}
\end{center}
\begin{center}
{\renewcommand{\arraystretch}{1.2}
\begin{tabular}{c|cccccccccc}
  & p & 0 & 1 & 2 & 3 & 4 & 5 & 6 & 7 & \ldots\\ \hline
q &   &   &   &   &   &   &   &   &   &\\
0 &   &$\fR^0_0$&${}^2\fR^0_0$&$\fR^2_1$&$\fC^3_1$&$\fH^4_1$&
${}^2\fH^4_1$&$\fH^6_2$&$\fC^7_4$&
$\ldots$\\
1&&$\fC^7_0$&$\fR^0_1$&${}^2\fR^0_1$&$\fR^2_2$&$\fC^3_2$&
$\fH^4_2$&${}^2\fH^4_2$&$\fH^6_4$&
$\ldots$\\
2&&$\fH^6_0$&$\fC^7_1$&$\fR^0_2$&${}^2\fR^0_2$&
$\fR^2_4$&$\fC^3_4$&$\fH^4_4$&${}^2\fH^4_4$&
$\ldots$\\
3&&${}^2\fH^4_0$&$\fH^6_1$&$\fC^7_2$&$\fR^0_4$&${}^2\fR^0_4$&
$\fR^2_8$&$\fC^3_8$&$\fH^4_8$&
$\ldots$\\
4&&$\fH^4_1$&${}^2\fH^4_1$&$\fH^6_2$&$\fC^7_4$&$\fR^0_8$&
${}^2\fR^0_8$&$\fR^2_{16}$&
$\fC^3_{16}$&$\ldots$\\
5&&$\fC^3_2$&$\fH^4_2$&${}^2\fH^4_2$&$\fH^6_4$&$\fC^7_8$&
$\fR^0_{16}$&${}^2\fR^0_{16}$&
$\fR^2_{32}$&$\ldots$\\
6&&$\fR^2_4$&$\fC^3_4$&$\fH^4_4$&${}^2\fH^4_4$&$\fH^6_8$&
$\fC^7_{16}$&$\fR^0_{32}$&
${}^2\fR^0_{32}$&$\ldots$\\
7&&${}^2\fR^0_4$&$\fR^2_8$&$\fC^3_8$&$\fH^4_8$&
${}^2\fH^4_8$&$\fH^6_{16}$&$\fC^7_{32}$&
$\fR^0_{64}$&$\ldots$\\
$\vdots$&&$\vdots$&$\vdots$&$\vdots$&$\vdots$&$\vdots$&$\vdots$&$\vdots$&
$\vdots$
\end{tabular}}
\end{center}
\end{figure}
On the other hand, in terms of minimal left ideal the modulo 8
periodicity\index{periodicity!modulo 8} looks like
\[
\dS_{n+8}\simeq\dS_n\otimes\dS_{16}.
\]
In virtue of the mapping $\gamma_{8,0}:\,\cl_{8,0}\rightarrow\M_2(\dO)$
\cite{MS96} (see also excellent review \cite{Bae01}) the latter relation
can be written in the form
\[
\dS_{n+8}\simeq\dS_n\otimes\dO^2,
\]
where $\dO$ is {\it an octonion algebra}.\index{algebra!octonion}
Since the algebra $\cl_{8,0}\simeq
\cl_{0,8}$ admits an octonionic representation, then in virtue of the
modulo 8 periodicity the octonionic representations can be defined for all
high dimensions and, therefore, on the system of real representations
of $\pin(p,q)$ we have a relation
\[
\fD^{m+2}\simeq\fD^{m}\otimes\fO,
\]
where $\fO$ is 
{\it an octonionic representation}\index{representation!octonionic}
of the group $\pin(p,q)$
($\fO\sim\fR^0_4$). 

As known
\cite{Rash,Port}, the odd--dimensional algebras
$\cl_{p,q}$ and $\C_n$ are isomorphic to direct sums of the two algebras
with even dimensionality if correspondingly
$p-q\equiv 1,5\pmod{8}$
and $p+q\equiv 1,3,5,7\pmod{8}$. Therefore, matrix representations of the
algebras $\cl_{p,q}$, $\C_{p+q}$ $(p+q=2m+1)$ are isomorphic to direct sums
of the full matrix algebras\index{algebra!matrix}
$\M_{2^m}(\K)\oplus\M_{2^m}(\K)$, 
where $\K\simeq\R,\,\K\simeq\BH,\K\simeq\C$. On the other hand, there exists
a homomorphic mapping of the algebras
$\cl_{p,q}$ and $\C_n$ onto one matrix algebra $\M_{2^m}(\K)$
with preservation of addition, multiplication and multiplication by the
number operations. Besides, in case of the field
$\F=\R$ and $p-q\equiv 3,7\pmod{8}$ the algebra $\cl_{p,q}$ is isomorphic 
to a full matrix algebra $\M_{2^m}(\C)$, therefore, representations of the
fundamental automorphisms of this algebra can be realized by means of
$\M_{2^m}(\C)$.
\begin{theorem}
If $p+q=2m+1$, then there exist the following homomorphisms:\\
1) $\F=\R$
\[
\epsilon:\;\cl_{p,q}\longrightarrow\M_{2^m}(\K),
\]
where $\K\simeq\R$ if $p-q\equiv 1\pmod{8}$ and $\K\simeq\BH$ if
$p-q\equiv 5\pmod{8}$.\\[0.2cm]
2) $\F=\C$
\[
\epsilon:\;\C_{p+q}\longrightarrow\M_{2^m}(\C),\quad\text{if}\quad
p-q\equiv 1,3,5,7\pmod{8}.
\]
\end{theorem}
\begin{proof} Let us start the proof with a more general case $\F=\C$.
In accordance with (\ref{e4}) the element $\omega$ belongs to a center of the
algebra $\C_n$ and, therefore, this element commutes with all the basis
elements of $\C_n$. Further, we see that the basis vectors
$\{e_1,e_2,\ldots,e_n\}$ generate a subspace
$\C^n\subset\C^{n+1}$. Thus, the algebra $\C_n$ in $\C^n$ is a subalgebra of
$\C_{n+1}$ and consists of the elements which do not contain the element
$\e_{n+1}$. The each element
$\cA\in\C_{n+1}$ can be decomposed as follows
\[
\cA=\cA^1+\cA^0,
\]
where the set $\cA^0$ contains all the elements with $\e_{n+1}$, and
$\cA^1$ does not contain $\e_{n+1}$, 
therefore, $\cA^1\in\C_n$. Multiplying
$\cA^0$ by $\epsilon\omega$, 
we see that the elements $\e_{n+1}$ are mutually annihilate. 
Hence it follows that $\epsilon\omega\cA^0\in\C_n$.
Denoting $\cA^2=\epsilon\omega\cA^0$ and taking into account that
$(\epsilon\omega)^2=1$, we obtain
\[
\cA=\cA^1+\varepsilon\omega\cA^2,
\]
where $\cA^1,\,\cA^2\in\C_n$. Let us define now a homomorphism
$\epsilon:\;\C_{n+1}
\rightarrow\C_n$, acting via the following law
\begin{equation}\label{3.7.1}
\epsilon:\;\cA^1+\varepsilon\omega\cA^2\longrightarrow\cA^1+\cA^2.
\end{equation}
It is obvious that the operations of addition, multiplication and
multiplication by the number are preserved. Indeed, if we take
\[
\cA=\cA^1+\varepsilon\omega\cA^2,\quad\cB=\cB^1+\cB^2,
\]
then in virtue of
$(\varepsilon\omega)^2=1$ and commutativity of
$\omega$ with all elements we have for the multiplication operation:
\begin{multline}
\cA\cB=(\cA^1\cB^1+\cA^2\cB^2)+\varepsilon\omega(\cA^1\cB^2+\cA^2\cB^1)
\stackrel{\epsilon}{\longrightarrow}\\
(\cA^1\cB^1+\cA^2\cB^2)+(\cA^1\cB^2+\cA^2\cB^1)=(\cA^1+\cA^2)(\cB^1+\cB^2),
\nonumber
\end{multline}
that is, an image of the product is equal to the product of factor images
in the same order. In particular case at
$\cA=\varepsilon\omega$ we have $\cA^1=0$ and
$\cA^2=1$, therefore,
\[
\varepsilon\omega\longrightarrow 1.
\]
Thus, a kernel of the homomorphism\index{kernel!of the homomorphism}
$\epsilon$ consists of the all elements
of the form
$\cA^1-\varepsilon\omega\cA^1$, which, as it is easy to see, under action
of $\epsilon$ are mapped into zero. It is clear that $\Ker\epsilon=\{
\cA^1-\varepsilon\omega\cA^1\}$ is a subalgebra of $\C_{n+1}$.
Moreover, the kernel of $\epsilon$ is 
a bilateral ideal\index{ideal!bilateral} of the algebra
$\C_{n+1}$. Therefore, the algebra ${}^{\epsilon}\C_n$, obtained in the
result of the mapping $\epsilon:\;\C_{n+1}\rightarrow\C_n$, is 
{\it a quotient algebra on the ideal}\index{algebra!quotient}
$\Ker\epsilon=\{\cA^1-\varepsilon
\omega\cA^1\}$:
\[
{}^{\epsilon}\C_n\simeq\C_{n+1}/\Ker\epsilon.
\] 
Since the algebra $\C_n$ at $n=2m$ is isomorphically mapped onto the full
matrix algebra $\M_{2^m}(\C)$, then in virtue of
$\epsilon:\;\C_{n+1}\rightarrow\C_n\subset\C_{n+1}$ we obtain a
homomorphic mapping of the algebra
$\C_{n+1}$ onto the matrix algebra
$\M_{2^m}(\C)$.

A homomorphism $\epsilon:\;\cl_{p,q}\rightarrow
\M_{2^m}(\K)$ is proved analogously. In this case a quotient algebra
has the form
\[
{}^{\epsilon}\cl_{p,q-1}\simeq\cl_{p,q}/\Ker\epsilon,
\]
or
\[
{}^{\epsilon}\cl_{q,p-1}\simeq\cl_{p,q}/\Ker\epsilon,
\]
where $\Ker\epsilon=\{\cA^1-\omega\cA^1\}$, since in accordance with (\ref{e3})
at $p-q\equiv 1,5\pmod{8}$ we have always $\omega^2=1$ and, therefore,
$\varepsilon=1$.\end{proof}

As known, complex Clifford algebras $\C_n$ are 
modulo 2 periodic\index{periodicity!modulo 2}
\cite{AtBSh} and, therefore, there exist two types of $\C_n$:
$n\equiv 0\s\pmod{2}$ and $n\equiv 1\s\pmod{2}$. We consider these two
types in the form of the following series:
\[
\begin{array}{cccccccccc}
\C_2 && \C_4 && \cdots && \C_{2k} && \cdots \\
& \C_3 && \C_5 && \cdots && \C_{2k+1} && \cdots
\end{array}
\]
Let us study the decomposition $\C_{2k+1}\simeq\C_{2k}\oplus\C_{2k}$
in more details. This decomposition may be represented by a following scheme
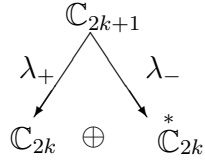
\[
\unitlength=0.5mm
\begin{picture}(70,50)
\put(35,40){\vector(2,-3){15}}
\put(35,40){\vector(-2,-3){15}}
\put(28.25,42){$\C_{2k+1}$}
\put(16,28){$\lambda_{+}$}
\put(49.5,28){$\lambda_{-}$}
\put(13.5,9.20){$\C_{2k}$}
\put(52.75,9){$\stackrel{\ast}{\C}_{2k}$}
\put(32.5,10){$\oplus$}
\end{picture}
\]
Here central idempotents
\[
\lambda^+=\frac{1+\varepsilon\e_1\e_2\cdots\e_{2k+1}}{2},\quad
\lambda^-=\frac{1-\varepsilon\e_1\e_2\cdots\e_{2k+1}}{2},
\]
where
\[
\varepsilon=\begin{cases}
1,& \text{if $k\equiv 0\pmod{2}$};\\
i,& \text{if $k\equiv 1\pmod{2}$},
\end{cases}
\]
satisfy the relations $(\lambda^+)^2=\lambda^+$, $(\lambda^-)^2=\lambda^-$,
$\lambda^+\lambda^-=0$. Thus, we have a decomposition of the initial
algebra $\C_{2k+1}$ into a direct sum of two mutually annihilating simple
ideals:\index{ideal!simple!annihilating}
$\C_{2k+1}\simeq\frac{1}{2}(1+\varepsilon\omega)\C_{2k+1}\oplus
\frac{1}{2}(1-\varepsilon\omega)\C_{2k+1}$. Each of the ideals
$\lambda^{\pm}\C_{2k+1}$ is isomorphic to the subalgebra 
$\C_{2k}\subset\C_{2k+1}$. According to Chisholm and Farwell \cite{CF97},
the idempotents $\lambda^+$ and $\lambda^-$ can be identified with the
helicity projection operators\index{operator!helicity projection}
which distinguish left and right handed
spinors.\index{spinor!left and right handed}
The Chisholm--Farwell notation for $\lambda^\pm$ we will widely
use below.Therefore,
in virtue of the isomorphism $\C_{2k+1}\simeq\C_{2k}\cup\C_{2k}$ and the
homomorphic mapping $\epsilon:\,\C_{2k+1}\rightarrow\C_{2k}$, the second series
(type $n\equiv 1\s\pmod{2}$) is replaced by a sequence of the quotient
algebras\index{algebra!quotient} ${}^\epsilon\C_{2k}$, that is,
\[
\begin{array}{cccccccccc}
\C_2 && \C_4 && \cdots && \C_{2k} && \cdots \\
& {}^\epsilon\C_2 && {}^\epsilon\C_4 && \cdots && {}^\epsilon\C_{2k} && \cdots
\end{array}
\]
Therefore, all the representations of $\pin(n,\C)$ are divided into the
following two series:
\[
\begin{array}{cccccccccc}
\fC_1 && \fC_2 && \cdots && \fC_{k} && \cdots \\
& {}^\epsilon\fC_1 && {}^\epsilon\fC_2 && \cdots && {}^\epsilon\fC_{k} && \cdots
\end{array}
\]

Further, over the field $\F=\R$ for the odd-dimensional algebras $\cl_{p,q}$,
$p-q\equiv 1,5\pmod{8}$, there exist two different decompositions:
$\cl_{p,q}\simeq\cl_{p,q-1}\oplus\cl_{p,q-1}$ and
$\cl_{p,q}\simeq\cl_{q,p-1}\oplus\cl_{q,p-1}$. These decompositions can be
represented by the following schemes:
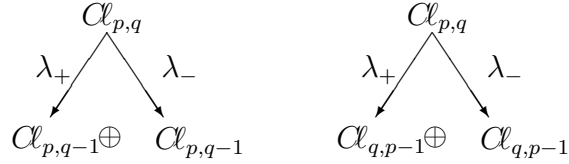
\[
\unitlength=0.5mm
\begin{picture}(70,50)
\put(35,40){\vector(2,-3){15}}
\put(35,40){\vector(-2,-3){15}}
\put(28.25,42){$\cl_{p,q}$}
\put(16,28){$\lambda_{+}$}
\put(49.5,28){$\lambda_{-}$}
\put(9.5,9.20){$\cl_{p,q-1}$}
\put(47.75,9){$\cl_{p,q-1}$}
\put(32.5,10){$\oplus$}
\end{picture}
\quad\quad
\unitlength=0.5mm
\begin{picture}(70,50)
\put(35,40){\vector(2,-3){15}}
\put(35,40){\vector(-2,-3){15}}
\put(28.25,42){$\cl_{p,q}$}
\put(16,28){$\lambda_{+}$}
\put(49.5,28){$\lambda_{-}$}
\put(9.5,9.20){$\cl_{q,p-1}$}
\put(47.75,9){$\cl_{q,p-1}$}
\put(32.5,10){$\oplus$}
\end{picture}
\]
Here central idempotents
\[
\lambda^+=\frac{1+\e_1\e_2\cdots\e_{p+q}}{2},\quad
\lambda^-=\frac{1-\e_1\e_2\cdots\e_{p+q}}{2}
\]
satisfy the relations $(\lambda^+)^2=\lambda^+$, $(\lambda^-)^2=\lambda^-$,
$\lambda^+\lambda^-=0$. Therefore, in virtue of the homomorphic mappings
$\epsilon:\,\cl_{p,q}\rightarrow{}^\epsilon\cl_{p,q-1}$ and
$\epsilon:\,\cl_{p,q}\rightarrow{}^\epsilon\cl_{q,p-1}$ we can replace
the double representations ${}^2\fR$, ${}^2\fH$ by quotient representations
${}^\epsilon\fR$, ${}^\epsilon\fH$. The distribution of the quotient
representations of $\pin(p,q)$ is given in the Table 7.
\begin{figure}
\begin{center}{\small {\bf Table 7.} Quotient representations of $\pin(p,q)$.}
\end{center}
\begin{center}
{\renewcommand{\arraystretch}{1.2}
\begin{tabular}{c|cccccccccc}
  & p & 0 & 1 & 2 & 3 & 4 & 5 & 6 & 7 & \ldots\\ \hline
q &   &   &   &   &   &   &   &   &   &\\
0 &   &$\fR^0_0$&${}^\epsilon\fR^0_0$&$\fR^2_1$&$\fC^3_1$&$\fH^4_1$&
${}^\epsilon\fH^4_1$&$\fH^6_2$&$\fC^7_4$&
$\ldots$\\
1&&$\fC^7_0$&$\fR^0_1$&${}^\epsilon\fR^0_1$&$\fR^2_2$&$\fC^3_2$&
$\fH^4_2$&${}^\epsilon\fH^4_2$&$\fH^6_4$&
$\ldots$\\
2&&$\fH^6_0$&$\fC^7_1$&$\fR^0_2$&${}^\epsilon\fR^0_2$&
$\fR^2_4$&$\fC^3_4$&$\fH^4_4$&${}^\epsilon\fH^4_4$&
$\ldots$\\
3&&${}^\epsilon\fH^4_0$&$\fH^6_1$&$\fC^7_2$&$\fR^0_4$&${}^\epsilon\fR^0_4$&
$\fR^2_8$&$\fC^3_8$&$\fH^4_8$&
$\ldots$\\
4&&$\fH^4_1$&${}^\epsilon\fH^4_1$&$\fH^6_2$&$\fC^7_4$&$\fR^0_8$&
${}^\epsilon\fR^0_8$&$\fR^2_{16}$&
$\fC^3_{16}$&$\ldots$\\
5&&$\fC^3_2$&$\fH^4_2$&${}^\epsilon\fH^4_2$&$\fH^6_4$&$\fC^7_8$&
$\fR^0_{16}$&${}^\epsilon\fR^0_{16}$&
$\fR^2_{32}$&$\ldots$\\
6&&$\fR^2_4$&$\fC^3_4$&$\fH^4_4$&${}^\epsilon\fH^4_4$&$\fH^6_8$&
$\fC^7_{16}$&$\fR^0_{32}$&
${}^\epsilon\fR^0_{32}$&$\ldots$\\
7&&${}^\epsilon\fR^0_4$&$\fR^2_8$&$\fC^3_8$&$\fH^4_8$&
${}^\epsilon\fH^4_8$&$\fH^6_{16}$&$\fC^7_{32}$&
$\fR^0_{64}$&$\ldots$\\
$\vdots$&&$\vdots$&$\vdots$&$\vdots$&$\vdots$&$\vdots$&$\vdots$&$\vdots$&
$\vdots$
\end{tabular}}
\end{center}
\end{figure}

\section{Fundamental automorphisms of\protect\newline odd-dimensional
Clifford algebras}
Let us consider now the automorphisms $\cA\rightarrow\cA^\star$,
$\cA\rightarrow\widetilde{\cA}$, $\cA\rightarrow\widetilde{\cA^\star}$,
$\cA\rightarrow\overline{\cA}$, $\cA\rightarrow\overline{\cA^\star}$,
$\cA\rightarrow\overline{\widetilde{\cA}}$,
$\cA\rightarrow\overline{\widetilde{\cA^\star}}$ of the algebras
$\C_{n+1}$ and $\cl_{p,q}$, $p-q\equiv 1,3,5,7\pmod{8}$.
We will examine the form of the 
fundamental automorphisms\index{automorphism!fundamental}
defined in $\C_{n+1}$ and $\cl_{p,q}$ after the homomorphic mappings
$\epsilon:\;\C_{n+1}\rightarrow\C_n\subset\C_{n+1}$ and
$\epsilon:\,\cl_{p,q}\rightarrow{}^\epsilon\cl_{p,q-1}$,
$\epsilon:\,\cl_{p,q}\rightarrow{}^\epsilon\cl_{q,p-1}$.
\begin{theorem}\label{tfactor}1) $\F=\C$.
Let $\cA\rightarrow\overline{\cA}$, $\cA\rightarrow\cA^\star$,
$\cA\rightarrow\widetilde{\cA}$ be the automorphisms of the odd--dimensional
complex Clifford algebra $\C_{n+1}$ ($n+1\equiv 1,3\s\pmod{4}$) corresponding
the discrete transformations $C,\,P,\,T$ 
(charge conjugation\index{conjugation!charge}, 
space inversion\index{space inversion},
time reversal)\index{time reversal}
and let ${}^\epsilon\C_n$ be a quotient algebra obtained in the
result of the homomorphic mapping $\epsilon:\;\C_{n+1}
\rightarrow\C_n$. Then over the field $\F=\C$ 
in dependence on the structure of
${}^\epsilon\C_n$ all the 
quotient representations\index{representation!quotient} of the
group $\pin(p,q)$ are divided in the following six classes:
\begin{eqnarray}
1)\;{}^\epsilon\fC_{a_1}&:&\{T,\,C\sim I\},\nonumber\\
2)\;{}^\epsilon\fC_{a_2}&:&\{T,\,C\},\nonumber\\
3)\;{}^\epsilon\fC_{b}&:&\{T,\,CP,\,CPT\},\nonumber
\end{eqnarray}
\begin{eqnarray}
4)\;{}^\epsilon\fC_{c}&:&\{PT,\,C,\,CPT\},\nonumber\\
5)\;{}^\epsilon\fC_{d_1}&:&\{PT,\,CP\sim IP,\,CT\sim IT\},\nonumber\\
6)\;{}^\epsilon\fC_{d_2}&:&\{PT,\,CP,\,CT\}.\nonumber
\end{eqnarray}
2) $\F=\R$. Real quotient representations are divided into four different
classes:
\begin{eqnarray}
7)\;{}^\epsilon\fR_{e_1}&:&\{T,\,C\sim I,\,CT\sim IT\},\nonumber\\
8)\;{}^\epsilon\fR_{e_2}&:&\{T,\,CP\sim IP,\,CPT\sim IPT\},\nonumber\\
9)\;{}^\epsilon\fH_{f_1}&:&\{T,\,C\sim C^\prime,\,CT\sim C^\prime T \},
\nonumber\\
10)\;{}^\epsilon\fH_{f_2}&:&\{T,\,CP\sim C^\prime P,\,CPT\sim C^\prime PT\}.
\nonumber
\end{eqnarray}
\end{theorem}
\begin{proof}
1) Complex representations.\\
Before we proceed to find an explicit form of the quotient representations
${}^\epsilon\fC$ it is necessary to consider in details a
structure of the quotient algebras
${}^\epsilon\C_n$ obtained in the result of the homomorphic mapping
$\epsilon:\,\C_{n+1}\rightarrow\C_n$. The structure of the quotient algebra
${}^\epsilon\C_n$ depends on the transfer of the automorphisms
$\cA\rightarrow\cA^\star,
\;\cA\rightarrow\widetilde{\cA},\;\cA\rightarrow\widetilde{\cA^\star},\;
\cA\rightarrow\overline{\cA}$, $\cA\rightarrow\overline{\cA^\star}$,
$\cA\rightarrow\overline{\widetilde{\cA}}$,
$\cA\rightarrow\overline{\widetilde{\cA^\star}}$ 
of the algebra $\C_{n+1}$ onto its subalgebra $\C_n$
under action of the
homomorphism $\epsilon$. 
The action of the homomorphism
$\epsilon$ is defined as follows
\[
\epsilon:\;\cA^1+\varepsilon\omega\cA^2\longrightarrow\cA^1+\cA^2,
\]
where $\cA^1,\,\cA^2\in\C_n$, $\omega=\e_{12\cdots n+1}$, and
\[
\varepsilon=\begin{cases}
1,& \text{if $n+1\equiv 1\s\!\!\pmod{4}$},\\
i,& \text{if $n+1\equiv 3\s\!\!\pmod{4}$};
\end{cases}
\]
so that $(\varepsilon\omega)^2=1$. At this point, 
$\varepsilon\omega\rightarrow 1$,
and the quotient algebra has a form
\[
{}^\epsilon\C_n\simeq\C_{n+1}/\Ker\epsilon,
\]
where $\Ker\epsilon=\left\{\cA^1-\varepsilon\omega\cA^1\right\}$ is a kernel
of the homomorphism\index{kernel!of the homomorphism} $\epsilon$. 
\begin{sloppypar}
First of all, for the
antiautomorphism
$\cA\rightarrow\widetilde{\cA}$ we see that elements
$\cA,\cB,\ldots\in\C_{n+1}$, which are mapped into one and the same element
$\cD\in\C_n$ (a kernel of the homomorphism $\epsilon$ if $\cD=0$), after the
transformation $\cA\rightarrow\widetilde{\cA}$ are must converted to the
elements $\widetilde{\cA},\widetilde{\cB},\ldots\in\C_{n+1}$, which are also
mapped into one and the same element $\widetilde{\cD}\in\C_n$. Otherwise,
the transformation $\cA\rightarrow\widetilde{\cA}$ is not transferred from
$\C_{n+1}$ to $\C_n$ as an unambiguous transformation. Therefore,
for the transfer of the antiautomorphism $\cA\rightarrow\widetilde{\cA}$ 
from $\C_{n+1}$ into $\C_n$ it is necessary that\end{sloppypar}
\begin{equation}\label{6.34}
\widetilde{\varepsilon\omega}=\varepsilon\omega.
\end{equation}
Indeed, since under action of $\epsilon$ the elements
1 and $\varepsilon\omega$ are equally mapped into the unit, then transformed
elements $\widetilde{1}$ and $\widetilde{\varepsilon\omega}$ are also
should be mapped into 1, but $\widetilde{1}=1\rightarrow 1$, and
$\widetilde{\varepsilon\omega}=\pm\varepsilon\omega\rightarrow\pm 1$ in
virtue of $\widetilde{\omega}=(-1)^{\frac{n(n-1)}{2}}\omega$, whence
\begin{equation}\label{6.34'}
\widetilde{\omega}=\begin{cases}
\omega,& \text{if $n+1\equiv 1\s\!\!\pmod{4}$};\\
-\omega,& \text{if $n+1\equiv 3\s\!\!\pmod{4}$}.
\end{cases}
\end{equation}
The condition (\ref{6.34}) is sufficient for the transfer of the
antiautomorphism
$\cA\rightarrow\widetilde{\cA}$ from $\C_{n+1}$ to $\C_n$. Indeed,
in this case at $\cA\rightarrow\widetilde{\cA}$ we have
\[
\cA^1-\cA^1\varepsilon\omega\longrightarrow\widetilde{\cA^1}-
\widetilde{\varepsilon\omega}\widetilde{\cA^1}=\widetilde{\cA^1}-
\varepsilon\omega\widetilde{\cA^1}.
\]
Thus, elements of the form $\cA^1-\cA^1\varepsilon\omega$, consisting of
the kernel of the homomorphism $\epsilon$, are converted at the transformation
$\cA\rightarrow\widetilde{\cA}$ into elements of the same form.
Therefore, {\it under action of the homomorphism $\epsilon$ the antiautomorphism
$\cA\rightarrow\widetilde{\cA}$ is transferred from $\C_{n+1}$ 
into $\C_n$ only at $n\equiv 0\s\pmod{4}$}.

In turn, for the transfer of the automorphism $\cA\rightarrow\cA^\star$
it is necessary that $(\varepsilon\omega)^\star=\varepsilon\omega$. However,
since the element $\omega$ is odd and $\omega^\star=(-1)^{n+1}
\omega$, then we have always
\begin{equation}\label{6.35}
\omega^\star=-\omega.
\end{equation}
Thus, {\it the automorphism $\cA\rightarrow\cA^\star$ is never transferred
from $\C_{n+1}$ into $\C_n$}.

Further, for the transfer of the antiautomorphism 
$\cA\rightarrow\widetilde{\cA^\star}$
from $\C_{n+1}$ into $\C_n$ it is necessary that
\begin{equation}\label{6.36}
\widetilde{(\varepsilon\omega)^\star}=\varepsilon\omega.
\end{equation}
It is easy to see that the condition (\ref{6.36}) is satisfied only at
$n+1\equiv 3\s\pmod{4}$, since in this case from the second equality of
(\ref{6.34'}) and (\ref{6.35}) it follows
\begin{equation}\label{6.36'}
\widetilde{(\varepsilon\omega)^\star}=\varepsilon\widetilde{\omega^\star}=
-\varepsilon\omega^\star=\varepsilon\omega.
\end{equation}
Therefore, {\it under action of the homomorphism $\epsilon$ the 
antiautomorphism
$\cA\rightarrow\widetilde{\cA^\star}$ is transferred from $\C_{n+1}$ into $\C_n$
only at $n\equiv 2\s\pmod{4}$}.

Let $n+1=p+q$. Defining in $\C_{n+1}$ the basis $\{\e_1,\ldots,\e_p,i\e_{p+1},
\ldots, i\e_{p+q}\}$, we extract the 
real subalgebra\index{subalgebra!real} $\cl_{p,q}$, where at
$p-q\equiv 3,7\s\pmod{8}$ we have a complex division ring
$\K\simeq\C$, and at $p-q\equiv 1\s\pmod{8}$ and $p-q\equiv 5\s\pmod{8}$
correspondingly a double real division ring $\K\simeq\R\oplus\R$ and a double
quaternionic division ring\index{ring!double!quaternionic}
$\K\simeq\BH\oplus\BH$. The product
$\e_1\e_2\cdots\e_pi\e_{p+1}\cdots i\e_{p+q}=i^q\omega\in\C_{n+1}$ sets
a volume element of the real subalgebra $\cl_{p,q}$. At this point we have a
condition $\overline{(i^q\omega)}=i^q\omega$, that is,
$(-i)^q\overline{\omega}=i^q\omega$, whence
\begin{equation}\label{6.37}
\overline{\omega}=(-1)^q\omega.
\end{equation}
When $q$ is even, from (\ref{6.37}) it follows that 
$\overline{\omega}=\omega$ and,
therefore, the pseudoautomorphism\index{pseudoautomorphism}
$\cA\rightarrow\overline{\cA}$ is transferred
at $q\equiv 0\s\pmod{2}$, and since $p+q$ is odd number, then we have always
$p\equiv 1\s\pmod{2}$. In more detail, at $n+1\equiv 3\s\pmod{4}$ the
pseudoautomorphism $\cA\rightarrow\overline{\cA}$ is transferred from $\C_{n+1}$
into $\C_n$ if the real subalgebra $\cl_{p,q}$ possesses the complex ring
$\K\simeq\C$, $p-q\equiv 3,7\s\pmod{8}$, and is not transferred
($\overline{\omega}=-\omega,\;q\equiv 1\s\pmod{2},\,p\equiv 0\s\pmod{2}$)
in the case of $\cl_{p,q}$ with double rings $\K\simeq\R\oplus\R$ and
$\K\simeq\BH\oplus\BH$, $p-q\equiv 1,5\s\pmod{8}$. In turn, at
$n+1\equiv 1\s\pmod{4}$ the pseudoautomorphism $\cA\rightarrow\overline{\cA}$
is transferred from $\C_{n+1}$ into $\C_n$ if the subalgebra $\cl_{p,q}$ has
the type $p-q\equiv 1,5\s\pmod{8}$ and is not transferred in the case
of $\cl_{p,q}$ with $p-q\equiv 3,7\s\pmod{8}$. 
Besides, in virtue of (\ref{6.35}) at
$n+1\equiv 3\s\pmod{4}$ with $p-q\equiv 1,5\s\pmod{8}$ 
and at $n+1\equiv 1\s\pmod{4}$ with 
$p-q\equiv 3,7\s\pmod{8}$ a pseudoautomorphism
$\cA\rightarrow\overline{\cA^\star}$ (a composition of the pseudoautomorphism
$\cA\rightarrow\overline{\cA}$ with the automorphism $\cA\rightarrow\cA^\star$)
is transferred from $\C_{n+1}$ into $\C_n$, since
\[
\overline{\varepsilon\omega^\star}=\varepsilon\omega.
\]
Further, in virtue of the second equality of (\ref{6.34'}) 
at $n+1\equiv 3\s\pmod{4}$ with
$p-q\equiv 1,5\s\pmod{8}$  
a pseudoantiautomorphism\index{pseudoantiautomorphism}
$\cA\rightarrow\overline{\widetilde{\cA}}$
(a composition of the pseudoautomorphism $\cA\rightarrow\overline{\cA}$ with
the antiautomorphism $\cA\rightarrow\widetilde{\cA}$) is transferred from
$\C_{n+1}$ into $\C_n$, since
\[
\overline{\widetilde{\varepsilon\omega}}=\varepsilon\omega.
\]
Finally, a pseudoantiautomorphism 
$\cA\rightarrow\overline{\widetilde{\cA^\star}}$ (a composition of the
pseudoautomorphism $\cA\rightarrow\overline{\cA}$ with the antiautomorphism
$\cA\rightarrow\widetilde{\cA^\star}$), 
corresponded to $CPT$--transformation (see Proposition \ref{prop2}),
is transferred from $\C_{n+1}$ into $\C_n$ at $n+1\equiv 3\pmod{4}$ and
$\cl_{p,q}$ with $p-q\equiv 3,7\pmod{8}$, since in this case in virtue of
(\ref{6.36'}) and (\ref{6.37}) we have
\[
\overline{\widetilde{(\varepsilon\omega^\star)}}=\varepsilon\omega.
\]
Also at $n+1\equiv 1\pmod{4}$ and $q\equiv 1\pmod{2}$ we obtain
\[
\overline{\widetilde{(\varepsilon\omega^\star)}}=-
\widetilde{(\varepsilon\omega^\star)}=-(\varepsilon\omega)^\star=
\varepsilon\omega,
\]
therefore, the transformation $\cA\rightarrow\overline{\widetilde{\cA^\star}}$
is transferred at $n+1\equiv 1\pmod{4}$ and $\cl_{p,q}$ with
$p-q\equiv 3,7\pmod{8}$.

The conditions for the transfer of the 
fundamental automorphisms\index{automorphism!fundamental} of the algebra
$\C_{n+1}$ into its subalgebra $\C_n$ under action of the homomorphism
$\epsilon$ allow us to define in evident way an explicit form of the quotient
algebras ${}^\epsilon\C_n$.\\[0.2cm]
1) The quotient algebra ${}^\epsilon\C_n$, $n\equiv 0\s\pmod{4}$.\\[0.1cm]
As noted previously, in the case $n+1\equiv 1\s\pmod{4}$ 
the antiautomorphism $\cA\rightarrow\widetilde{\cA}$ and
pseudoautomorphism $\cA\rightarrow\overline{\cA}$ are transferred from
$\C_{n+1}$ into $\C_n$ if the subalgebra
$\cl_{p,q}\subset\C_{n+1}$ possesses the double rings $\K\simeq\R\oplus\R$,
$\K\simeq\BH\oplus\BH$ ($p-q\equiv 1,5\s\pmod{8}$), and also the
pseudoautomorphism $\cA\rightarrow\overline{\cA^\star}$ and
pseudoantiautomorphism $\cA\rightarrow\overline{\widetilde{\cA^\star}}$
are transferred if
$\cl_{p,q}$ has the complex ring $\K\simeq\C$ ($p-q\equiv 3,7\s\pmod{8}$).
It is easy to see that in dependence on the type of $\cl_{p,q}$ the structure
of the quotient algebras\index{algebra!quotient}
${}^\epsilon\C_n$ of this type is divided into two
different classes:\\[0.1cm]
{\bf a}) The class of quotient algebras ${}^\epsilon\C_n$ containing the
antiautomorphism
$\cA\rightarrow\widetilde{\cA}$ and pseudoautomorphism $\cA\rightarrow
\overline{\cA}$. It is obvious that in dependence on a division ring
structure of the subalgebra
$\cl_{p,q}\subset\C_{n+1}$ this class is divided into two subclasses:
\begin{description}
\item[$a_1$)] ${}^\epsilon\C_n$ with $\cA\rightarrow\widetilde{\cA}$,
$\cA\rightarrow\overline{\cA}$ when $\cl_{p,q}$ has the ring
$\K\simeq\R\oplus\R$, $p-q\equiv 1\s\pmod{8}$.
\item[$a_2$)] ${}^\epsilon\C_n$ with $\cA\rightarrow\widetilde{\cA}$,
$\cA\rightarrow\overline{\cA}$ when $\cl_{p,q}$ has the ring
$\K\simeq\BH\oplus\BH$, $p-q\equiv 5\s\pmod{8}$.
\end{description}
{\bf b}) The class of quotient algebras ${}^\epsilon\C_n$ containing the
transformations
$\cA\rightarrow\widetilde{\cA}$, $\cA\rightarrow
\overline{\cA^\star}$, $\cA\rightarrow\overline{\widetilde{\cA^\star}}$ 
if the subalgebra $\cl_{p,q}\subset\C_{n+1}$ has the
complex ring\index{ring!complex}
$\K\simeq\C$, $p-q\equiv 3,7\s\pmod{8}$.\\[0.2cm]
2) The quotient algebra ${}^\epsilon\C_n$, $n\equiv 2\s\pmod{4}$.\\[0.1cm]
In the case $n+1\equiv 3\s\pmod{4}$ the antiautomorphism
$\cA\rightarrow\widetilde{\cA^\star}$, pseudoautomorphism
$\cA\rightarrow\overline{\cA}$ and pseudoantiautomorphism
$\cA\rightarrow\overline{\widetilde{\cA^\star}}$
are transferred from $\C_{n+1}$ into $\C_n$
if the subalgebra $\cl_{p,q}\subset\C_{n+1}$
possesses the complex ring $\K\simeq\C$ ($p-q\equiv 3,7\s\pmod{8}$), and also
the pseudoautomorphism $\cA\rightarrow\overline{\cA^\star}$ and
pseudoantiautomorphism $\cA\rightarrow\overline{\widetilde{\cA}}$ are
transferred if $\cl_{p,q}$ has the double rings $\K\simeq\R\oplus\R$,
$\K\simeq\BH\oplus\BH$ ($p-q\equiv 1,5\s\pmod{8}$). In dependence on the type of
$\cl_{p,q}\subset\C_{n+1}$ all the quotient algebras ${}^\epsilon\C_n$ of this
type are divided into following two classes:\\[0.1cm]
{\bf c}) The class of quotient algebras ${}^\epsilon\C_n$ containing the
transformations
$\cA\rightarrow\widetilde{\cA^\star}$,
$\cA\rightarrow\overline{\cA}$, 
$\cA\rightarrow\overline{\widetilde{\cA^\star}}$
if the subalgebra $\cl_{p,q}$ has the ring
$\K\simeq\C$, $p-q\equiv 3,7\s\pmod{8}$.\\[0.1cm]
{\bf d}) The class of quotient algebras ${}^\epsilon\C_n$ containing the
antiautomorphism
$\cA\rightarrow\widetilde{\cA^\star}$, pseudoautomorphism
$\cA\rightarrow\overline{\cA^\star}$ and pseudoautomorphism
$\cA\rightarrow\overline{\widetilde{\cA}}$. At this point, in dependence on the
division ring structure of $\cl_{p,q}$ we have two subclasses
\begin{description}
\item[$d_1$)] ${}^\epsilon\C_n$ with $\cA\rightarrow\widetilde{\cA^\star}$,
$\cA\rightarrow\overline{\cA^\star}$ and $\cA\rightarrow
\overline{\widetilde{\cA}}$ at $\cl_{p,q}$ with the ring $\K\simeq\R\oplus\R$,
$p-q\equiv 1\s\pmod{8}$.
\item[$d_2$)] ${}^\epsilon\C_n$ with $\cA\rightarrow\widetilde{\cA^\star}$,
$\cA\rightarrow\overline{\cA^\star}$ and $\cA\rightarrow
\overline{\widetilde{\cA}}$ at $\cl_{p,q}$ with the ring $\K\simeq\BH\oplus\BH$,
$p-q\equiv 5\s\pmod{8}$.
\end{description}
Thus, we have 6 different classes of the 
quotient algebras\index{algebra!quotient} ${}^\epsilon\C_n$.
Further, in accordance with Proposition \ref{prop1},
the automorphism $\cA\rightarrow\cA^\star$ corresponds to 
space inversion\index{space inversion}
$P$, the antiautomorphisms
$\cA\rightarrow\widetilde{\cA}$ and $\cA\rightarrow\widetilde{\cA^\star}$
set correspondingly time reversal $T$\index{time reversal}
and full reflection\index{full reflection}
$PT$, and in accordance with Theorem \ref{tpseudo}
the pseudoautomorphism $\cA\rightarrow\overline{\cA}$ corresponds to
charge conjugation $C$.\index{conjugation!charge}
Taking into account these relations
we come to the classification presented in Theorem for complex
quotient representations.\index{representation!quotient!complex}\\[0.2cm]
2) Real representations.\\
Let us define 
real quotient representations\index{representation!quotient!real}
of the group $\pin(p,q)$. First
of all, in the case of types $p-q\equiv 3,7\pmod{8}$ 
these representations are equivalent
to complex representations. Further, when
$p-q\equiv 1,5\pmod{8}$ we have the real algebras $\cl_{p,q}$ with the
rings $\K\simeq\R\oplus\R$, $\K\simeq\BH\oplus\BH$ and, therefore, there
exist homomorphic mappings $\epsilon:\,\cl_{p,q}\rightarrow\cl_{p,q-1}$,
$\epsilon:\,\cl_{p,q}\rightarrow\cl_{q,p-1}$. In this case the quotient
algebra has a form
\[
{}^\epsilon\cl_{p,q-1}\simeq\cl_{p,q}/\Ker\epsilon
\]
or
\[
{}^\epsilon\cl_{q,p-1}\simeq\cl_{p,q}/\Ker\epsilon,
\]
where $\Ker\epsilon=\left\{\cA^1-\omega\cA^1\right\}$ is a kernel of $\epsilon$, 
since in accordance with
\[
\omega^2=\begin{cases}
-1& \text{if $p-q\equiv 2,3,6,7\pmod{8}$},\\
+1& \text{if $p-q\equiv 0,1,4,5\pmod{8}$}
\end{cases}
\]
at $p-q\equiv 1,5\pmod{8}$ we have always $\omega^2=1$ and, therefore,
$\varepsilon=1$. Thus, for the transfer of the antiautomorphism
$\cA\rightarrow\widetilde{\cA}$ from $\cl_{p,q}$ into $\cl_{p,q-1}$
($\cl_{q,p-1}$) it is necessary that
\[
\widetilde{\omega}=\omega.
\]
In virtue of the relation $\widetilde{\omega}=(-1)^{\frac{(p+q)(p+q+1)}{2}}
\omega$ we obtain
\begin{equation}\label{Real1}
\widetilde{\omega}=\begin{cases}
+\omega& \text{if $p-q\equiv 1,5\pmod{8}$},\\
-\omega& \text{if $p-q\equiv 3,7\pmod{8}$}.
\end{cases}
\end{equation}
Therefore, for the algebras over the field $\F=\R$ the antiautomorphism
$\cA\rightarrow\widetilde{\cA}$ is transferred at the mappings
$\cl_{p,q}\rightarrow\cl_{p,q-1}$, $\cl_{p,q}\rightarrow\cl_{q,p-1}$, where
$p-q\equiv 1,5\pmod{8}$.

In turn, for the transfer of the automorphism $\cA\rightarrow\cA^\star$
it is necessary that $\omega^\star=\omega$. However, since the element
$\omega$ is odd and $\omega^\star=(-1)^{p+q}\omega$, then we have always
\begin{equation}\label{Real2}
\omega^\star=-\omega.
\end{equation}
Thus, the automorphism $\cA\rightarrow\cA^\star$ is never transferred from
$\cl_{p,q}$ into $\cl_{p,q-1}$ ($\cl_{q,p-1}$).

Further, for the transfer of the antiautomorphism 
$\cA\rightarrow\widetilde{\cA^\star}$ it is necessary that
\[
\widetilde{\omega^\star}=\omega.
\]
From (\ref{Real1}) and (\ref{Real2}) for the types $p-q\equiv 1,5\pmod{8}$
we obtain
\begin{equation}\label{Real3}
\widetilde{\omega^\star}=\omega^\star=-\omega.
\end{equation}
Therefore, under action of the homomorphism $\epsilon$ the antiautomorphism
$\cA\rightarrow\widetilde{\cA^\star}$ is never transferred from $\cl_{p,q}$
into $\cl_{p,q-1}$ ($\cl_{q,p-1}$).

As noted previously, for the real representations of $\pin(p,q)$ the
pseudoautomorphism $\cA\rightarrow\overline{\cA}$ is reduced into identical
transformation $\sI$ for $\fR^{0,2}_m$ and to particle--antiparticle
conjugation $C^\prime$ for $\fH^{4,6}_m$. The volume element $\omega$
of $\cl_{p,q}$ (types $p-q\equiv 1,5\pmod{8}$) can be represented by the
product $\e_1\e_2\cdots\e_p\e^{\p}_{p+1}\e^{\p}_{p+2}\cdots\e^{\p}_{p+q}$,
where $\e^{\p}_{p+j}=i\e_{p+j}$, $\e^2_j=1$, $(\e^{\p}_{p+j})^2=-1$.
Therefore, for the transfer of $\cA\rightarrow\overline{\cA}$ from $\cl_{p,q}$
into $\cl_{p,q-1}$ ($\cl_{q,p-1}$) we have a condition
\[
\overline{\omega}=\omega,
\]
and in accordance with (\ref{6.37}) it follows that the pseudoautomorphism
$\cA\rightarrow\overline{\cA}$ is transferred at $q\equiv 0\pmod{2}$. 
Further, in virtue of the relation (\ref{Real2}) the pseudoautomorphism
$\cA\rightarrow\overline{\cA^\star}$ is transferred at $q\equiv 1\pmod{2}$,
since in this case we have
\[
\overline{\omega^\star}=\omega.
\]
Also from (\ref{Real1}) it follows that the pseudoantiautomorphism
$\cA\rightarrow\overline{\widetilde{\cA}}$ is transferred at
$p-q\equiv 1,5\pmod{8}$ and $q\equiv 0\pmod{2}$, since
\[
\overline{\widetilde{\omega}}=\omega.
\]\begin{sloppypar}\noindent
Finally, the pseudoantiautomorphism 
$\cA\rightarrow\overline{\widetilde{\cA^\star}}$ ($CPT$--transformation)
in virtue of (\ref{Real3}) and (\ref{6.37}) is transferred from $\cl_{p,q}$
into $\cl_{p,q-1}$ ($\cl_{q,p-1}$) at $p-q\equiv1,5\pmod{8}$ and
$q\equiv 1\pmod{2}$.\end{sloppypar}

Now we are in a position that allows us to classify the 
real quotient algebras\index{algebra!quotient!real}
${}^\epsilon\cl_{p,q-1}$ (${}^\epsilon\cl_{q,p-1}$).\\
1) The quotient algebra ${}^\epsilon\cl_{p,q-1}$ (${}^\epsilon\cl_{q,p-1}$),
$p-q\equiv 1\pmod{8}$.\\
In this case the initial algebra $\cl_{p,q}$ has the double real division
ring $\K\simeq\R\oplus\R$ and its subalgebras $\cl_{p,q-1}$ and
$\cl_{q,p-1}$ are of the type $p-q\equiv 0\pmod{8}$ or $p-q\equiv 2\pmod{8}$
with the ring $\K\simeq\R$. Therefore, in accordance with Theorem 
\ref{tpseudo} for all such quotient algebras the pseudoautomorphism
$\cA\rightarrow\overline{\cA}$ is equivalent to the identical transformation
$\sI$. The antiautomorphism $\cA\rightarrow\widetilde{\cA}$ in this case
is transferred into $\cl_{p,q-1}$ ($\cl_{q,p-1}$) at any $p-q\equiv 1\pmod{8}$.
Further, in dependence on the number $q$ we have two different classes of
the quotient algebras of this type:
\begin{description}
\item[$e_1$)] ${}^\epsilon\cl_{p,q-1}$ (${}^\epsilon\cl_{q,p-1}$) with
$\cA\rightarrow\widetilde{\cA}$, $\cA\rightarrow\overline{\cA}$,
$\cA\rightarrow\overline{\widetilde{\cA}}$, $p-q\equiv 1\pmod{8}$,
$q\equiv 0\pmod{2}$.
\item[$e_2$)] ${}^\epsilon\cl_{p,q-1}$ (${}^\epsilon\cl_{q,p-1}$) with
$\cA\rightarrow\widetilde{\cA}$, $\cA\rightarrow\overline{\cA^\star}$,
$\cA\rightarrow\overline{\widetilde{\cA^\star}}$, $p-q\equiv 1\pmod{8}$,
$q\equiv 1\pmod{2}$.
\end{description}
2) The quotient algebras ${}^\epsilon\cl_{p,q-1}$ (${}^\epsilon\cl_{q,p-1}$),
$p-q\equiv 5\pmod{8}$.\\
In this case the initial algebra $\cl_{p,q}$ has the double quaternionic
division ring $\K\simeq\BH\oplus\BH$ and its subalgebras $\cl_{p,q-1}$ and
$\cl_{q,p-1}$ are of the type $p-q\equiv 4\pmod{8}$ or $p-q\equiv 6\pmod{8}$
with the ring $\K\simeq\BH$. Therefore, in this case the pseudoautomorphism
$\cA\rightarrow\overline{\cA}$ is equivalent to the particle--antiparticle
conjugation $C^\prime$. As in the previous case the antiautomorphism
$\cA\rightarrow\widetilde{\cA}$ is transferred at any $p-q\equiv 5\pmod{8}$.
For this type in dependence on the number $q$ there are two different classes:
\begin{description}
\item[$f_1$)] ${}^\epsilon\cl_{p,q-1}$ (${}^\epsilon\cl_{q,p-1}$) with
$\cA\rightarrow\widetilde{\cA}$, $\cA\rightarrow\overline{\cA}$,
$\cA\rightarrow\overline{\widetilde{\cA}}$, $p-q\equiv 5\pmod{8}$,
$q\equiv 0\pmod{2}$.
\item[$f_2$)] ${}^\epsilon\cl_{p,q-1}$ (${}^\epsilon\cl_{q,p-1}$) with
$\cA\rightarrow\widetilde{\cA}$, $\cA\rightarrow\overline{\cA^\star}$,
$\cA\rightarrow\overline{\widetilde{\cA^\star}}$, $p-q\equiv 5\pmod{8}$,
$q\equiv 1\pmod{2}$.
\end{description}
\end{proof}

In accordance with Theorem \ref{tfactor}, in the case of odd--dimensional spaces
$\R^{p,q}$ and $\C^{p+q}$ 
the algebra homomorphisms\index{homomorphism!algebra}
$\cl_{p,q}\rightarrow
\cl_{p,q-1},\,\cl_{p,q}\rightarrow\cl_{q,p-1}$ and $\C_{p+q}\rightarrow
\C_{p+q-1}$ induce group homomorphisms\index{homomorphism!group}
$\pin(p,q)\rightarrow\pin(p,q-1),\,
\pin(p,q)\rightarrow\pin(q,p-1)$, $\pin(p+q,\C)\rightarrow\pin(p+q-1,\C)$.
\begin{theorem}\label{todd}\begin{sloppypar}\noindent
1) If $\F=\C$ and $\pin^{a,b,c,d,e,fg}(n+1,\C)\simeq\pin^{a,b,c,d,e,f,g}(n,\C)\cup
\pin^{a,b,c,d,e,f,g}(n,\C)$ are universal coverings of the complex
orthogonal groups\index{group!orthogonal!complex} $O(n+1,\C)$, 
then in the result of the homomorphic mapping 
$\epsilon:\,\pin(n+1,\C)\rightarrow\pin(n,\C)$ we have the following
quotient groups:\end{sloppypar}
\begin{eqnarray}
\pin^b(n,\C)&\simeq&\frac{(\spin_+(n,\C)\odot\dZ_2\otimes\dZ_2)}{\dZ_2},
\nonumber\\
\pin^{b,d}(n,\C)&&\nonumber\\
\pin^{b,e,g}(n,\C)&\simeq&\frac{(\spin_+(n,\C)\odot C^{b,e,g})}{\dZ_2}
\nonumber
\end{eqnarray}
if $n+1\equiv 1\pmod{4}$ and
\begin{eqnarray}
\pin^{c,d,g}(n,\C)&\simeq&\frac{(\spin_+(n,\C)\odot C^{c,d,g})}{\dZ_2},
\nonumber\\
\pin^{a,b,c}(n,\C)&\simeq&\frac{(\spin_+(n,\C)\odot C^{a,b,c})}{\dZ_2},
\nonumber\\
\pin^{c,e,f}(n,\C)&\simeq&\frac{(\spin_+(n,\C)\odot C^{c,e,f})}{\dZ_2}
\nonumber
\end{eqnarray}\begin{sloppypar}\noindent
if $n+1\equiv 3\pmod{4}$.\\[0.2cm]
2) If $\F=\R$ and $\pin^{a,b,c,d,e,f,g}(p,q)\simeq
\pin^{a,b,c,d,e,f,g}(p,q-1)\cup\omega
\pin^{a,b,c,d,e,f,g}(p,q-1)$, $\pin^{a,b,c,d,e,f,g}(p,q)\simeq
\pin^{a,b,c,d,e,f,g}(q,p-1)\cup\omega
\pin^{a,b,c,d,e,f,g}(q,p-1)$ are 
universal coverings\index{covering!universal} of the orthogonal
groups\index{group!orthogonal} $O(p,q)$
then in the results of the homomorphic mappings $\epsilon:\,\pin(p,q)
\rightarrow\pin(p,q-1)$ and $\epsilon:\,\pin(p,q)\rightarrow
\pin(q,p-1)$ we have the following quotient groups:\index{group!quotient}
\end{sloppypar}
\begin{eqnarray}
\pin^b(p,q-1)&\simeq&\frac{(\spin_+(p,q-1)\odot\dZ_2\otimes\dZ_2)}{\dZ_2},
\nonumber\\
\pin^b(q,p-1)&\simeq&\frac{(\spin_+(q,p-1)\odot\dZ_2\otimes\dZ_2)}{\dZ_2},
\nonumber\\
\pin^{a,b,c}(p,q-1)&\simeq&\frac{(\spin_+(p,q-1)\odot C^{a,b,c})}{\dZ_2},
\nonumber\\
\pin^{a,b,c}(q,p-1)&\simeq&\frac{(\spin_+(q,p-1)\odot C^{a,b,c})}{\dZ_2}
\nonumber
\end{eqnarray}
if $p-q\equiv 1\pmod{8}$ and
\begin{eqnarray}
\pin^{b,d,f}(p,q-1)&\simeq&\frac{(\spin_+(p,q-1)\odot C^{b,d,f})}{\dZ_2},
\nonumber\\
\pin^{b,d,f}(q,p-1)&\simeq&\frac{(\spin_+(q,p-1)\odot C^{b,d,f})}{\dZ_2}
\nonumber\\
\pin^{b,e,g}(p,q-1)&\simeq&\frac{(\spin_+(p,q-1)\odot C^{b,e,g})}{\dZ_2},
\nonumber\\
\pin^{b,e,g}(q,p-1)&\simeq&\frac{(\spin_+(q,p-1)\odot C^{b,e,g})}{\dZ_2}
\nonumber
\end{eqnarray}
if $p-q\equiv 5\pmod{8}$.
\end{theorem}
\begin{proof} 
As it has been shown in Theorem \ref{tfactor}, over the field $\F=\C$
for the case $n+1\equiv 1\pmod{4}$ the antiautomorphism
$\cA\rightarrow\widetilde{\cA}$ and pseudoautomorphism
$\cA\rightarrow\overline{\cA}$ are transferred from $\C_{n+1}$ to $\C_n$
if the real subalgebra\index{subalgebra!real}
$\cl_{p,q}\subset\C_{n+1}$ possesses the double
rings $\K\simeq\R\oplus\R$, $\K\simeq\BH\oplus\BH$, $p-q\equiv 1,5\pmod{8}$.
Further, the quotient algebra\index{algebra!quotient}
${}^\epsilon\C_n$ generates the quotient
group\index{group!quotient}
$\pin^\epsilon(n,\C)$. In the case of $\K\simeq\R\oplus\R$ we
have $\pin^b(n,\C)$, since the pseudoautomorphism 
$\cA\rightarrow\overline{\cA}$ is equivalent to the identical
transformations for $\K\simeq\R\oplus\R$. The automorphism group
of ${}^\epsilon\C_n$ is 
$\Aut({}^\epsilon\C_n)=\{\Id,\widetilde{\phantom{cc}}\}\simeq
\{1,T\}$ with the multiplication table
\begin{equation}\label{Cayley1}{\renewcommand{\arraystretch}{1.4}
\begin{tabular}{|c||c|c|}\hline
  & $1$ & $T$ \\ \hline\hline
$1$ & $1$ & $T$ \\ \hline
$T$ & $T$ & $1$ \\ \hline
\end{tabular}\;\;\sim\;\;
\begin{tabular}{|c||c|c|}\hline
  & $\Id$ & $\widetilde{\phantom{cc}}$ \\ \hline\hline
$\Id$ & $\Id$ & $\widetilde{\phantom{cc}}$ \\ \hline
$\widetilde{\phantom{cc}}$ & $\widetilde{\phantom{cc}}$ & $\Id$ \\ \hline
\end{tabular}.
}
\end{equation}
It is easy to see that $\{1,T\}\sim\{\Id,\widetilde{\phantom{cc}}\}
\simeq\{\sI,\sE\}\simeq\dZ_2$. Therefore, 
the double covering\index{covering!double} $C^b$
of the reflection group is isomorphic to $\dZ_2\otimes\dZ_2$.

In the case of $\K\simeq\BH\oplus\BH$ we have the quotient group
$\pin^{b,d}(n,\C)$. At this point, a set of the discrete transformations
of the space $\C^n$ associated with ${}^\epsilon\C_n$ is defined by a
three-element set $\{1,T,C\}\sim\{\sI,\sE,\Pi\}$, where
$\{\sI,\sE,\Pi\}$ is an automorphism set of ${}^\epsilon\C_n$, $\sE$ and
$\Pi$ are the matrices of $\cA\rightarrow\widetilde{\cA}$ and
$\cA\rightarrow\overline{\cA}$, respectively. It is easy to see that
the set $\{1,T,C\}\sim\{\sI,\sE,\Pi\}$ does not form a group.
\begin{sloppypar}
Further, when the subalgebra $\cl_{p,q}\subset\C_{n+1}$ has the complex
ring $\K\simeq\C$ ($p-q\equiv 3,7\pmod{8}$), the transformations
$\cA\rightarrow\widetilde{\cA}$,
$\cA\rightarrow\overline{\cA^\star}$ and
$\cA\rightarrow\overline{\widetilde{\cA^\star}}$ are transferred from
$\C_{n+1}$ to $\C_n$, $n+1\equiv 1 \pmod{4}$. In this case we have
the quotient group $\pin^\epsilon(n,\C)\sim\pin^{b,e,g}(n,\C)$.
The automorphism group\index{group!automorphism} of ${}^\epsilon\C_n$ is 
$\Aut({}^\epsilon\C_n)=\{\Id,\widetilde{\phantom{cc}},\overline{\star},
\overline{\widetilde{\star}}\}\sim\{1,T,CP,CPT\}$ with the multiplication
table\end{sloppypar}
\begin{equation}\label{Cayley3}
{\renewcommand{\arraystretch}{1.4}
\begin{tabular}{|c||c|c|c|c|}\hline
   & $1$ & $T$ & $CP$ & $CPT$ \\ \hline\hline
$1$& $1$ & $T$ & $CP$ & $CPT$ \\ \hline
$T$& $T$ & $1$ &$CPT$ & $CP$ \\ \hline
$CP$ & $CP$ & $CPT$ & $1$ & $T$ \\ \hline
$CPT$ & $CPT$ & $CP$ & $T$ & $1$ \\ \hline
\end{tabular}\;\;\sim\;\;
\begin{tabular}{|c||c|c|c|c|}\hline
   & $\Id$ & $\widetilde{\phantom{cc}}$ & $\overline{\star}$ &
$\overline{\widetilde{\star}}$ \\ \hline\hline
$\Id$ & $\Id$ & $\widetilde{\phantom{cc}}$ & $\overline{\star}$ &
$\overline{\widetilde{\star}}$ \\ \hline
$\widetilde{\phantom{cc}}$ & $\widetilde{\phantom{cc}}$ & $\Id$ &
$\overline{\widetilde{\star}}$ & $\overline{\star}$ \\ \hline
$\overline{\star}$ & $\overline{\star}$ & $\overline{\widetilde{\star}}$ &
$\Id$ & $\widetilde{\phantom{cc}}$ \\ \hline
$\overline{\widetilde{\star}}$ & $\overline{\widetilde{\star}}$ &
$\overline{\star}$ & $\widetilde{\phantom{cc}}$ & $\Id$ \\ \hline
\end{tabular}.
}
\end{equation}\begin{sloppypar}\noindent
Therefore, $\Aut({}^\epsilon\C_n)\simeq\dZ_2\otimes\dZ_2$, and we see that
$\Aut({}^\epsilon\C_n)=\{\Id,\widetilde{\phantom{cc}},\overline{\star},
\overline{\widetilde{\star}}\}$ is a subgroup of the extended
automorphism group\index{group!automorphism!extended}
$\Ext(\C_n)=\{\Id,\star,\widetilde{\phantom{cc}},
\widetilde{\star},\overline{\phantom{cc}},\overline{\star},
\overline{\widetilde{\phantom{cc}}},\overline{\widetilde{\star}}\}$
considered in the section 3. When $T^2=(CP)^2=(CPT)^2=\pm 1$ and
$T(CP)=-(CP)T$ we have a group $\sAut({}^\epsilon\C_n)=\{\sI,\sE,\sK,\sF\}
\subset\sExt(\C_n)$, where $\sExt(\C_n)=\{\sI,\sW,\sE,\sC,\Pi,\sK,\sS,\sF\}$.
Therefore, $\pin^{b,e,g}(n,\C)\simeq(\spin_+(n,\C)\odot C^{b,e,g})/\dZ_2$,
where $C^{b,e,g}$ are the four double coverings of the Gauss-Klein
viergruppe $\dZ_2\otimes\dZ_2$; that is, $C^{b,e,g}$ are the groups
$\dZ_2\otimes\dZ_2\otimes\dZ_2$ (when $b=e=g=+$), $D_4$ (dihedral group,
when there are two pluses and one minus in the triple $b,e,g$), 
$\dZ_4\otimes\dZ_2$ (when there are two minuses and one plus in $b,e,g$),
and $Q_4$ (quaternions, when $b=e=g=-$).
Since $\sAut({}^\epsilon\C_n)=\{\sI,\sE,\sK,\sF\}$ is a subgroup of
$\sExt(\C_n)$, then we can define a structure of $\sAut({}^\epsilon\C_n)$
using Theorem \ref{tautext}. First of all, we see that there are four
different types of $\sAut({}^\epsilon\C_n)$:\end{sloppypar}
\begin{eqnarray}
\sAut^1({}^\epsilon\C_n)&=&\{\sI,\cE_{j_1j_2\cdots j_k},
\cE_{\beta_1\beta_2\cdots\beta_b},\cE_{d_1d_2\cdots d_g}\},\nonumber\\
\sAut^2({}^\epsilon\C_n)&=&\{\sI,\cE_{j_1j_2\cdots j_k},
\cE_{\alpha_1\alpha_2\cdots\alpha_a},\cE_{c_1c_2\cdots c_s}\},\nonumber\\
\sAut^3({}^\epsilon\C_n)&=&\{\sI,\cE_{i_1i_2\cdots i_{p+q-k}},
\cE_{\beta_1\beta_2\cdots\beta_b},\cE_{c_1c_2\cdots c_s}\},\nonumber\\
\sAut^4({}^\epsilon\C_n)&=&\{\sI,\cE_{i_1i_2\cdots i_{p+q-k}},
\cE_{\alpha_1\alpha_2\cdots\alpha_a},\cE_{d_1d_2\cdots d_g}\}.\nonumber
\end{eqnarray}
All the elements $\sE=\cE_{j_1j_2\cdots i_k}$, 
$\sK=\cE_{\beta_1\beta_2\cdots\beta_b}$, $\sF=\cE_{d_1d_2\cdots d_g}$ of
$\sAut^1({}^\epsilon\C_n)$ are even products, that is, $k\equiv 0\pmod{2}$,
$b\equiv 0\pmod{2}$ and $g\equiv 0\pmod{2}$. The element $\sK$ commutes
with $\sF$ at $u\equiv 0\pmod{2}$ and commutes with $\sE$ at 
$u(m+v)\equiv 0\pmod{2}$. In turn, from (\ref{CPT55}) it follows that
$\sF$ commutes with $\sE$ at $m(v+u)\equiv 0\pmod{2}$. In this case
we have an Abelian group $\sAut({}^\epsilon\C_n)\simeq\dZ_2\otimes\dZ_2$ if
the elements $\sE$, $\sK$, $\sF$ have positive squares ($b=e=g=+$).
And also we have the groups $\sAut({}^\epsilon\C_n)\simeq\dZ_4$ with the
signatures $(b,e,g)$, where among $b,e,g$ there are two minuses and one
plus. Otherwise, we have non-Abelian groups $Q_4/\dZ_2$ and
$D_4/\dZ_2$. In the case of $\sAut^2({}^\epsilon\C_n)$ the elements
$\sE$, $\sK$, $\sF$ are both even and odd: $k\equiv 0\pmod{2}$,
$a\equiv 1\pmod{2}$, $s\equiv 1\pmod{2}$. The element
$\sK=\cE_{\alpha_1\alpha_2\cdots\alpha_a}$ commutes with
$\sF=\cE_{c_1c_2\cdots c_s}$ at $m\equiv 0\pmod{2}$ and commutes with
$\sE=\cE_{j_1j_2\cdots j_k}$ at $m(u+l)\equiv 0\pmod{2}$. In turn,
from (\ref{CPT54}) it follows that $\sF$ commutes with $\sE$ at
$u(l+m)\equiv 0\pmod{2}$. At these conditions $\sAut^2({}^\epsilon\C_n)$
coincides with the Abelian groups $\dZ_2\otimes\dZ_2$, otherwise we have
the groups $Q_4/\dZ_2$, $D_4/\dZ_2$. It is easy to see that
$\sAut^3({}^\epsilon\C_n)$ and $\sAut^4({}^\epsilon\C_n)$ have the
analogous structure.

Further, when the subalgebra $\cl_{p,q}\subset\C_{n+1}$ has the complex
ring $\K\simeq\C$ for $n+1\equiv 3\pmod{4}$, the transformations
$\cA\rightarrow\widetilde{\cA^\star}$, $\cA\rightarrow\overline{\cA}$
and $\cA\rightarrow\overline{\widetilde{\cA^\star}}$ are transferred
from $\C_{n+1}$ to $\C_n$. In this case we have 
the quotient group\index{group!quotient}
$\pin^\epsilon(n,\C)\sim\pin^{c,d,g}(n,\C)$. 
The automorphism group\index{group!automorphism}
of the quotient algebra ${}^\epsilon\C_n$ is 
$\Aut({}^\epsilon\C_n)=\{\Id,\widetilde{\star},\overline{\phantom{cc}},
\overline{\widetilde{\star}}\}\sim\{1,PT,C,CPT\}$ with the following
multiplication table:
\[
{\renewcommand{\arraystretch}{1.4}
\begin{tabular}{|c||c|c|c|c|}\hline
   & $1$ & $PT$ & $C$ & $CPT$ \\ \hline\hline
$1$& $1$ & $PT$ & $C$ & $CPT$ \\ \hline
$PT$& $PT$ & $1$ & $CPT$ & $C$ \\ \hline
$C$ & $C$ & $CPT$ & $1$ & $PT$ \\ \hline
$CPT$ & $CPT$ & $C$ & $PT$ & $1$ \\ \hline
\end{tabular}\;\sim\;
\begin{tabular}{|c||c|c|c|c|}\hline
    & $\Id$ & $\widetilde{\star}$ & $\overline{\phantom{cc}}$ &
$\overline{\widetilde{\star}}$ \\ \hline\hline
$\Id$& $\Id$ & $\widetilde{\star}$ & $\overline{\phantom{cc}}$ &
$\overline{\widetilde{\star}}$ \\ \hline
$\widetilde{\star}$ & $\widetilde{\star}$ & $\Id$ &
$\overline{\widetilde{\star}}$ & $\overline{\phantom{cc}}$ \\ \hline
$\overline{\phantom{cc}}$ & $\overline{\phantom{cc}}$ &
$\overline{\widetilde{\star}}$ & $\Id$ & $\widetilde{\star}$ \\ \hline
$\overline{\widetilde{\star}}$ & $\overline{\widetilde{\star}}$ &
$\overline{\phantom{cc}}$ & $\widetilde{\star}$ & $\Id$ \\ \hline
\end{tabular}.
}
\]
From the table it follows that $\Aut({}^\epsilon\C_n)=\{\Id,
\widetilde{\star},\overline{\phantom{cc}},\overline{\widetilde{\star}}\}
\simeq\dZ_2\otimes\dZ_2\subset\Ext(\C_n)$. When $(PT)^2=C^2=(CPT)^2=\pm 1$
and $(PT)C=-C(PT)$ we have a group $\sAut({}^\epsilon\C_n)=
\{\sI,\sC,\Pi,\sF\}\subset\sExt(\C_n)$. Therefore,
$\pin^{c,d,g}(n,\C)\simeq(\spin_+(n,\C)\odot C^{c,d,g})/\dZ_2$, where
$C^{c,d,g}$ are the four double coverings of $\dZ_2\otimes\dZ_2$.
As in the previous case, since $\sAut({}^\epsilon\C_n)=\{\sI,\sC,\Pi,\sF\}$
is the subgroup of $\sExt(\C_n)$, then we can define a group structure of
$\sAut({}^\epsilon\C_n)$ using Theorem \ref{tautext}. In this case we see that
there are four different types of $\{\sI,\sC,\Pi,\sF\}$:
\begin{eqnarray}
\sAut^1({}^\epsilon\C_n)&=&\{\sI,\cE_{i_1i_2\cdots i_{p+q-k}},
\cE_{\alpha_1\alpha_2\cdots\alpha_a},\cE_{d_1d_2\cdots d_g}\},\nonumber\\
\sAut^2({}^\epsilon\C_n)&=&\{\sI,\cE_{i_1i_2\cdots i_{p+q-k}},
\cE_{\beta_1\beta_2\cdots\beta_b},\cE_{c_1c_2\cdots c_s}\},\nonumber\\
\sAut^3({}^\epsilon\C_n)&=&\{\sI,\cE_{j_1j_2\cdots j_k},
\cE_{\alpha_1\alpha_2\cdots\alpha_a},\cE_{c_1c_2\cdots c_s}\},\nonumber\\
\sAut^4({}^\epsilon\C_n)&=&\{\sI,\cE_{j_1j_2\cdots j_k},
\cE_{\beta_1\beta_2\cdots\beta_b},\cE_{d_1d_2\cdots d_g}\}.\nonumber
\end{eqnarray}
In the group $\sAut^1({}^\epsilon\C_n)$ the element
$\Pi=\cE_{\alpha_1\alpha_2\cdots\alpha_a}$ commutes with
$\sF=\cE_{d_1d_2\cdots d_g}$ at $l\equiv 0\pmod{2}$ (see (\ref{CPT42})) and
also commutes with $\sC=\cE_{i_1i_2\cdots i_{p+q-k}}$ at
$l(m+v)\equiv 0\pmod{2}$. In turn, from (\ref{CPT56}) it follows that
$\sF\in\sAut^1({}^\epsilon\C_n)$ commutes with $\sC$ at 
$v(m+l)\equiv 0\pmod{2}$. Therefore, at the conditions
$l,l(m+v),v(m+l)\equiv 0\pmod{2}$ we have the groups $\dZ_2\otimes\dZ_2$
and $\dZ_4$, otherwise $\sAut^1({}^\epsilon\C_n)$ coincides with
$Q_4/\dZ_2$, $D_4/\dZ_2$. It is easy to verify that
$\sAut^2({}^\epsilon\C_n)$, $\sAut^3({}^\epsilon\C_n)$ and
$\sAut^4({}^\epsilon\C_n)$ have the analogous structure.

When the subalgebra $\cl_{p,q}\subset\C_{n+1}$ has the double ring
$\K\simeq\R\oplus\R$, where $p-q\equiv 1\pmod{8}$ and
$n+1\equiv 3\pmod{4}$, the transformations
$\cA\rightarrow\widetilde{\cA^\star}$, $\cA\rightarrow\overline{\cA^\star}$
and $\cA\rightarrow\overline{\widetilde{\cA}}$ are transferred from
$\C_{n+1}$ to $\C_n$. In accordance with Theorem \ref{tpseudo} the
pseudoautomorphism $\cA\rightarrow\overline{\cA}$ is reduced to the
identical transformation in case of the real ring $\K\simeq\R$. Therefore,
in this case the transformations $\cA\rightarrow\overline{\cA^\star}$ and
$\cA\rightarrow\overline{\widetilde{\cA}}$ are reduced to
$\cA\rightarrow\cA^\star$ and $\cA\rightarrow\widetilde{\cA}$, respectively.
Thus, we have the quotient group $\pin^\epsilon(n,\C)\sim\pin^{a,b,c}(n,\C)$
and the automorphism group of ${}^\epsilon\C_n$ is
$\Aut({}^\epsilon\C_n)=\{\Id,\star,\widetilde{\phantom{cc}},
\widetilde{\star}\}\sim\{1,P,T,PT\}$ 
(reflection group\index{group!reflection}, see Proposition
\ref{prop1}) with the multiplication table
\begin{equation}\label{Cayley2}
{\renewcommand{\arraystretch}{1.4}
\begin{tabular}{|c||c|c|c|c|}\hline
   & $1$ & $P$ & $T$ & $PT$ \\ \hline\hline
$1$& $1$ & $P$ & $T$ & $PT$ \\ \hline
$P$& $P$ & $1$ & $PT$ & $T$ \\ \hline
$T$ & $T$ & $PT$ & $1$ & $P$ \\ \hline
$PT$ & $PT$ & $T$ & $P$ & $1$ \\ \hline
\end{tabular}\;\sim\;
\begin{tabular}{|c||c|c|c|c|}\hline
   & $\Id$ & $\star$ & $\widetilde{\phantom{cc}}$ & 
$\widetilde{\star}$ \\ \hline\hline
$\Id$ & $\Id$ & $\star$ & $\widetilde{\phantom{cc}}$ & 
$\widetilde{\star}$ \\ \hline
$\star$ & $\star$ & $\Id$ & $\widetilde{\star}$ & 
$\widetilde{\phantom{cc}}$ \\ \hline
$\widetilde{\phantom{cc}}$ & $\widetilde{\phantom{cc}}$ &
$\widetilde{\star}$ & $\Id$ & $\star$ \\ \hline
$\widetilde{\star}$ & $\widetilde{\star}$ & $\widetilde{\phantom{cc}}$ &
$\star$ & $\Id$ \\ \hline
\end{tabular}.
}
\end{equation}
The structure of $\Aut({}^\epsilon\C_n)=\{\Id,\star,\widetilde{\phantom{cc}},
\widetilde{\star}\}=\{\sI,\sW,\sE,\sC\}$ is well studied (see
Theorem \ref{tautr}). Thus, in this case we come again to the
D\c{a}browski groups\index{group!D\c{a}browski} $\pin^{a,b,c}(n,\C)
\simeq(\spin_+(n,\C)\odot C^{a,b,c})/\dZ_2$, where $C^{a,b,c}$ are the four
double coverings of $\dZ_2\otimes\dZ_2$ (see Theorem \ref{tgroupr}).

Further, when the subalgebra $\cl_{p,q}\subset\C_{n+1}$ has the double
quaternionic ring $\K\simeq\BH\oplus\BH$, where $p-q\equiv 5\pmod{8}$ and
$n+1\equiv 3\pmod{4}$, the transformations 
$\cA\rightarrow\widetilde{\cA^\star}$, $\cA\rightarrow\overline{\cA^\star}$
and $\cA\rightarrow\overline{\widetilde{\cA}}$ are transferred from
$\C_{n+1}$ to $\C_n$. In this case we have the quotient group
$\pin^\epsilon(n,\C)\sim\pin^{c,e,f}(n,\C)$. The automorphism group of
the quotient algebra ${}^\epsilon\C_n$ is 
$\Aut({}^\epsilon\C_n)=\{\Id,\widetilde{\star},\overline{\star},
\overline{\widetilde{\phantom{cc}}}\}\sim\{1,PT,CP,CT\}$ with the
following multiplication table:
\[
{\renewcommand{\arraystretch}{1.4}
\begin{tabular}{|c||c|c|c|c|}\hline
   & $1$ & $PT$ & $CP$ & $CT$ \\ \hline\hline
$1$& $1$ & $PT$ & $CP$ & $CT$ \\ \hline
$PT$& $PT$ & $1$ & $CT$ & $CP$ \\ \hline
$CP$ & $CP$ & $CT$ & $1$ & $PT$ \\ \hline
$CT$ & $CT$ & $CP$ & $PT$ & $1$ \\ \hline
\end{tabular}\;\sim\;
\begin{tabular}{|c||c|c|c|c|}\hline
   & $\Id$ & $\widetilde{\star}$ & $\overline{\star}$ &
$\overline{\widetilde{\phantom{cc}}}$ \\ \hline\hline
$\Id$ & $\Id$ & $\widetilde{\star}$ & $\overline{\star}$ &
$\overline{\widetilde{\phantom{cc}}}$ \\ \hline
$\widetilde{\star}$ & $\widetilde{\star}$ & $\Id$ &
$\overline{\widetilde{\phantom{cc}}}$ & $\overline{\star}$ \\ \hline
$\overline{\star}$ & $\overline{\star}$ &
$\overline{\widetilde{\phantom{cc}}}$ & $\Id$ &
$\widetilde{\star}$ \\ \hline
$\overline{\widetilde{\phantom{cc}}}$ & $\overline{\widetilde{\phantom{cc}}}$ &
$\overline{\star}$ & $\widetilde{\star}$ & $\Id$ \\ \hline
\end{tabular}.
}
\]\begin{sloppypar}\noindent
From this table it follows that $\Aut({}^\epsilon\C_n)=\{\Id,
\widetilde{\star},\overline{\star},\overline{\widetilde{\phantom{cc}}}\}
\simeq\dZ_2\otimes\dZ_2\subset\Ext(\C_n)$. When $(PT)^2=(CP)^2=(CT)^2=\pm 1$
and $(PT)(CT)=-(CP)(PT)$ we have a group $\sAut({}^\epsilon\C_n)=
\{\sI,\sC,\sK,\sS\}\subset\sExt(\C_n)$. Therefore, 
$\pin^{c,e,f}(n,\C)\simeq(\spin_+(n,\C)\odot C^{c,e,f})/\dZ_2$.
Since $\sAut({}^\epsilon\C_n)=\{\sI,\sC,\sK,\sS\}$ is the subgroup of
$\sExt(\C_n)$, then using Theorem \ref{tautext} we see that there are four
different types of $\{\sI,\sC,\sK,\sS\}$:\end{sloppypar}
\begin{eqnarray}
\sAut^1({}^\epsilon\C_n)&=&\{\sI,\cE_{i_1i_2\cdots i_{p+q-k}},
\cE_{\beta_1\beta_2\cdots\beta_b},\cE_{c_1c_2\cdots c_s}\},\nonumber\\
\sAut^2({}^\epsilon\C_n)&=&\{\sI,\cE_{i_1i_2\cdots i_{p+q-k}},
\cE_{\alpha_1\alpha_2\cdots\alpha_a},\cE_{d_1d_2\cdots d_g}\},\nonumber\\
\sAut^3({}^\epsilon\C_n)&=&\{\sI,\cE_{j_1j_2\cdots j_k},
\cE_{\beta_1\beta_2\cdots\beta_b},\cE_{d_1d_2\cdots d_g}\},\nonumber\\
\sAut^4({}^\epsilon\C_n)&=&\{\sI,\cE_{j_1j_2\cdots j_k},
\cE_{\alpha_1\alpha_2\cdots\alpha_a},\cE_{c_1c_2\cdots c_s}\}.\nonumber
\end{eqnarray}
It is easy to see that a structure of $\sAut({}^\epsilon\C_n)=
\{\sI,\sC,\sK,\sS\}$ is analogous to the structure of other subgroups of
$\sExt(\C_n)$ such as $\{\sI,\sE,\sK,\sF\}$, $\{\sI,\sC,\Pi,\sF\}$ and
$\{\sI,\sW,\sE,\sC\}$.

Let us consider now the quotient groups $\pin^\epsilon(p,q-1)$,
$\pin^\epsilon(q,p-1)$ over the field $\F=\R$. First of all, when the
algebra $\cl_{p,q}$ at $q\equiv 0\pmod{2}$ has the double division ring
$\K\simeq\R\oplus\R$, $p-q\equiv 1\pmod{8}$, the transformations
$\cA\rightarrow\widetilde{\cA}$, $\cA\rightarrow\overline{\cA}$ and
$\cA\rightarrow\overline{\widetilde{\cA}}$ are transferred from $\cl_{p,q}$
to $\cl_{p,q-1}$ or $\cl_{q,p-1}$, where $\cl_{p,q-1}$ and 
$\cl_{q,p-1}$ are of the type $p-q\equiv 0\pmod{8}$ or
$p-q\equiv 2\pmod{8}$. However, in accordance with Theorem \ref{tpseudo},
the pseudoautomorphism $\cA\rightarrow\overline{\cA}$ is equivalent to
$\Id$ for the ring $\K\simeq\R$. Therefore, the transformations
$\cA\rightarrow\overline{\cA}$ and 
$\cA\rightarrow\overline{\widetilde{\cA}}$ are reduced to $\Id$ and
$\cA\rightarrow\widetilde{\cA}$, respectively. For that reason an
automorphism group of the quotient algebras ${}^\epsilon\cl_{p,q-1}$ and
${}^\epsilon\cl_{q,p-1}$ is equivalent to $\{\Id,\widetilde{\phantom{cc}}\}
\simeq\{1,T\}\simeq\dZ_2$ with the multiplication table (\ref{Cayley1}).
Thus, we have the quotient groups $\pin^b(p,q-1)\simeq(\spin_+(p,q-1)\odot
\dZ_2\otimes\dZ_2)/\dZ_2$ and $\pin^b(q,p-1)\simeq(\spin_+(q,p-1)\odot
\dZ_2\otimes\dZ_2)/\dZ_2$. In the case when $q\equiv 1\pmod{2}$
the automorphisms $\cA\rightarrow\widetilde{\cA}$,
$\cA\rightarrow\overline{\cA^\star}$ and 
$\cA\rightarrow\overline{\widetilde{\cA^\star}}$ are transferred from
$\cl_{p,q}$ to $\cl_{p,q-1}$ or $\cl_{q,p-1}$. Over the ring $\K\simeq\R$
the transformations $\cA\rightarrow\overline{\cA^\star}$ and
$\cA\rightarrow\overline{\widetilde{\cA^\star}}$ are reduced to
$\cA\rightarrow\cA^\star$ and $\cA\rightarrow\widetilde{\cA^\star}$,
respectively. Therefore, the automorphism groups of ${}^\epsilon\cl_{p,q-1}$
(${}^\epsilon\cl_{q,p-1}$) is equivalent to 
$\{\Id,\star,\widetilde{\phantom{cc}},\widetilde{\star}\}\simeq
\{1,P,T,PT\}$ with the multiplication table (\ref{Cayley2}). Thus, in this
case we have the quotient groups $\pin^{a,b,c}(p,q-1)\simeq(\spin_+(p,q-1)
\odot C^{a,b,c})/\dZ_2$ and $\pin^{a,b,c}(q,p-1)\simeq(\spin_+(q,p-1)\odot
C^{a,b,c})/\dZ_2$, where $C^{a,b,c}$ are the four double coverings
of the Gauss-Klein group $\dZ_2\otimes\dZ_2$.

Further, when the algebra $\cl_{p,q}$ at $q\equiv 0\pmod{2}$ has the
double quaternionic division ring $\K\simeq\BH\oplus\BH$, the automorphisms
$\cA\rightarrow\widetilde{\cA}$, $\cA\rightarrow\overline{\cA}$ and
$\cA\rightarrow\overline{\widetilde{\cA}}$ are transferred from $\cl_{p,q}$
to $\cl_{p,q-1}$ or $\cl_{q,p-1}$, where subalgebras $\cl_{p,q-1}$ and
$\cl_{q,p-1}$ are of the type $p-q\equiv 4\pmod{8}$ or $p-q\equiv 6\pmod{8}$.
In this case we have the quotient groups $\pin^\epsilon(p,q-1)\sim
\pin^{b,d,f}(p,q-1)$ and $\pin^\epsilon(q,p-1)\sim\pin^{b,d,f}(q,p-1)$.
The automorphism group\index{group!automorphism}
of the quotient algebras ${}^\epsilon\cl_{p,q-1}$ and
${}^\epsilon\cl_{q,p-1}$ is
$\{\Id,\widetilde{\phantom{cc}},\overline{\phantom{cc}},
\overline{\widetilde{\phantom{cc}}}\}\sim\{1,T,C,CT\}$ with the following
multiplication table:
\[
{\renewcommand{\arraystretch}{1.4}
\begin{tabular}{|c||c|c|c|c|}\hline
   & $1$ & $T$ & $C$ & $CT$ \\ \hline\hline
$1$& $1$ & $T$ & $C$ & $CT$ \\ \hline
$T$& $T$ & $1$ & $CT$ & $C$ \\ \hline
$C$ & $C$ & $CT$ & $1$ & $T$ \\ \hline
$CT$ & $CT$ & $C$ & $T$ & $1$ \\ \hline
\end{tabular}\;\sim\;
\begin{tabular}{|c||c|c|c|c|}\hline
   & $\Id$ & $\widetilde{\phantom{cc}}$ & $\overline{\phantom{cc}}$ &
$\overline{\widetilde{\phantom{cc}}}$ \\ \hline\hline
$\Id$ & $\Id$ & $\widetilde{\phantom{cc}}$ & $\overline{\phantom{cc}}$ &
$\overline{\widetilde{\phantom{cc}}}$ \\ \hline
$\widetilde{\phantom{cc}}$ & $\widetilde{\phantom{cc}}$ & $\Id$ &
$\overline{\widetilde{\phantom{cc}}}$ & $\overline{\phantom{cc}}$ \\ \hline
$\overline{\phantom{cc}}$ & $\overline{\phantom{cc}}$ &
$\overline{\widetilde{\phantom{cc}}}$ & $\Id$ & 
$\widetilde{\phantom{cc}}$ \\ \hline
$\overline{\widetilde{\phantom{cc}}}$ & $\overline{\widetilde{\phantom{cc}}}$ &
$\overline{\phantom{cc}}$ & $\widetilde{\phantom{cc}}$ & $\Id$ \\ \hline
\end{tabular}.
}
\]
From this table we see that $\{\Id,\widetilde{\phantom{cc}},
\overline{\phantom{cc}},\overline{\widetilde{\phantom{cc}}}\}\simeq
\dZ_2\otimes\dZ_2\subset\Ext(\cl_{p,q})$. When $T^2=C^2=(CT)^2=\pm 1$ and
$TC=-CT$ we have a group $\{\sI,\sE,\Pi,\sS\}\subset\sExt(\cl_{p,q})$.
It is easy to see that a structure of $\{\sI,\sE,\Pi,\sS\}$ is analogous to
the structure of other subgroups of $\sExt$ such as $\{\sI,\sE,\sK,\sF\}$,
$\{\sI,\sC,\Pi,\sF\}$, $\{\sI,\sW,\sE,\sC\}$ and $\{\sI,\sC,\sK,\sS\}$.
Therefore, $\pin^{b,d,f}(p,q-1)\simeq(\spin_+(p,q-1)\odot C^{b,d,f})/\dZ_2$
and $\pin^{b,d,f}(q,p-1)\simeq(\spin_+(q,p-1)\odot C^{b,d,f})/\dZ_2$,
where $C^{b,d,f}$ are the four double coverings of $\dZ_2\otimes\dZ_2$.

In the case when $q\equiv 1\pmod{2}$ the automorphisms
$\cA\rightarrow\widetilde{\cA}$, $\cA\rightarrow\overline{\cA^\star}$ and
$\cA\rightarrow\overline{\widetilde{\cA^\star}}$ are transferred from
$\cl_{p,q}$ to $\cl_{p,q-1}$ or $\cl_{q,p-1}$. Therefore, the automorphism
group of ${}^\epsilon\cl_{p,q-1}$ (${}^\epsilon\cl_{q,p-1}$) is equivalent
to $\{\Id,\widetilde{\phantom{cc}},\overline{\star},
\overline{\widetilde{\star}}\}\sim\{1,T,CP,CPT\}$ with the multiplication
table (\ref{Cayley3}). Thus, in this case we have the quotient groups
$\pin^{b,e,g}(p,q-1)\simeq(\spin_+(p,q-1)\odot C^{b,e,g})/\dZ_2$ and
$\pin^{b,e,g}(q,p-1)\simeq(\spin_+(q,p-1)\odot C^{b,e,g})/\dZ_2$.
\end{proof}

\section*{Acknowledgements}
I am deeply grateful to Prof. J. S. R. Chisholm, to Prof. L. D\c{a}browski
and Prof. A. Trautman for sending me their interesting works.

\end{document}